\documentclass[12pt]{article}
\usepackage[english]{babel}
\usepackage{inputenc}
\usepackage{csquotes}
\usepackage{float}
\usepackage{bm}
\usepackage[top=1in,bottom=1in,left=1in,right=1in,
marginparwidth=1.75cm]{geometry}
\usepackage{amsmath,amssymb}
\usepackage{listings}

\newcommand{\beginsupplement}{%
        \setcounter{table}{0}
        \renewcommand{\thetable}{S\arabic{table}}%
        \setcounter{figure}{0}
        \renewcommand{\thefigure}{S\arabic{figure}}%
     }
\usepackage{booktabs}
\usepackage{xcolor}
\usepackage{graphicx}
\usepackage{caption}
\usepackage{subcaption}
\usepackage{soul}
\usepackage{comment}
\usepackage{setspace}
\usepackage{bbm}
\usepackage[round]{natbib}

\usepackage{todonotes}
\usepackage{pdfpages}
 \usepackage{booktabs}
\usepackage{multirow}
\usepackage{longtable}
\usepackage{lscape} 
\usepackage{authblk}
\newcommand{\E}{\text{E}}

\title{\Large Instrumental Variable Approach to Estimating Individual Causal Effects in N-of-1 Trials: Application to ISTOP Study}
\author[1]{Kexin Qu}
\author[1]{Christopher H. Schmid}
\author[1]{Tao Liu}
\affil[1]{Brown University, School of Public Health}
\date{}  
\begin{document}

\maketitle

\begin{abstract}
    An N-of-1 trial is a multiple crossover trial conducted in a single individual to provide evidence to directly inform personalized treatment decisions. Advances in wearable devices greatly improved the feasibility of adopting these trials to identify optimal individual treatment plans, particularly when treatments differ among individuals and responses are highly heterogeneous. Our work was motivated by the I-STOP-AFib Study, which examined the impact of different triggers on atrial fibrillation (AF) occurrence. We described a causal framework for “N-of-1” trial using potential treatment selection paths and potential outcome paths. Two estimands of individual causal effect were defined: 1) the effect of continuous exposure, and 2) the effect of an individual’s observed behavior. We addressed three challenges: 1) imperfect compliance to the randomized treatment assignment; 2) binary treatments and binary outcomes, which led to the “non-collapsibility” issue of estimating odds ratios; and 3) serial inference in the longitudinal observations. We adopted the Bayesian IV approach where the study randomization was the IV as it impacted the patient’s choice of exposure but not directly the outcome.  Estimations were obtained through a system of two parametric Bayesian models to estimate the individual causal effect.  Our model got around the non-collapsibility and non-consistency by modeling the confounding mechanism through latent structural models and by inferring with Bayesian posterior of functionals. Autocorrelation present in the repeated measurements was also accounted for. The simulation study showed our method largely reduced bias and greatly improved the coverage of the estimated causal effect, compared to existing methods (ITT, PP, and AT). We applied the method to I-STOP-AFib Study to estimate the individual effect of alcohol on AF occurrence.
\end{abstract}

\newpage

\section*{List of Main Notations}
We adopted the $do(X=x_0)$ operator  introduced by 
\cite{pearl2009causal} to denote an action of forcing a random variable $X$ at the fixed value of $x_0$. 

\begin{itemize}
    \item $\mathbf R_{1:t}$: observed randomization sequence from period 1 to $t$.
    \item $\mathbf{X}_{11:tj}$: observed treatment selection sequence from day $1$ of period 1 to day $j$ of period $t$.
     \item $\mathbf{X}_{1:t}(\mathbf{r}_{11:tj})$ = $\mathbf{X}_{11:tj}(do (\mathbf R_{1:t} = \mathbf r_{1:t}))$: potential treatment selection sequence from day $1$ of period 1 to day $j$ of period $t$.
    \item $\mathbf{Y}_{11:tj}$: observed outcome path from day $1$ of period 1 to day $j$ of period $t$.
     \item $\mathbf{Y}_{11:tj}(\mathbf{x}_{11:tj})$ = $\mathbf{Y}_{11:tj}(do( \mathbf X_{11:tj} = \mathbf x_{11:tj}))$: potential outcome path from day $1$ of period 1 to day $j$ of period $t$ under a fixed treatment selection path $\mathbf x_{11:tj}$.
     \item  $\mathbf{Y}_{11:tj}(\mathbf{X}_{11:tj}(\mathbf{r}_{1:t}))$ = $\mathbf{Y}_{11:tj}(do( \mathbf X_{11:tj} = \mathbf X_{11:tj}(do (\mathbf{R}_{1:t} = \mathbf r_{11:tj}))$: potential outcome path from day $1$ of period 1 to day $j$ of period $t$ under a fixed randomization sequence $\mathbf r_{1:t}$.
    
\end{itemize}
\newpage
\section{Introduction}
%\subsection{N-of-1 trial, wearable device, and mHealth }

N-of-1 trials use a personalized clinical research design where a single participant receives alternative treatments multiple times in a randomized sequence to compare treatments and identify optimal strategies for individual care \citep{kravitz2004evidence,duan2013single}.  
Often referred to as a trial of ``me'', the ``N-of-1'' study design differs from traditional randomized controlled trials (RCTs) by allowing individualized treatment selection, personalized data collection, and direct interpretability for individualized medical care \citep{duan2013single}.
N-of-1 trials are useful when responses to interventions are known to vary substantially among individuals, as evidenced by studies showing that many prescribed medications do little to help certain participants \citep{schork2015personalized,kravitz2004evidence,goldenberg1996randomized}. 
Because traditional RCTs focus on estimating the average treatment effect in a population, they may fail to identify the best treatment for each individual. The N-of-1 trial is an important tool for individuals to systematically compare and quantify the effects of alternative treatments and to make an appropriate, personalized healthcare decision. \citep{kravitz2008ever, gabler2011n, cook1996randomized, shadish2011characteristics, kratochwill2014enhancing}

Rapid advances in smartphones and wearable devices have created many opportunities for N-of-1 research in the era of personalized mHealth \citep[e.g.,][]{bulte2013single, declercq2020multisced, manolov2017can}. 
Many tools have emerged for collecting and analyzing single-case data, including mobile apps such as \textit{StudyU} \citep{konigorski2022studyu} and \textit{StudyMe} \citep{zenner2021studyme} and SCD-MVA \citep{moeyaert2021scd}.  
These developments hold great potential for addressing relevant questions in the field of precision medicine \citep{schork2023exploring}.

Data from N-of-1 studies are time series in which repeated measurements are taken during multiple treatment periods constructed as a crossover design for each participant.
In an N-of-1 trial, one is interested in drawing inferences about the participant-specific treatment effect, which is estimated by aggregating the participant’s repeated measurements.

%When treatments are randomized over time, traditional analyses typically use a multilevel modeling approach \cite[e.g.,][]{marcus2021individualized, kravitz2018effect} based on the ``intention-to-treat'' (ITT), ``as treated'' (AT), or ``per protocol'' (PP) principles \citep{friedman2015fundamentals,gupta2011intention}. 
%However, with imperfect adherence to randomization, none of these methods produces the causal effect of comparative treatments.  
%In the current study, we implement study randomization as an instrumental variable (IV) and propose a latent Bayesian structural estimation method to estimate individual causal effects of exposure to self-selected triggers. 

The causal inference literature has primarily focused on quantifying causal treatment effects at a population level among a well-defined group of people \citep{rubin1974estimating, rubin2005causal, little2000causal}.  
A fundamental problem of causal inference is to compare \textit{potential outcomes} under different actions, e.g., treatment versus control conditions, that cannot occur simultaneously. This poses the challenge that all possible outcomes are not observed at the same time \citep{rubin1978bayesian, rubin1980randomization, rubin1990comment, rubin1978bayesian, holland1986statistics, imbens2015causal}. 
Commonly used causal inference methods rely on randomization or assume that some versions of ignorability conditions hold \citep{rubin1978bayesian, pearl2000models, hernan2006instruments}, such that participants under different exposures are comparable conditional on a set of observed pretreatment variables. 
Another class of methods that can be used when the ignorability condition does not hold relies on finding a valid instrumental variable (IV), a variable that varies externally and affects outcome only indirectly through its impact on exposure.
We provide a brief review of IV methods in Section \ref{sec:review.iv}. 

N-of-1 trials provide an opportunity to apply causal inference at an individual level, where individuals serve as their own controls.  
Research on individual causal inference in N-of-1 studies is limited but growing. 
\cite{daza2018causal} proposed a counterfactual framework to estimate the average period treatment effect (APTE) using g-formula and inverse-probability-weighting (IPW) methods; see also \cite{daza2019person} and \cite{daza2022model}. 
\cite{bojinov2019time} used full potential paths to define causal estimands in the context of sequential randomization with perfect compliance.
\cite{van2018robust} proposed a targeted maximum likelihood estimation method for learning the optimal treatment rule for a single participant in an adaptive design, with recent developments in doubly robust estimation \citep{malenica2021adaptive}.
\cite{neto2016towards} adopted an instrumental variable approach to estimate the time-invariant personalized causal treatment effect and developed a randomization test of the null hypothesis of no causal effect as well as confidence intervals for the causal effects. 
Their work relies on the mechanism-based account of causation by  \cite{pearl2000models} instead of Rubin’s potential outcome framework \citep{rubin1978bayesian,rubin1977assignment,rubin1976inference}. 
A key distinction between our work and the study by \cite{neto2016towards} is in the design of the N-of-1 study. In their study, randomization takes place at each time point, whereas in a typical N-of-1 trial, randomization occurs at the beginning of each period, with all subsequent measurements in each period following the same randomized treatment. Our proposed causal framework describes a typical N-of-1 trial.

This paper was motivated by the I-STOP-AFib study \citep{marcus2021individualized}, which was conducted to evaluate the impact of various potential triggers (alcohol use, smoking, etc.) on the occurrence of atrial fibrillation (AF), a condition characterized by rapid heart rhythms that increases the risk of stroke and heart failure. 
I-STOP-AFib used a mobile application developed on the \textit{Eureka} platform \citep{Eureka2022} to collect data and implemented an N-of-1 study design in one study arm. During the primary study phase, a study participant chose a trigger of their interest. The exposure (denoted as “ON” or “OFF” condition) to the chosen trigger was randomized over 6 weeks, with each condition lasting one week. More detail about the I-STOP-AFib study is provided in Section~\ref{sec:Afib.study}. 

Obtaining a causal effect of the chosen trigger for each individual in the I-STOP-AFib study presents several challenges. 
First, individuals did not always comply with their exposure condition. 
This is similar to noncompliance in RCTs, where a study participant chooses to undertake treatment different from their assigned one. 
Noncompliance violates the assumption that participants receive interventions randomly.
One approach to analyzing such data compares the outcomes grouped by the randomization assignment regardless of the actual treatments received, which is referred to as the \textit{intention-to-treat} (ITT) analysis \cite[e.g.,][] {marcus2021individualized, kravitz2018effect}. 
A limitation of ITT is that it evaluates only the causal effect of \textit{treatment randomization}, not the direct causal effect comparing the alternative treatments themselves \citep{sheiner1995intention,sheiner2002intent, fergusson2002post,heritier2003inclusion, hollis1999meant,wertz1995intention,moncur2009clinical, little1996intent,kleinman1998bayesian,gupta2011intention}. 
Other approaches such as as-treated (AT) analysis (comparing participants' outcomes according to the treatment actually taken ignoring the randomization) and 
per-protocol (PP) analysis (focusing only on those who strictly complied with the treatment assignment) do not have causal interpretations in general because their definitions rely on a post-randomization treatment choice.

The second challenge is that the study outcomes are not independent. The non-interference assumption among study outcomes (usually satisfied in RCTs) is not plausible in our time series data because a participant's outcome and treatment selection at a given time can be affected by past values. Repeated measurements lead to temporal interference and autocorrelation \citep{wang2021causal, daza2018causal, daza2022model}, features which we also need to account for in a model to properly quantify the uncertainty of parameter estimates. To define a causal estimand for a single participant requires consideration of the entire treatment and outcome paths in order to define the marginal causal effect as an average over all possible historical paths
\cite[e.g.,][]{bojinov2019time,daza2022model}. 

Third, when study outcomes are binary, e.g., occurrence of AF in our study, treatment effects quantified by odds ratios (OR) or risk ratios (RR) are not collapsible \citep{freedman2008randomization}. 
The issue of non-collapsibility has been discussed in depth in the literature \cite[e.g.,][]{wooldridge2010econometric,freedman2008randomization,gail1984biased,austin2007performance, mood2010logistic}.  
In a study with multiple participants, the individual treatment effects, even if they are the same, can be different from the population (marginal) treatment effect, a phenomenon that arises because OR and RR are nonlinear functions of probabilities 
\citep{greenland1999confounding}.

In this paper, we propose a two-stage Bayesian latent structural model, utilizing study treatment randomization as an IV, to analyze data from the I-STOP-AFib study. Our paper makes the following contributions to the literature. First, we introduce two causal estimands, both relevant in the context of our study: the causal effect of continuous exposure to a potential trigger (e.g., alcohol) and the causal effect of an individual’s observed exposure or usual behavior on AF, in contrast with the counterfactual if the individual had fully abstained from the exposure. Second, our Bayesian IV structural model is developed by positing probit models that link the treatment selection and AF outcomes with two corresponding latent variables.
The latent probit models an error term to account for confounding due to self-selection of exposure (noncompliance) and a longitudinal error term to describe the auto-correlation in outcomes.
We propose to use the Bayesian method to directly obtain the posterior distribution of the functions of model parameters to solve the issue of non-collapsibility.
Finally, we describe how to extend the Bayesian hierarchical model structure to analyze data from multiple N-of-1 studies in a meta-analysis that enables estimating an average treatment effect and improving estimates of individual effects when individuals are exchangeable.

\section{Brief Review of IV methods}\label{sec:review.iv}
Instrumental variables have been used for estimating causal effects when the selection mechanism of exposure is nonrandom.
In a simplified setting, let $R$ denote an IV (e.g.,\ randomization assignment indicator), $X$ a treatment or exposure of interest, $Y$ an outcome, and $U$ other pretreatment factors (measured and unmeasured) that affect both the outcome and actual treatment selection; see Supplemental Figure \ref{fig:iv_DAG}.
For $R$ to be a valid IV, three conditions are needed: (i) $R$ has a causal effect on $X$, (ii) $R$ has no \textit{direct} effect on the outcome $Y$; it affects $Y$ only through $X$ (exclusion restriction), and (iii) $R$ does not share common causes with the outcome $Y$ (i.e., no confounding for the effect of $R$ on $Y$) \citep{hernan2006instruments}.  

IV analysis enables us to estimate the causal effects of $X$ on $Y$, even when we do not observe all the data on confounders $U$.  
IV estimation methods for continuous outcomes are well established, e.g., the conventional two-stage least-squares IV estimator  \citep{kent2006multivariate}. 
IV estimation for binary outcomes poses challenges due to the need for additional assumptions about the data-generating process or specific modeling assumptions \citep{vansteelandt2011instrumental,clarke2012instrumental}.
Depending on targeting parameters and model assumptions \citep{clarke2012instrumental,didelez2010assumptions}, the proposed IV estimators for binary outcomes can be grouped into four classes: (a) bounds estimation \citep{balke1997bounds, chesher2010instrumental}; (b) fully parametric estimation \citep{heckman1999local,kleibergen1998bayesian,kleibergen2003bayesian}; (c) causal effect among a subpopulation (i.e., among treated) using semiparametric structural mean models (SMMs); and (d) local causal effect estimation based on a monotonicity assumption and principal stratification \citep{frangakis2002principal, angrist1995identification, didelez2010assumptions, palmer2008adjusting, clarke2012instrumental}.   
%Without assuming monotonicity, the Wald OR or RR estimators target the population COR or CRR, but are not consistent when both the outcome and the exposure variable are dichotomous.
Fully parametric methods fully describe the data-generating processes and are widely used in economics \citep{heckman1999local,kleibergen1998bayesian,kleibergen2003bayesian}. %cite 25, 26 in Gang li

In this paper, we consider a fully parametric Bayesian estimation approach.  
\cite{kleibergen2003bayesian} conducted theoretical comparisons between Bayesian and classical frequentist IV approaches for continuous outcomes, emphasizing their efficiencies and robustness. 
\cite{lopes2014bayesian} provided a comprehensive review of parametric Bayesian IV estimation for continuous outcomes. 
We choose the Bayesian approach for its major advantage of being flexible for analyzing data with complex structures and allowing for inferences on the posterior distributions of nonlinear functions of model parameters, such as odds ratios and risk ratios.

\section{Motivating Example: I-STOP-AFib study}\label{sec:Afib.study}

The I-STOP-AFib study~\citep{marcus2022individualized} included a collection of N-of-1 studies investigating the potential impact of participants' self-selected triggers on paroxysmal atrial fibrillation. AF is a common arrhythmia that affects otherwise healthy high-functioning individuals and has a lifetime risk of 1 in 4. It is estimated that by 2050, 16 million Americans will be affected by AF \citep{go2001prevalence}. Despite its prevalence, behaviors or exposures that may trigger AF episodes have not been systematically studied.

The I-STOP-AFib study included two phases. A primary phase of 10 weeks compared the use of N-of-1 trigger testing to surveillance followed by an optional phase in which members of both groups could test triggers in additional N-of-1 trials. 
Participants in the surveillance arm only tracked their outcomes in the primary phase. During N-of-1 trials, participants tested one trigger of their choice (e.g., alcohol, caffeine, exercises, large meals, cold food). 
The study aimed to: (a) examine the comparative effectiveness of use of N-of-1 trigger tests versus symptom surveillance (data tracking alone) for reducing AF frequency and severity (as well as for changing life quality); and (b) evaluate whether particular self-selected triggers increased AF among participants in the N-of-1 arm.

\begin{figure}[hbt!]
    \centering
    \includegraphics[width=0.95\textwidth]{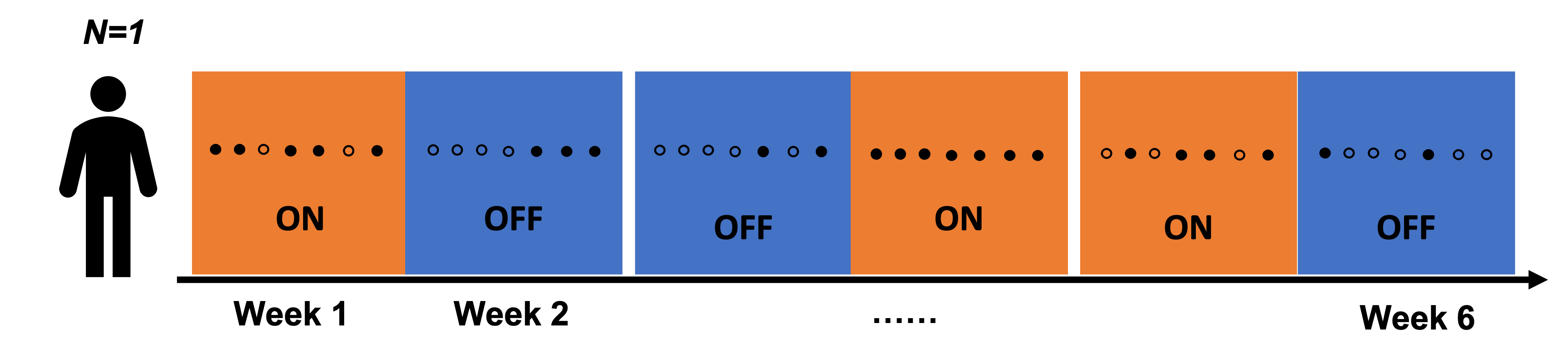}
    \caption{An example sequence of the 6-week primary period in the I-STOP-AFib trial.  Trigger on: trigger exposure; trigger off: trigger elimination. Circles represent daily AF outcomes. Filled circles are days on which an AF event occurred and unfilled circles are days with no AF event. }
    \label{fig:design}
\end{figure}

This paper focuses on the second aim, i.e., to estimate the causal effect of a self-selected trigger on AF for individual participants. 
We use the data from participants who chose to test alcohol in the N-of-1 arm in the primary phase where compliance was measured. 
This arm consists of multiple N-of-1 trials, where in each trial, a participant selected a trigger of interest and tested it over six periods each lasting one week. 
The six periods were randomly assigned to three periods of trigger exposure (trigger `ON' condition where participants were allowed to expose themselves to the trigger) and three periods without exposure (trigger `OFF' condition where participants were supposed to avoid the trigger) in blocks of two weeks such that each condition occurred once in each block (see Figure \ref{fig:design}). 
Participants reported their trigger exposure as well as AF occurrence through the Eureka mobile application \citep{Eureka2022}, along with repeated measurements of other covariates of interest such as mood or sleep quality. 

\begin{figure}[!ht]
     \centering
      \includegraphics[width=0.8\textwidth]{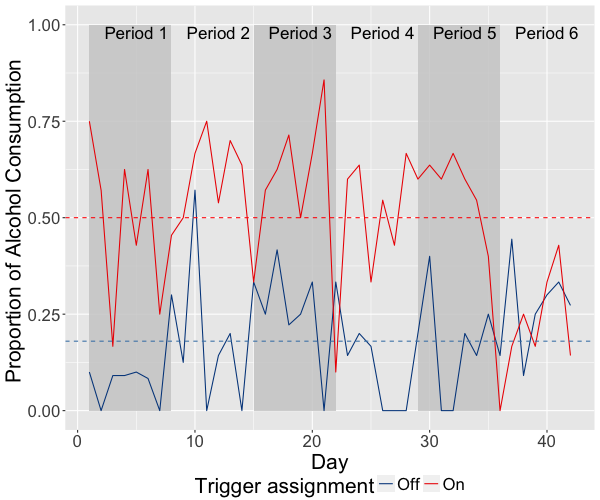}
      \label{42day}
        \caption{Proportions of participants who consumed alcohol on a given day on which they were randomized to trigger "ON”  (red, allowed to drink alcohol) or to trigger “OFF” (blue, not allowed to drink alcohol) over 42 days of 6 periods. Each participant was randomized to 3 ‘ON’ periods and 3 ‘OFF’ periods in random order. Dashed lines represent the average proportions of alcohol consumption over the 42 day trial period.}
        \label{fig:comp_time}
\end{figure}

Alcohol was the most common trigger chosen and we analyzed data from the 51 individuals who chose to test alcohol during the primary phase of the study. 
%Self-selection of exposure complicated the trigger's causal effect estimation. 
%We define participants to be `compliant' when they chose not to consume alcohol when they were suggested to do so or to drink when asked to abstain. 
As shown in Figure \ref{fig:comp_time}, study randomization to trigger ON and OFF conditions had a differential impact on participants' alcohol consumption but did not fully determine their drinking behaviors. The average daily alcohol consumption rate during trigger ON and OFF periods varied from 10-80\% and from 10-60\%, respectively. The overall average rates are at 50\% and 20\% represented by two dashed lines.

\section{Methods}

We introduce notation and outline our assumptions in the context of the I-STOP-AFib study. 
To facilitate introduction of models, we refer to the relationship between a trigger (exposure) and outcome as the \textit{scientific process} and the process of how a study participant chooses an exposure as the \textit{selection process} \citep{rubin2008objective,clarke2012instrumental}. 
We outline the model assumptions for the two processes separately.  

\subsection{Notation and Definition} \label{sec:notation}

We begin with one N-of-1 trial, aiming to infer the causal effect of a self-selected trigger on the individual participant. 
Later in Section~\ref{sec:meta.analysis}, we extend this approach to a meta-analysis of data from multiple N-of-1 trials.
Our formulation of potential outcome paths is based on \cite{bojinov2019time}, and we extend it to account for potential treatment selection paths with imperfect compliance to randomization. 
Conceptualizing treatments and potential outcomes as paths in longitudinal analyses was introduced by \cite{robins1999estimation} and later applied by other researchers \citep[][etc.]{brodersen2015inferring, blackwell2018make}. 

(a) \textbf{Selection process. } 
Let $\mathbf{R}_{1:t} = (R_1, \dots, R_t) \in \{0, 1\}^t$ denote a vector of treatment randomization assignment over the first $t$ periods of $T$ potential periods of a trial.
For the I-STOP-AFib primary phase, each N-of-1 trial had a total of $T= 6$ periods, each lasting for a week. 
To ensure balance, the I-STOP-AFib used block randomization (block size = 2) to randomly assign three trigger ON ($R_t=1$) and three trigger OFF ($R_t=0$) conditions. 
$R_t\overset{\mathrm{iid}}{\sim} \textrm{Bernoulli}(\pi_R=0.5)$ in each block for ISTOP-AFib. 
Denote one of the potential realizations from period 1 to period $t$ by $\mathbf{r}_{1:t}$.
% and its support by $\mathbf\Omega(\mathbf{r}_{1:d})$

Let $J$ denote the number of days in each period (assume each period is of the same length and no missing data. We discuss how to handle missing observations later). 
For the I-STOP-AFib trial, $J=7$ days, i.e., data on alcohol consumption and AF occurrence were collected daily for seven days during each period. 
%We use $\mathbf R_{1:(tJ)} = (\mathbf R_1, \dots, \mathbf R_t)$ to denote a treatment assignment path up to the end of the $t$-th period, where $\mathbf R_t =S_t\mathbf{1}_J$ with $\mathbf{1}_J$ being a $J$-vector of 1's. 
%Denote one of the realizations from day 1 to day $d$ by $\mathbf{r}_{1:d}$ and its support by $\mathbf\Omega(\mathbf{r}_{1:d})$ consists of all sub-sequences from day 1 to day $d$ in the support $\mathbf\Omega(\mathbf{r}_{1:tJ}) \in \{\mathbf0_J,\mathbf1_J\}^{t}$. \kxnote{check l216}
%Let $\mathbf R_{t}\in \{\mathbf{0}_{J\times1}, \mathbf{1}_{J_t\times1}\}$ denote a vector of treatment randomizations during period $t$ of $J_t$ measurements, $t = 1,...,T$. 
%The primary N-of-1 phase of I-STOP-AFib had $T= 6$ periods/weeks, and within each period, exposures and outcomes $J=7$ measurements were collected for over seven days.
%$\mathbf R_t =s_t\mathbf{1} + (1-s_t)\mathbf{0}$ where $s_t \sim Bernoulli(\pi_R)$.  A typical N-of-1 trial usually has $\pi_R=0.5$.  $\mathbf R_t=\mathbf0$ if assigned to trigger off (refrain from alcohol) and $\mathbf R_t=\mathbf 1$ if assigned to trigger on (access to alcohol). 
%The random \textit{treatment assignment path} is
%$$\mathbf R_{1:t} = (\mathbf R_1,...,\mathbf R_t)$$
Let $\{\mathbf X_{tj}(\bullet)\}$ denote the \textit{set} of potential treatment selections (alcohol uses) on day $j$ of period $t$, 
    $$\{\mathbf X_{tj}(\bullet)\} = \{ X_{tj}(\mathbf r_{1:t}), \mathbf r_{1:t} \in \{0,1\}^t\},$$ 
where $X_{tj}(\mathbf{r}_{1:t}) = X_{tj}(do (\mathbf R_{1:t} = \mathbf r_{1:t})) = \{{0,1}\}$ if hypothetically/counterfactually (to what we observed) we force/fix the treatment assignment to $\mathbf r_{1:t}$.  
The $do(X=x)$ operator introduced by 
\cite{pearl2009causal} indicates an action of fixing a random variable $X$ at one value of $x$. 
Throughout the paper, we omit the $do()$ operator for simplicity, and let small letters denote a realized (observed) event and capital letters denote random variables or potential events.
$X_{tj}(\mathbf{r}_{1:t})$ denotes the potential treatment on day $j$ of period $t$ and is assumed to depend on the treatment assignments through period $t$ only, but not on future ones. 

\begin{figure}[H]
    \centering
    \includegraphics[width=0.9\textwidth]{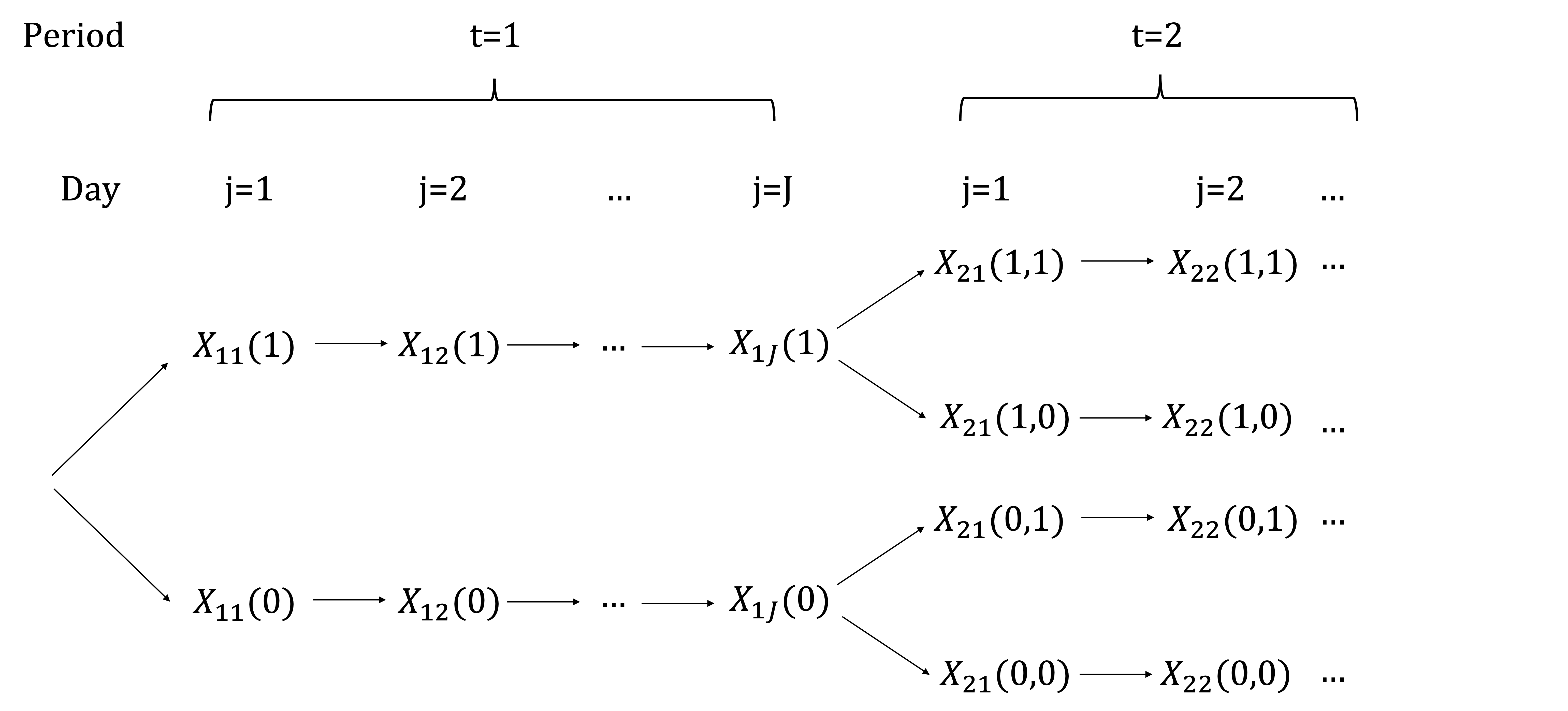}
    \caption{All potential treatments up to day 2 of period 2. }
    \label{fig:treatment_paths}
\end{figure}

Let $\mathbf X_{11:tj}(\mathbf r_{1:t})$ denote the \textit{potential treatment selection path} up to day $j$ of period $t$ given randomization assignment $\mathbf{r}_{1:t}$,  
$$\mathbf X_{11:tj}(\mathbf r_{1:t}) = \{X_{11}(\mathbf r_{1:1}), X_{12}(\mathbf r_{1:1}),\dots, X_{tj}(\mathbf r_{1:t})\}.$$
Let $\mathbf X_{11:tj}(\bullet) = \{\mathbf X_{11:tj}(\mathbf r_{1:t}), \mathbf r_{1:t}\in \{0,1\}^t\}$ denote the set of all potential treatment selections given the $2^t$ possible randomization paths through period $t$ (See example in Figure \ref{fig:treatment_paths}). 

\begin{figure}[H]
    \centering
    \includegraphics[width=0.9\textwidth]{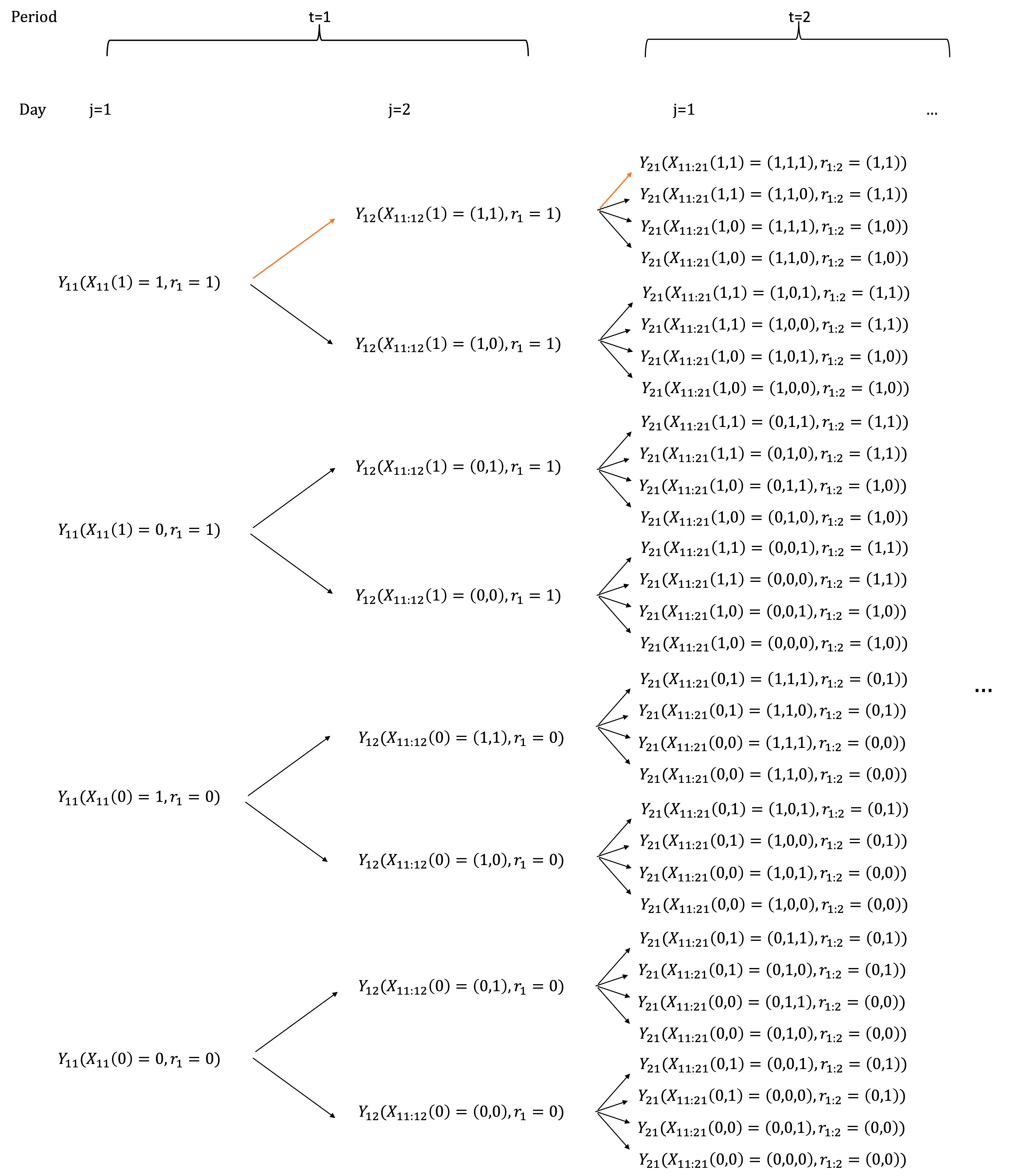}
    \caption{All potential outcomes through day 1 of period 2. The path in orange is the observed outcome path $\mathbf Y_{11:21}(\mathbf x_{11:21},\mathbf r_{1:2})$ for $\mathbf x_{11:21} = (1,1,1)$ and $\mathbf r_{1:2} = (1,1)$. }
    \label{fig:outcome_paths}
\end{figure}

(b) \textbf{Scientific process. } 

An outcome measure follows a treatment selection. We use $ \{\mathbf Y_{tj}(\bullet,\bullet)\}$ to denote the set of \textit{potential outcomes} of AF on day $j$ of period $t$,  
    $$\mathbf\{ Y_{tj}(\bullet,\bullet)\} = \{Y_{tj}(\mathbf x_{11:tj},\mathbf r_{1:t}), \mathbf x_{11:tj} \in \mathbf X_{11:tj}(\bullet), \mathbf r_{1:t} \in \{0,1\}^t \}$$
where $Y_{tj}(\mathbf x_{11:tj},\mathbf r_{1:t}) = Y_{tj}(do (\mathbf X_{11:tj} = \mathbf x_{11:tj},\mathbf R_{1:t} = \mathbf r_{1:t}))$ denotes the potential AF outcome on day $j$ of period $t$ following a particular treatment path $\mathbf x_{11:tj}$  and randomization sequence $\mathbf r_{1:t}$. Figure \ref{fig:outcome_paths} depicts an example of all potential outcomes through day 1 of period 2. 

\textit{Assumption 1}. We assume that the outcomes of $Y_{tj}$ are affected by the randomization sequence only through its impact on the treatment path, which implies that $Y_{tj}(\mathbf x_{11:tj}, \mathbf r_{1:t}) =  Y_{tj}(\mathbf x_{11:tj}, \mathbf r'_{1:t})$ for any given treatment path  $x_{11:tj}$ and $\mathbf r_{1:t}\ne \mathbf r'_{1:t}$. 
So we can write $\{\mathbf Y_{tj}(\bullet,\bullet)\} = \mathbf \{Y_{tj}(\bullet)\} = \{ Y_{tj}(\mathbf x_{11:tj}), \mathbf x_{11:tj} \in \mathbf X_{11:tj}(\bullet)\}$, and the collection of all \textit{potential outcome paths} through day $j$ of period $t$ is denoted by
 $$\{\mathbf Y_{11:tj}(\bullet)\} = \{\mathbf Y_{11}(\bullet), \mathbf Y_{12}(\bullet), \dots,\mathbf Y_{tj}(\bullet)\}.$$

\subsection{Causal Estimands}
For two treatment paths $\mathbf x_{11:tj}$ and $\mathbf x'_{11:tj}$, the causal odds ratio that characterizes the difference in the potential outcomes on day $j$ of period $t$ is defined as
\begin{equation}
   \tau_{tj}(\mathbf x_{11:tj},\mathbf x'_{11:tj}) := \frac{\E( Y_{tj}(\mathbf x_{11:tj}))}{1-\E( Y_{tj}(\mathbf x_{11:tj}))} \cdot \frac{1-\E( Y_{tj}(\mathbf x'_{11:tj}))}{\E( Y_{tj}(\mathbf x'_{11:tj}))}. 
            \label{ce1}
\end{equation}
The temporal average causal odds ratio over all periods is defined as
\begin{equation}
    \overline{\tau}(\mathbf{x}, \mathbf{x}') = \frac{\bar{p}(\mathbf x)}{1-\bar{p}(\mathbf x)}\cdot \frac{1-\bar{p}(\mathbf x')}{\bar{p}(\mathbf x')},
             \label{ce2}
\end{equation}
where 
    $$\bar{p}(\mathbf x)= \frac{1}{TJ}{\sum_{t=1}^T\sum_{j=1}^{J}\E( Y_{tj}(\mathbf x_{11:tj}))}$$ 
denotes the average event (AF) rate over the treatment path $\mathbf{x}=\mathbf{x}_{11:TJ}$. 
%
%Assuming there are no period and direct time effects (Assumptions 2 and 3), the potential outcome at one time point would be independent of the treatment path, i.e., (\ref{ce1}) is simplified to 
%\begin{equation}
%   \tau_t(\mathbf x_{1:t},\mathbf x'_{1:t})  =  \tau_t(x_{tj}, x'_{tj}) = \frac{\E( Y_{tj}( x_{tj}))}{1-\E( Y_{tj}( x_{tj}))} \frac{1-\E( Y_{tj}( x'_{tj}))}{\E( Y_{tj}( x'_{tj}))}, \forall j = 1,...,J_t.
%            \label{ce3}
%\end{equation}
%If we further assume causal effects are the same across all periods,
%\begin{equation}
%    \overline{\tau}(\mathbf{x}_{1:T}) =  \tau_t(\mathbf{x_t}, \mathbf{x'_t}) = \tau_t(x_{tj}, x'_{tj}),  \forall t = 1,...,T, j = 1,...,J_t.
%         \label{ce4}
%\end{equation}
Similarly, for two realized treatment assignment paths $\mathbf r_{1:t}$ and $\mathbf r'_{1:t}$, the causal odds ratio, in terms of the impact of treatment assignment on the outcome, is defined as
\begin{equation}
   \theta_{tj}(\mathbf r_{1:t},\mathbf r'_{1:t}) := \frac{\E( Y_{tj}(\mathbf X_{11:tj}(\mathbf r_{1:t})))}{1-\E( Y_{tj}(\mathbf X_{11:tj}(\mathbf r_{1:t})))} \frac{1-\E( Y_{tj}(\mathbf X_{11:tj}(\mathbf r'_{1:t})))}{\E( Y_{tj}(\mathbf X_{11:tj}(\mathbf r'_{1:t})))}. 
            \label{itt1}
\end{equation}
For example, for $\theta_{12}(1,0)$, the expectation $\E(Y_{12}(\mathbf{X}_{11:12}(1))$ involves all potential outcomes under $r_1=1$ (top four outcomes in the middle column in Figure \ref{fig:outcome_paths}), and the expectation $\E(Y_{12}(\mathbf{X}_{11:12}(0))$ involves all potential outcomes under $r_1=0$ (bottom four outcomes in the middle column in Figure \ref{fig:outcome_paths}).
%and can be simplified to 
%\begin{equation}
%   \theta_t(\mathbf r_{1:t},\mathbf r'_{1:t})  = \theta(r_t, r'_{tj}) =  \frac{\E( Y_{tj}( X_{tj}(r_t)))}{1-\E( Y_{tj}( X_{tj}(r_t)))} \frac{1-\E( Y_{tj}(  X_{tj}(r'_{tj})))}{\E( Y_{tj}(  X_{tj}(r'_{tj})))}, \forall t = 1,...,T, j = 1,...,J_t.
%            \label{itt2}
%\end{equation}
%
The corresponding temporal average causal odds ratio is defined as 
\begin{equation}
    \overline{\theta}(\mathbf{r}_{1:T}, \mathbf{r}'_{1:T}) = \frac{\bar{p}_\theta(\mathbf r_{1:T})}{1-\bar{p}_\theta(\mathbf r_{1:T})}\frac{1-\bar{p}_\theta(\mathbf r'_{1:T})}{\bar{p}_\theta(\mathbf r'_{1:T})},
             \label{itt3}
\end{equation}
where $$\bar{p}_\theta(\mathbf r_{1:T})= \frac{1}{TJ} \sum_{t=1}^{T}\sum_{j=1}^{J}\E( Y_{tj}(\mathbf{X}_{11:tj}(\mathbf r_{1:t}))).$$
The two types of odds ratios are related by
\begin{align*}
    \theta_{tj}(\mathbf r_{1:t},\mathbf r'_{1:t}) 
    &=  \tau_{tj}(\mathbf{X}_{11:tj} (\mathbf r_{1:t}),\mathbf{X}_{11:tj} (\mathbf r'_{1:t})), \\
    \overline{\theta}(\mathbf{r}_{1:T}, \mathbf{r}'_{1:T})
    & = \overline\tau(\mathbf{X}_{11:TJ} (\mathbf r_{1:T}),\mathbf{X}_{11:TJ} (\mathbf r'_{1:T})). 
\end{align*}
%A third type of estimand, which is a hybrid of estimands  (\ref{ce1})  and (\ref{itt1})  contrasts a treatment assignment path and a treatment path and is defined as 
%\begin{equation}
%   \gamma_t(\mathbf r_{1:t},\mathbf x_{1:t}) := \frac{\E( Y_{tj}(\mathbf X_{1:t}(\mathbf r_{1:t}))}{1-\E( Y_{tj}(\mathbf X_{1:t}(\mathbf r_{1:t})))} \frac{1-\E( Y_{tj}(\mathbf x_{1:t}))}{\E( Y_{tj}(\mathbf x_{1:t}))}, \forall j = 1,...,J_t,
%            \label{dh1}
%\end{equation}
%and can be simplified to 
%\begin{equation}
%   \gamma_t(\mathbf r_{1:t},\mathbf x_{1:t})  = \gamma(r_t, x_{tj}) = \frac{\E( Y_{tj}( X_{tj}(r_t))}{1-\E( Y_{tj}( X_{tj}(r_t)))} \frac{1-\E( Y_{tj}(x_{tj})}{\E( Y_{tj}(x_{tj}))}, \forall t = 1,...,T, j = 1,...,J_t.
%            \label{dh2}
%\end{equation}

\begin{figure}[H]
    \centering
    \includegraphics[width=0.9\textwidth]{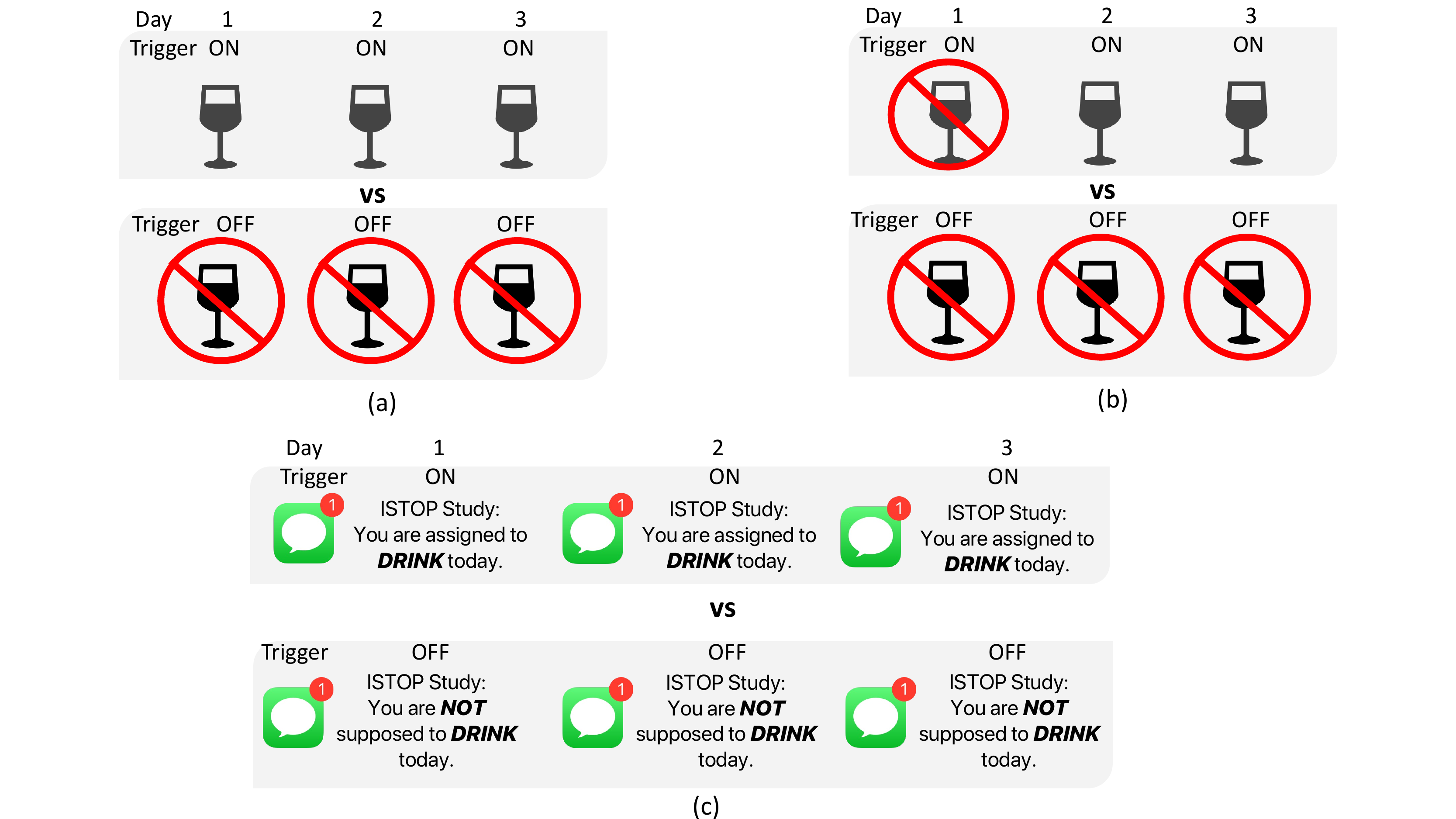}
    \caption{An illustration of three types of estimands for an individual's drinking behavior over a period of three consecutive days. Estimand (a) contrasts drinking vs abstinence; Estimand (b) contrasts drinking habit vs abstinence; Estimand (c) is the Intention-to-Treat estimand and contrasts the two drinking assignments. }
    \label{fig:estimands}
\end{figure}

With the definitions outlined above, we propose to define and estimate two types of causal effects of alcohol on developing AF: 
%Estimand (a) which quantifies the alcohol effect, and Estimand (b) which quantifies the causal effect of the drinking behaviors/habits. 
Estimand (a) is defined as $\overline\tau (\mathbf{1}, \mathbf{0})$, which contrasts the average outcomes between the two counterfactual conditionals in which a participant is continuously exposed to alcohol (i.e., drinks daily, denoted by DD) versus is never exposed to alcohol; see Figure \ref{fig:estimands}.  
Estimand (b) is as defined $\overline\tau (\mathbf{X}_{11:TJ}(\mathbf{1}), \mathbf{0})$, the causal effect of drinking habit on AF between the counterfactuals when someone is allowed to drink versus when they completely abstain from alcohol. The second estimand is useful in evaluating the impact of an individual's usual drinking habit  (denoted by UD) versus abstinence. 

Of note, the ITT type estimand is defined as $\overline\theta (\mathbf{1}, \mathbf{0}) = \overline\tau (\mathbf{X}_{11:TJ}(\mathbf{1}), \mathbf{X}_{11:TJ}(\mathbf{0}))$. 

%We assume a linear model between the latent potential outcome $Y^*(R)$ and the assignment $R$, 
%$$Y^*_{tj}(r) = \gamma_0 + \gamma_1 r + \gamma_w w_{tj} + e^r_{tj},\quad e^r_{rj}\sim N(0,1) $$
%such that E$(Y_{rj}(r)) =\text{P}(Y_{tj}(r)=1) = \gamma_0 + \gamma_1r + \gamma_w w_{tj}$. Likewise, $\text{E}(Y(X(1))$ is estimated by $\frac{1}{L}\sum_{tj}(\hat \gamma_0 + \hat \gamma_1 + \hat \gamma_w w_{tj})$. 

\subsection{Bayesian estimation model}
%%%%%%%%%%%%%%%%%%%%%%%%%%%%%%%%%%%%%%%%%%%%%%
The fundamental Stable Unit Treatment Value Assumption (SUTVA) in causal inference posits that the treatment a participant is assigned does not influence the potential outcomes of other participants for a study with multiple participants. 
However, in an N-of-1 trial, treatments are assigned at random in a crossover fashion on the same individual so SUTVA would require that treatment assignment at one time cannot affect potential outcomes at a different time and in particular at a later time. 
But because of the potential for treatment carryover effects, SUTVA may very well be invalid in an N-of-1 trial. 
Consequently, we make the following assumptions.

\textit{Assumption No-carryover (NCO).} We assume that treatment exposures on previous days do not have a direct lagged impact on the outcomes of the later days. 
In other words, expectations in Equation (\ref{ce1}) and (\ref{itt1}) on the treatment path are restricted to the current visit.
However, we allow the outcomes of adjacent days to be correlated. We use a latent outcome model to describe the occurrence of AF and assume that the error terms in latent outcomes can be fully described using an auto-correlation (AR) structure. 

\textit{Assumption Carryover (CO).} We relax Assumption NCO by allowing previous treatments to exert carryover effects on AF during the subsequent days. 
Because it is widely accepted that alcohol is usually eliminated from the human body within 12 hours, we assume that this carryover effect is limited to a one-day timeframe. 

\textit{Assumption 2.} We assume that treatment assignments have a direct impact only on treatment selection on the day of the assignment and not on later days.

\begin{figure}[H]
    \centering \includegraphics[width=0.7\textwidth]{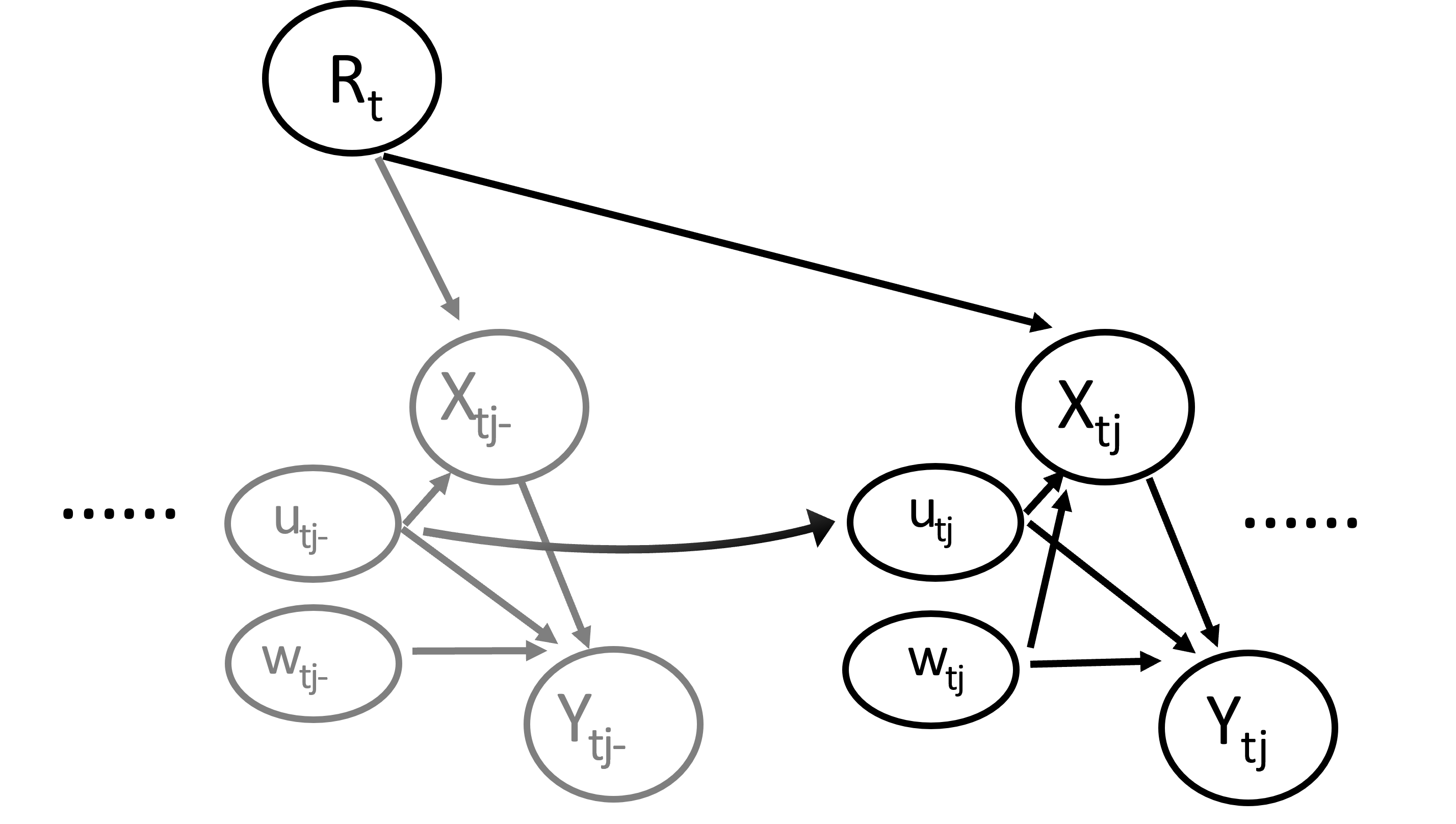}
    \caption{Causal diagram describing a causal framework for one N-of-1 trial at day $j$ of period $t$. $tj-$ denotes the visit on the previous day, e.g., $X_{tj-} = X_{tj-1} $ for $j\ne 1$ and $X_{tj-} = X_{(t-1)J}$ for $j=1$. The arrow from $X_{tj-)}\rightarrow$ Y$_{tj}$ is modeled in model CO but is absent in model NCO. Variables involved in IV analysis at period $t$ and time $j$ are shown in black and those involved at time $j-1$ are shown in grey. R$_t \rightarrow$ X$_{tj}$ is the selection process and X$_{tj} \rightarrow$ Y$_{tj}$ is the scientific process. Two consecutive time points are connected in model NCO through the unmeasured confounding effect $u_{tj}$. $w_{tj}$ is the measured confounding effect.}
    \label{fig:dag}
\end{figure}

Combined with Assumption 2, Assumption NCO leads to model NCO and Assumption CO leads to model CO, which will be introduced in section \ref{section: MD}. 
Figure \ref{fig:dag} illustrates the causal framework for a single N-of-1 trial under models NCO and CO which will be introduced below.

We denote the measured and unmeasured confounders at day $j$ of period $t$ by $w_{tj}$ and $u_{tj}$, respectively. $u_{tj}$ is assumed to have zero mean.

%%%%%%%%%%%%%%%%%%%%%%%%%%%%%%%%%%
%%%

\subsubsection{Models} \label{section: MD}

\textbf{Model NCO}

Using assumptions 1 and 2, we propose the following structural models to describe the selection and scientific processes. 
%We use the subscribe $tj$ to denote observations on day $j$ in period $t$. 
   \begin{align}
        \Pr(X_{tj}(r_{t})=1)  
        &= h^{-1}\{ \alpha_0 + \alpha_1 r_{t} + \alpha_w^\top \mathbf f(w_{tj}) + u_{tj} \}, \nonumber \\
       \Pr(Y_{tj}(x_{tj})=1)  
       &= h^{-1}\{\beta_0 + \beta_1 x_{tj} +  \beta_w^\top \mathbf{g}(w_{tj}) + su_{tj}\},
       \label{O1}
    \end{align} 
where $\mathbf f(\cdot)$ and $\mathbf {g}(\cdot)$ denote any general functions including variable selection, and $h(\cdot)$ is a link function.  
$w_{tj}$ denotes the set of all possible measured confouders at period $t$ day $j$. 
We limit the coefficient $s$ in the second equation to be either 1 or $-1$, i.e., $s$ = \{1,-1\} and denotes the sign of the product of the unknown effect of the confounders to $X_{tj}$ and $Y_{tj}$.
In Appendix \ref{section:confoundingcoef}, we show that $u_{tj}$ represents a summary of unmeasured confounders. For Gaussian models it is adequate to assume the same $u_{tj}$ for both equations and the magnitude of the coefficient is not important.

We further introduce two latent quantities $X_{ij}^\dagger(\cdot)$ and $Y_{ij}^\dagger(\cdot)$ such that 
\begin{align*}
    Y_{tj}(\cdot) = \mathbbm{1}(Y_{tj}^\dagger(\cdot)>0), \quad 
    X_{tj}(\cdot) = \mathbbm{1}(X_{tj}^\dagger(\cdot)>0), 
\end{align*}
assuming   
\begin{align}
     X_{tj}^\dagger(r_{t}) &=  \alpha_{0} + \alpha_{1}r_{t} + \alpha_w^\top \mathbf{f}(w_{tj}) + u_{tj} +  \epsilon_{tj},\nonumber \\ 
     Y_{tj}^\dagger(x_{tj}) &=  \beta_{0} + \beta_{1}x_{tj}  + \beta_w^\top \mathbf{g}(w_{tj}) + su_{tj} + \eta_{tj}, 
     \label{m1}
\end{align}
where $u_{tj} \sim N(0, \sigma_u^2)$, 
$\epsilon_{tj} \overset{\mathrm{iid}} \sim N(0,1)$, $\eta_{tj} \overset{\mathrm{iid}} \sim N(0, 1)$ are mutually independent. $\mathbbm{1}(\cdot)$ is the indicator function. 
We assume that $u_{tj}$ are auto-correlated with an AR(1) structure and $\mathrm{corr}(u_{tj}, u_{t'j'}) = \rho_u^{|(t-1)J+j-(t'-1)J-j'|}$.

The latent models \eqref{m1} suggest that for given $t,j$, the latent treatments $X_{tj}^\dagger(r_{t})$ and outcomes $Y_{tj}^\dagger(x_{tj})$ are correlated due to the existence of $u_{tj}$, which captures the confounding effect that relates the potential self-selected treatments to the potential outcomes. 
We can quantify the magnitude of residual confounding (after adjusting for $w_{tj}$) using $$\rho = \mathrm{corr}(X_{tj}^\dagger(r),Y_{tj}^\dagger(x)) = \frac{\mathrm{Cov}(X_{tj}^\dagger(r),Y_{tj}^\dagger(x_{tj}))}{\sqrt{V(X_{tj}^\dagger(r))V(Y_{tj}^\dagger(x))}} = s\sigma_u^2/(1+\sigma_u^2).$$
Longitudinally, the potential treatment selections and potential outcomes are correlated due to $\{u_{tj}\}$, which we assume capture the ``lagged'' effects through periods, as illustrated in Figure \ref{fig:dag}. 

To facilitate model estimation, divide both sides of Equations \eqref{m1} by $c=\sqrt{1+\sigma^2_u}$. 
Using the superscript $*$ to denote the resulting re-scaled outcomes and model parameters, we rewrite the latent model \eqref{m1} as 
\begin{align}
     X_{tj}^*(r_{t}) &=  \alpha_{0}^* + \alpha_{1}^*r_{t} + \alpha_w^{*\top} \mathbf{f}(w_{tj}) +  \tilde\epsilon_{tj}^*,\nonumber \\ 
     Y_{tj}^*(x_{tj}) &=  \beta_{0}^* + \beta_{1}^*x_{tj}  + \beta_w^{*\top} \mathbf{g}(w_{tj}) + \tilde\eta_{tj}^*, 
     \label{m1.rescale}
\end{align}
where  $\tilde\epsilon^*_{tj} =  (u_{tj}+ \epsilon_{tj})/ c$ and  $\tilde\eta^*_{tj} =  (su_{tj}+ \eta_{tj}) /c$. 
For given $t,j$, the two error terms have a binormal distribution,  
\begin{align}
  \begin{pmatrix}
    \tilde\epsilon^*_{tj} \\
    \tilde\eta^*_{tj}
\end{pmatrix}
\sim 
N\left(\mathbf 0, 
    \begin{pmatrix}
        1 &  \rho  \\
        \rho  & 1
    \end{pmatrix}\right). 
    \label{m2}  
\end{align}

Re-scaling does not change the cross-sectional correlation $\rho$ or the longitudinal AR(1) correlation structure. 
Further, since we have
\begin{align*}
    Y_{tj}(\cdot) = \mathbbm{1}(Y_{tj}^\dagger(\cdot)>0) = \mathbbm{1}(Y_{tj}^*(\cdot)>0) , \quad 
    X_{tj}(\cdot) = \mathbbm{1}(X_{tj}^\dagger(\cdot)>0) = \mathbbm{1}(X_{tj}^*(\cdot)>0) , 
\end{align*}
the equations (\ref{O1}) and \eqref{m1.rescale} imply a latent probit structural model,
\begin{align}
    &\Pr(X_{tj}(r_{t})=1 \mid w_{tj}) = \Pr(X_{tj}^*(r_{t})>0\mid w_{tj}) = \Phi(\alpha_{0}^* + \alpha_{1}^*r_{t}  + \alpha_w^{*\top} \mathbf{g}(w_{tj})), \nonumber\\
    &\Pr(Y_{tj}(x_{tj})=1 \mid w_{tj}) = \Pr(Y^*_{tj}(x_{tj})>0\mid w_{tj}) = \Phi(\beta_{0}^* + \beta_{1}^*x_{tj}  + \beta_w^{*\top} \mathbf{g}(w_{tj})), \label{eq:probit} 
\end{align}
where $\Phi$ is the standard normal CDF. 
Note that $X_{tj}(r_{t})$ and $Y_{tj}(x_{tj})$ are correlated. Model (\ref{eq:probit}) only describes the marginal distributions of the two potential outcomes. 
%That means $\Pr(Y_{tj}(x)1\mid w_{tj}) =\Pr(Y^*(x)_{tj}>0\mid w_{tj}) = \Phi(\beta_{0} + \beta_{1}x  + \beta_w^\top \mathbf{g}(w_{tj}))$ and $\Pr(X_{tj}(r)=1\mid w_{tj})=\Pr(X_{tj}^*(r)>0\mid w_{tj}) = \Phi(\alpha_{0} + \alpha_{1}x  + \alpha_w^\top \mathbf{g}(w_{tj}))$.
%The correlation comes from the unmeasured confounding effect $u_{tj}$ in the errors, i.e., we assume $\epsilon^*_{tj} =  u_{tj}+ \epsilon_{tj}$ and $\eta^*_{tj} =  u_{tj}+ \eta_{tj}$. We impose an AR(1) structure on $\bm u$, i.e., cor$(u_{k},u_{k'}) = \rho_{u}^{|k-k'|}$ for $k\ne k'$. Therefore, both outcomes $\bm Y$  and treatment selections $\bm X$ are correlated through $\bm u$ in time, respectively.

Note that the model parameters $\alpha_1^*$ and $\beta_1^*$ in \eqref{eq:probit} are not the same as $\alpha_1$ and $\beta_1$ in the structural model \eqref{O1} (they differ by a factor $c$). 
However, we are not particularly interested in estimating $\alpha_1$ and $\beta_1$, which describe the corresponding effect of `randomization on alcohol consumption' and effect of `alcohol consumption on AF' conditional on $w_{tj}$ and $u_{tj}$ at the specific time of period $t$ and day $j$.
Let us temporarily disregard the time-based correlation and treat the various measurements obtained from the N-of-1 study participant as measurements from multiple study participants from a conventional trial;
we can consider $\alpha_1$ and $\beta_1$ as conditional effects specific to time, similar to individual-level effects in a study involving multiple participants.
%Then, the difference between $(\alpha_1^*, \beta_1^*)$ and $(\alpha_1, \beta_1)$ is partially explained by non-collapsibility due to $u_{tj}$. 
Instead of focusing on those parameters, we propose to estimate marginal/participant-level causal effects (similar to the population-level effects in a multi-participant study) that are more directly related to $\alpha_1^*$ and $\beta_1^*$, which we elaborate on in section~\ref{sec:estimation}.  

% The observed data model naturally follows from Eq.(\ref{m1.rescale})
% \begin{align}
%      X_{tj}^* &=  \alpha_{0}^* + \alpha_{1}^*r_t + \alpha_w^{*\top} \mathbf{f}(w_{tj}) +  \tilde\epsilon_{tj}^*,\nonumber \\ 
%      Y_{tj}^* &=  \beta_{0}^* + \beta_{1}^*X_{tj}  + \beta_w^{*\top} \mathbf{g}(w_{tj}) + \tilde\eta_{tj}^*.
%      \label{m1.rescale.odm}
% \end{align}

\textbf{Model CO}

Model CO extends model NCO by adding explicit carryover through a lagged treatment effect. Here we consider only 1-day carryover given the short effect of alcohol, although the model could easily be generalized to additional lagged treatments (see \cite{liao2023analysis}). 
The following model generalizes equation (\ref{O1}) by introducing an additional $\beta_2$ term to describe the effect of alcohol on the previous day, $X_{tj-}$.    
   \begin{align}
        \Pr(X_{tj}(r_{t})=1\mid w_{tj}, u_{tj})  
        &= \Phi( \alpha_0 + \alpha_1 r_{t} + \alpha_w^\top \mathbf f(w_{tj}) + u_{tj} ), \nonumber \\
       \Pr(Y_{tj}(X_{tj} = x_{tj}, X_{tj-}=x_{tj-})=1\mid w_{tj}, u_{tj})  
       &= \Phi (\beta_0 + \beta_1 x_{tj} + \beta_2x_{tj-} +  \beta_w^\top \mathbf{g}(w_{tj}) + su_{tj}).\label{eq.str.model.carryover}
    \end{align} 
Similar to model NCO, the latent models are also rescaled to  achieve the binormal distribution as specified in equation (\ref{m2}).

\subsubsection{Causal Effect Estimation } \label{sec:estimation}
\textbf{Model NCO}

For a given treatment path $\mathbf x = \mathbf x_{11:TJ}$, we calculate the average period event rate as
\begin{align*}
\bar{p}(\mathbf x) 
    & = \frac{1}{TJ}\sum_{t=1}^{T}\sum_{j=1}^{J}\E( Y_{tj}(\mathbf x_{11:tj}))
      %= \frac{1}{TJ}\sum_{t=1}^{T}\sum_{j=1}^{J}\E\E\{ Y_{tj}(\mathbf x_{11:tj}) \mid w_{tj}\} \\
     = \frac{1}{TJ}\sum_{t=1}^{T}\sum_{j=1}^{J}  \Phi(\beta_{0}^* + \beta_{1}^*x_{tj}  +  \beta_w^{*\top} \mathbf{g}(w_{tj})). 
     % = \frac{1}{L} \int_{w_k} \Phi(\beta_{0}^* + \beta_{1}^*x_{k}  +  \beta_w^{*\top} \mathbf{g}(w_{k})) \mathrm{d}F(w_k)\\
    %\approx \frac{\sum_{t=1}^{T}\sum_{j=1}^{J} \Phi(\beta_{0}^* + \beta_{1}^*x_{tj}  + \beta_w^{*\top} \mathbf{g}(w_{tj})) }{TJ}. 
\end{align*}
So the period average causal effect of continuous exposure to alcohol is given by 
\begin{align*}
    \overline\tau (\mathbf{1}, \mathbf{0}) 
    & =  \frac{\bar{p}(\mathbf{1})}{1-\bar{p}(\mathbf 1)}\cdot \frac{1-\bar{p}(\mathbf 0)}{\bar{p}(\mathbf 0)} \\
    & = \frac{\sum_{t=1}^{T}\sum_{j=1}^{J} \Phi(\beta_{0}^* + \beta_{1}^* + \beta_w^{*\top} \mathbf{g}(w_{tj}))}{1-\sum_{t=T}^{J}\sum_{j=1}^{J} \Phi(\beta_{0}^* + \beta_{1}^* + \beta_w^{*\top} \mathbf{g}(w_{tj}))} \cdot
    \frac{1-\sum_{t=1}^{T}\sum_{j=1}^{J} \Phi(\beta_{0}^* + \beta_w^{*\top} \mathbf{g}(w_{tj}))}{\sum_{t=1}^{T}\sum_{j=1}^{J} \Phi(\beta_{0}^* + \beta_w^{*\top} \mathbf{g}(w_{tj}))} . 
\end{align*}

Another causal estimand of interest is the causal effect of usual alcohol use or drinking behavior. 
For the randomization sequence $\mathbf r_{1:T} = \mathbf 1$, the potential treatment path is $\mathbf{X}_{11:TJ}(\mathbf r_{1:T} = \mathbf 1)$ with the treatment selection on day $j$ of period $t$. $X_{tj}(\mathbf 1) = \mathbbm{1}(X_{tj}^*(\mathbf 1)>0).$
\begin{align*}
    & \overline\tau (\mathbf{X}_{11:TJ}(\mathbf{1}), \mathbf{0}) 
      = \frac{\bar{p}(\mathbf{X}_{11:TJ}(\mathbf{1}))}{1-\bar{p}(\mathbf{X}_{11:TJ}(\mathbf{1}))}   \frac{1-\bar{p}(\mathbf 0)}{\bar{p}(\mathbf 0)}\\
    & = \frac{\sum_{t=1}^{T}\sum_{j=1}^{J} \Phi(\beta_{0}^* + \beta_{1}^*X_{tj}(\mathbf 1) + \beta_w^{*\top} \mathbf{g}(w_{tj}))}{1-\sum_{t=1}^{T}\sum_{j=1}^{J} \Phi(\beta_{0}^* + \beta_{1}^*X_{tj}(\mathbf 1) + \beta_w^{*\top} \mathbf{g}(w_{tj}))}\frac{1-\sum_{t=1}^{T}\sum_{j=1}^{J} \Phi(\beta_{0}^* + \beta_w^{*\top} \mathbf{g}(w_{tj}))}{\sum_{t=1}^{T}\sum_{j=1}^{J} \Phi(\beta_{0}^* + \beta_w^{*\top} \mathbf{g}(w_{tj}))} . 
\end{align*}

The period average causal effect of intention-to-treat, which contrasts the event rate under two randomization sequence $\mathbf{r} = \mathbf{1}$ and $\mathbf{r} = \mathbf{0}$, is estimated by
\begin{align*}
    & \overline\theta (\mathbf{1}, \mathbf{0})
      = \overline\tau (\mathbf{X}_{11:TJ}(\mathbf{1}), \mathbf{X}_{11:TJ}(\mathbf{0})) 
      = \frac{\bar{p}(\mathbf X_{11:TJ}(\mathbf{1}))}{1-\bar{p}(\mathbf X_{11:TJ}(\mathbf{1}))}   \frac{1-\bar{p}(\mathbf X_{11:TJ}(\mathbf{0}))}{\bar{p}(\mathbf X_{11:TJ}(\mathbf{0}))}\\
    & = \frac{\sum_{t=1}^{T}\sum_{j=1}^{J} \Phi(\beta_{0}^* + \beta_{1}^*X_{tj}(\mathbf 1) + \beta_w^{*\top} \mathbf{g}(w_{tj}))}{1-\sum_{t=1}^{T}\sum_{j=1}^{J} \Phi(\beta_{0}^* + \beta_{1}^*X_{tj}(\mathbf 1) + \beta_w^{*\top} \mathbf{g}(w_{tj}))}\frac{1-\sum_{t=1}^{T}\sum_{j=1}^{J} \Phi(\beta_{0}^* + \beta_{1}^*X_{tj}(\mathbf 0) + \beta_w^{*\top} \mathbf{g}(w_{tj}))}{\sum_{t=1}^{T}\sum_{j=1}^{J} \Phi(\beta_{0}^* + \beta_{1}^*X_{tj}(\mathbf 0) +\beta_w^{*\top} \mathbf{g}(w_{tj}))} . 
\end{align*}

%The marginal (log) odds ratio is different from the expectation of the time-specific (log) odds ratio because it is a non-linear function of probabilities. This refers to the non-collapsibility phenomenon discussed in section 1.3 under binary outcome. Therefore, the marginal (log) odds ratio is estimated by plugging in marginal probabilities that are averaged over all time points. Without time-varying covariates, the estimates are reduced to log$\hat\tau(\bm1,\bm0) = \text{log}\frac{\Phi(\hat\beta_0 + \hat\beta_1)} {1-\Phi(\hat\beta_0 +\hat\beta_1)}- \text{log}\frac{\Phi(\hat\beta_0)}{1-\Phi(\hat\beta_0)}$ and log$\hat\gamma(\bm1,\bm0) = \text{log}\frac{\Phi(\hat\delta_0 + \hat\delta_1)} {1-\Phi(\hat\delta_0 +\hat\delta_1)}- \text{log}\frac{\Phi(\hat\beta_0)}{1-\Phi(\hat\beta_0)}$. 

\textbf{Model CO}

Estimations of the three estimands using model CO remain the same as model NCO, except that now the average period event rate also accounts for the alcohol selection on the previous days, i.e.,  
\begin{align*}
\bar{p}(\mathbf x_{11:TJ}) 
    & =\frac{1}{TJ}\sum_{t=1}^{T}\sum_{j=1}^{J}  \Phi(\beta_{0}^* + \beta_{1}^*x_{tj}  + \beta_{2}^*x_{tj-} +   \beta_w^{*\top} \mathbf{g}(w_{tj})) 
\end{align*}
and 
\begin{align*}
   \bar{p}(\mathbf X_{11:TJ}(\mathbf{r_{1:T}})) 
    &= \sum_{t=1}^{T}\sum_{j=1}^{J} \Phi(\beta_{0}^* + \beta_{1}^*X_{tj}(\mathbf r_{1:t}) +\beta_{2}^*X_{tj-}(\mathbf r_{1:t}) +  \beta_w^{*\top} \mathbf{g}(w_{tj})).
\end{align*}
\subsubsection{Joint Posterior Density Using Data Augmentation } \label{sec:likelihood}
\textbf{Model NCO}

Similar to parametric Bayesian approaches \citep[e.g.][]{koop2005semiparametric, chib2009estimation,li2015bayesian} that assume a bivariate normal distribution for the errors of the two processes, we employ a probit Bayesian approach that uses the data augmentation method combined with Gibbs sampling \citep{albert1993bayesian} to draw inferences on the estimands. 

Equation \eqref{m2} implies
\begin{align}
    Y_{tj}^* \mid X_{tj}^*, r_t, w_{tj}  
        \sim N\left(\beta_{0}^* + \beta_{1}^*\mathbbm{1}(X_{tj}^*>0) + \beta_w^*{}^\top \mathbf{g}(w_{tj}) + \rho(X_{tj}^*-\alpha_{0}^* - \alpha_{1}^*r_t - \alpha_w^*{}^\top \mathbf{f}(w_{tj})), (1-\rho^2) \right ).
             \label{ygivenx}
\end{align}

%As the binormal distribution of $(\epsilon^*_{tj}, \eta^*_{tj})$ is fixed $\forall t,j$, conditioning on $X^*_{t(j-1})$ makes $Y^*_{tj}$ independent of $Y^*_{t(j-1)}$. For simplicity, a single index $i$ replaces $tj$ to denote any observation in a sequence of a trial with length $L$. 

%Assuming AR(1) gives $\bm u \sim N(0, \Sigma_u)$ where
% $$\Sigma_u = \rho\begin{pmatrix}
%    1& \rho_u & \hdots & \rho_u^{L-1}\\
%     \rho_u &\ddots& & \rho_u^{L-2}  \\
%    \vdots &&\ddots & \vdots \\
%   \rho_u^{L-1}&\rho_u^{L-2}&\hdots& 1 %\end{pmatrix}.$$
 
Longitudinally, we assume that $\{u_{tj}\}$ are normally distributed with an AR(1) correlation structure, which implies that 
\begin{align}
    &\begin{pmatrix}
        X_{tj}^* \\
        X_{tj-}^* 
    \end{pmatrix} \mid r_t, w_{tj},r_{t-}, w_{tj-} \sim N\left(
    \begin{pmatrix}
        \alpha_{0}^* + \alpha_{1}^*r_{t} + \alpha_w^*{}^\top \mathbf{f}(w_{tj}) \\
        \alpha_{0}^* + \alpha_{1}^*r_{t-} + \alpha_w^*{}^\top \mathbf{f}(w_{tj-})
    \end{pmatrix}, 
    \begin{pmatrix}
        1 & \rho\rho_u  \\
        \rho\rho_u   & 1 
    \end{pmatrix}\right) \nonumber\\
    & \Rightarrow X_{tj}^* \mid X_{tj-}^*, r_t, w_{tj},r_{t-}, w_{tj-}  
        \nonumber\\
        & \qquad \sim N\left(
       \alpha_{0}^* + \alpha_{1}^*r_t + \alpha_w^*{}^\top \mathbf{f}(w_{tj}) + \rho\rho_u(X_{tj-}^*-\alpha_{0}^* - \alpha_{1}^*r_{t-} - \alpha_w^*{}^\top \mathbf{f}(w_{tj-})), 1-(\rho\rho_u)^2\right ),
            \label{xgivenx-1}
\end{align}
where $r_{t-}$ = $r_t$ for $j>1$ and $r_{t-} = r_{t-1}$ for $j=1$. 

Let $\theta = (\alpha_0^*, \alpha_1^*, \alpha_w^*, \beta_0^*, \beta_1^*, \beta_w^*, \rho,\rho_u)$, 
$\mathbf{Y^*}$=\{$Y_{tj}^*, t=1,...,T, j=1,...,J$\}, $\mathbf {X^*}$= \{$X_{tj}^*, t=1,...,T, j=1,...,J$\}, 
$\mathbf {R}$= \{$r_t, t=1,...,T, j=1,...,J$\}, and
$\mathbf {w}$= \{$w_{tj}, t=1,...,T, j=1,...,J$\}.
The joint posterior of the unobserved ($\mathbf{\theta, Y^*, X^*}$) is written as follows, 
\begin{align}
    & p(\theta, \bm{Y}^*, \bm X^*\mid \bm Y,\bm X,\bm R, \bm w) 
    \propto  p(\bm Y, \bm X \mid \bm{Y}^*,\bm X^*,\theta,\bm R, \bm w)  p(\bm Y^*\mid \bm X^*, \theta,\bm R, \bm w)p(\bm X^* \mid \theta,\bm R, \bm w)\pi(\theta)\notag\\
    = &\prod_{t=1}^{T}\prod_{j=1}^{J} p(Y_{tj}\mid Y^*_{tj})p(X_{tj}\mid X^*_{tj})p(Y^*_{tj}\mid X^*_{tj}, r_t, w_{tj},\theta)\notag\\
    &\times\prod_{t=2}^{T}\prod_{j=1}^J p( X^*_{tj}\mid X^*_{tj-},r_t, r_{t-},w_{tj},w_{tj-},\theta)
    \prod_{j=2}^{J}
        p( X^*_{1j}\mid X^*_{1j-},r_{1},w_{1j},w_{1j-},\theta) \notag\\
        &\quad \quad \times p( X^*_{11}|r_{1},w_{11},\theta)\pi(\theta), \label{jointpost}
    \end{align}
where $\pi(\theta)$ is a prior, $p(Y^*_{tj}\mid X^*_{tj},r_t,w_{tj},\theta)$ and $p(X^*_{tj}\mid X^*_{tj-},r_t,r_{t-},w_{tj}, w_{tj-},\theta)$ are given in Eq.(\ref{ygivenx}) and Eq.(\ref{xgivenx-1}) respectively, 
$$(X^*_{11}|r_{1},w_{11},\theta)  \sim N(\alpha_0^* + \alpha_1^*r_{1} + \alpha_w^{*}\top\mathbf{f}(w_{11}),1),$$
and $(Y_{tj}\mid Y^*_{tj})$ and $(X_{tj}\mid X^*_{tj})$ have degenerate distributions, %with a point mass at $Y_{tj} = \mathbf{1}(Y^*_{tj}\ge 0)$ and $X_{tj}=\mathbf{1}(X^*_{tj}\ge 0)$, respectively. 
\begin{align*}
p(Y_{tj}\mid Y^*_{tj})     &= \mathbbm{1}(Y_{tj}=1)\mathbbm{1}(Y^*_{tj}\ge 0)+\mathbbm{1}(Y_{tj}=0)\mathbbm{1}(Y^*_{tj}<0),\\
%p(Y^*_{i}\mid X^*_{i},\theta)  &= \phi(Y^*_{i};\beta_{0} + \beta_{1}\mathbf{1}(X_{i}^*>0) + \beta_w^\top \mathbf{g}(W_{i}) + \rho(X_{i}^*-\alpha_{0} - \alpha_{1}r_{i} - \alpha_w^\top \mathbf{f}(W_{i})),\\ &\; \; \; \; \; \;  1-\rho^2 )\\
p(X_{tj}\mid X^*_{tj})  &= \mathbbm{1}(X_{tj}=1)\mathbbm{1}(X^*_{tj}\ge 0)+\mathbbm{1}(X_{tj}=0)\mathbbm{1}(X^*_{tj}<0). %,\\
%p(X^*_{i}\mid X^*_{i-1},r_{i},r_{i-1},W_{i},W_{i-1},\theta)  &= \phi(X^*_{i};
%       \alpha_{0} + \alpha_{1}r_{i} + \alpha_w^\top \mathbf{f}(W_{i}) + \rho\rho_u(X_{i-1}^*-\alpha_{0} - \alpha_{1}R_{i-1} - \alpha_w^\top \mathbf{f}(W_{i-1})), \\
%  &\; \; \; \; \; \; 1-(\rho\rho_u)^2 )\\
%p(X^*_{11}\mid r_{11}, w_{11}, \theta)      &= \phi(X^*_{11}; \alpha_{0}^* + \alpha_{1}^*r_{11} + \alpha_w^*{}^\top \mathbf{f}(w_{11}) , 1 ). 
\end{align*}

\textbf{Model CO}
The joint posterior remains the same as in model NCO except that $Y_{tj}$ now depends on both $X_{tj}$ and $X_{tj-}$, i.e., $p(Y^*_{tj}\mid X^*_{tj}, w_{tj},\theta)$ in equation (\ref{jointpost}) is replaced with $p(Y^*_{tj}\mid X^*_{tj}, X^*_{tj-}, w_{tj},\theta)$, where  $Y_{tj}^* \mid X_{tj}^*, X_{tj-}^*, r_{t}, w_{tj} \sim $
$$    N\left(\beta_{0}^* + \beta_{1}^*\mathbbm{1}(X_{tj}^*>0) + \beta_{2}^*\mathbbm{1}(X_{tj-}^*>0)+
        \beta_w^*{}^\top \mathbf{g}(w_{tj}) + \rho(X_{tj}^*-\alpha_{0}^* - \alpha_{1}^*r_{t} - \alpha_w^*{}^\top \mathbf{f}(w_{tj})), (1-\rho^2) \right ).
$$
\subsection{Meta-analysis of multiple N-of-1 trials} \label{sec:meta.analysis}
We can extend the individual participant model to a multiple-participant model to analyze data from all N-of-1 trials, which can be considered as meta-analysis of N-of-1 trials. 
We show how to extend model NCO for meta-analysis, with model CO being extended in a similar fashion. 
The model structure remains mostly the same as the single-participant model except for the parameters $\alpha^*_{0,i}, \alpha^*_{1,i},\beta^*_{0,i},\beta^*_{1,i}$ are allowed to vary across participants and be related through an imposed structure:
 $$\alpha^*_{0,i} \sim N(a^*_0,\sigma_{a_0}^2), \quad \alpha^*_{1,i} \sim N(a^*_1,\sigma_{a_1}^2)
 \quad \alpha^*_{w,i} \sim N(a^*_w,\sigma_{a_w}^2)$$
 $$\beta^*_{0,i} \sim N(b^*_0,\sigma_{b_0}^2), \quad \beta^*_{1,i} \sim N(b^*_1,\sigma_{b_1}^2),
 \quad \beta^*_{w,i} \sim N(b^*_w,\sigma_{b_w}^2)$$ %\quad \beta^*_{w,i} \sim N(b^*_{w}, \sigma^2_w)$$
 where each participant is indexed by $i$. 
Let $\mathbf{x}_{i,11:TJ}$ denote a treatment path received by participant $i$  and $\{\mathbf{x}_{i,11:TJ}\}^N_{i=1}$ denote the set of all treatment paths by $N$ participants, i.e., $\{\mathbf{x}_{i,11:TJ}\}^N_{i=1} = \{\mathbf{x}_{1,11:TJ}, \mathbf{x}_{2,11:TJ}, ..., \mathbf{x}_{N,11:TJ}\}$.  
The average period event rate among the participants is given by
 \begin{align}
 \overline{ \overline{p}}(\{\mathbf{x}_{i,11:TJ}\}^N_{i=1}) 
   & = \frac{1}{N}\sum_{i=1}^{N}\bar p_{i}(\mathbf{x}_{i,11:TJ}) \nonumber\\
    & = \frac{1}{NTJ}\sum_{i=1}^{N}\sum_{t=1}^{T}\sum_{j=1}^{J}\E(Y_{itj}(\mathbf{x}_{i,11:TJ})) \label{eq:meta.E}\\ 
    &= \frac{\sum_{i=1}^{N}\sum_{t=1}^{T}\sum_{j=1}^{J}\Phi(\beta_{0,i}^* + \beta_{1,i}^*x_{itj}  +  \beta_{w,i}^{*\top} \mathbf{g}(w_{itj}))}{NTJ}. \nonumber
\end{align}
where $\bar p_{i}(\{\mathbf{x}_{i,11:TJ}\}^N_{i=1})$ is the average period event rate for participant $i$.
Assuming all participants receive the same treatment path, 
$\mathbf{x}_{i,11:TJ} = \mathbf{1}$ or $\mathbf{0}, i=1,...,N$, the period average causal effect of continuous exposure to alcohol among participants is estimated by
\begin{align*}
    \overline{\overline\tau} (\mathbf{1}, \mathbf{0}) 
    & =  \frac{\overline{\overline{p}}(\mathbf{1})}{1-\overline{\overline{p}}(\mathbf 1)}\frac{1-\overline{\overline{p}}(\mathbf 0)}{\overline{\overline{p}}(\mathbf 0)}. \\
\end{align*}
The period average causal effect of drinking behavior among participants is estimated by
\begin{align*}
    & \overline{\overline{\tau}} (\{\mathbf{X}_{i,11:TJ}(\mathbf{1})\}^N_{i=1}, \mathbf{0}) 
      = \frac{\overline{\overline{p}}(\{\mathbf{X}_{i,11:TJ}(\mathbf{1})\}^N_{i=1})}{1-\overline{\overline{p}}(\{\mathbf{X}_{i,11:TJ}(\mathbf{1})\}^N_{i=1})}   \frac{1-\overline{\overline{p}}(\mathbf 0)}{\overline{\overline{p}}(\mathbf 0)},\\
\end{align*}
where $\{\mathbf{X}_{i,11:TJ}(\mathbf{1})\}^N_{i=1}$ denotes all potential treatment paths for $N$ participants when assigned to trigger on conditions throughout day $J$ of period $T$.  
The period average causal effect of intention-to-treat among participants is estimated by
\begin{align*}
    & \overline{\overline\theta} (\mathbf{1}, \mathbf{0})
      = \overline{\overline{\tau}} (\{\mathbf{X}_{i,11:TJ}(\mathbf{1})\}^N_{i=1}, \{\mathbf{X}_{i,11:TJ}(\mathbf{0})\}^N_{i=1}) 
      = \frac{\overline{\overline{p}}(\{\mathbf{X}_{i,11:TJ}(\mathbf{1})\}^N_{i=1})}{1-\overline{\overline{p}}(\{\mathbf{X}_{i,11:TJ}(\mathbf{1})\}^N_{i=1})}   \frac{1-\overline{\overline{p}}(\{\mathbf{X}_{i,11:TJ}(\mathbf{0})\}^N_{i=1})}{\{\mathbf{X}_{i,11:TJ}(\mathbf{0})\}^N_{i=1})}.\\ 
\end{align*}

\begin{comment}
 \begin{align*}
     P(Y(0)=1) &= \E(Y(0))  = \E\E(Y(0)|w) = \E(\Pr(Y(0)=1|w)) = \E(Y^*(0)|w)  \\
     &= \E(\Phi(b_{0} + b_{w}w ))\\
     &= \int_w \int_{\bm\beta} \int_{\bm b}\Phi(b_{0} + b_w^*{}^\top \mathbf{f}(w_{tj}))\phi(w)\pi(\bm b|\bm\beta)\pi(\bm\beta|w,Y)d\bm b d\bm\beta dw\\
    %\widehat {P}(Y(0)=1) 
    &\approx \frac{1}{NL}\sum_{ij}\Phi(\hat b_0 + \hat b_w w_{ij})
             &\stepcounter{equation}\tag{\theequation}\label{p0_pop}
 \end{align*}
 
  \begin{align*}
     P(Y(1)=1) &= \E(Y(0))  = \E\E(Y(1)|w) = \E(\Pr(Y(1)=1|w)) = \E(Y^*(1)|w)  \\
     &= \E(\Phi(b_{0} + b_1 + b_{w}w ))\\
     &= \int_w \int_{\bm\beta} \int_{\bm b}\Phi(b_{0} + b_1 + b_{w}w )\phi(w)\pi(\bm b|\bm\beta)\pi(\bm\beta|w,Y)d\bm b d\bm\beta dw\\
     &\approx \frac{1}{NL}\sum_{ij}\Phi(\hat b_0 + \hat b_1 + \hat b_w w_{ij})
             &\stepcounter{equation}\tag{\theequation}\label{p1_pop}
 \end{align*}

 Without time-varying or participant-varying covariates, the estimated probabilities $\hat P(Y(0)=1)$ and $\hat P(Y(1)=1)$ are reduced to $\Phi(\hat b_0 )$ and  $\Phi(\hat b_0 + \hat b_1)$, respectively.
\end{comment}

Similar to single participant analyses, we assumed assumption NCO in the estimation model, i.e., expectations in Equation (\ref{eq:meta.E}) on the treatment path $\mathbf{x}_{i,11:tj}$ are restricted to the current visit $\mathbf{x}_{i,tj}$ of participant $i$.

\subsection{Jags Implementation}

All models were fit using Markov chain Monte Carlo (MCMC) implemented in JAGS 4-10 called from R 4.2.2 by the `\textsf{rjags}' package. JAGS code for model CO is provided in Appendix \ref{section: appd_code}.  We ran three MCMC chains for 5000 iterations each with a burn-in of 15000 iterations and saved every 4th iteration.
We monitored the chains using the Gelman-Rubin
diagnostic \citep{gelman1992inference}.
All parameters were monitored. We assumed convergence if the ratio of between-chain to within-chain variance is less than 1.01 for each parameter monitored.
Trace plots were also checked for visual verification.

To assess model adequacy, we performed posterior predictive checks \citep{gelman2013bayesian}. Essentially, this entails comparing the observed data to the new samples generated from the posterior distributions. Specifically, let $y^{rep}_s$ be a sample from the posterior predictive distribution of $f(y|\bm \theta_s)$ where $\bm \theta_s$ is a set of parameters from the $s$th iteration of MCMC simulation and T($y^{rep}_s,\bm \theta_s$) be a test quantity based on posterior predictive $y^{rep}_s$. We can compare T($y^{rep}_s,\bm \theta_s$) with T($y^{obs}_s,\bm\theta_s$) that is based on the observed data $y^{obs}$. To describe the degree of the discrepancy, we calculate the Bayesian p-value as $Pr(T (y^{rep}, \bm \theta ) > T (y^{obs}, \bm \theta ) + \frac{1}{2}T (y^{rep}, \bm \theta ) = T (y^{obs},\bm \theta))$, denoted by $pB$. An extreme Bayesian p-value suggests inconsistency of the assumed model with the observed data. We examined the following test quantities: (a) Overall-goodness-of-fit (deviance), \\
    $ T_1(y, \bm \theta) = -2\sum_{t,j=1}^{T,J} \left[\mathbbm{1}(y_{tj} =1)\text{log}(\hat \pi_{tj}|\bm \theta)  +\mathbbm{1}(y_{tj} =0)\text{log}(1-\hat \pi_{tj}|\bm \theta)\right],$
    where $\hat \pi_{tj}$ is the estimated probability of $Y_{tj}=1$ at $tj$th time given $\bm \theta$; 
(b) The number of events in the observed sequence,
    $T_2(y,\bm \theta)= \sum_{t,j=1}^{T,J} \mathbbm{1}(y_{tj} =1 )$; 
and (c) the number of outcome changes in the observed sequence:
    $T_3(y,\bm \theta)= \sum_{t,j=1}^{T,J-1} \mathbbm{1}(y_{tj} \ne y_{t(j+1)} ).$

\section{Numerical Illustration}
\subsection{Simulation Studies}
We conducted simulations focusing on one individual trial. Assuming no measured confounders, 1-day carryover effect and $X_{tj}$ and $Y_{tj}$ are positively correlated (i.e.,  $s=1$), a trial with $T$ periods and $J$ days within each period can be simulated according to the following latent model, 
\begin{align*}
     X_{tj}^\dagger &=  \alpha_{0} + \alpha_{1}r_t + u_{tj} +  \epsilon_{tj},\nonumber \\ 
     Y_{tj}^\dagger &=  \beta_{0} + \beta_{1}X_{tj}  + \beta_{2}X_{tj-} + u_{tj} + \eta_{tj}, \\
     u_{tj}&\sim N(0,\sigma^2_u),\\
     \epsilon_{tj}, \eta_{tj} &\sim N(0,1), \text{mutually independent.}
\end{align*}
Equivalently, we used the following re-scaled latent models to generate the data,
\begin{align}
     &X_{tj}^* =  \alpha_{0}^* + \alpha^*_{1}r_t +  u^*_{tj}+ \epsilon^*_{tj},\quad  X_{tj} = \mathbbm{1}(X_{tj}^*>0),\nonumber\\
     &Y_{tj}^* =  \beta_{0}^* + \beta_{1}^*X_{tj}   + \beta_{2}^*X_{tj-1} + u^*_{tj}+ \eta^*_{tj}, \quad  Y_{tj} = \mathbbm{1}(Y_{tj}^*>0),   \label{eq:sim_gen}\\
    &\epsilon^*_{tj}, \eta^*_{tj} \sim N(0,1-\rho), \nonumber\\
    &\bm u^* = (u^*_{11},u^*_{12},...,u^*_{TJ}) \sim N(0, \Sigma_u), \quad
        \Sigma_u = \rho\begin{pmatrix}
    1& \rho_u & \hdots & \rho_u^{L-1} \nonumber\\
     \rho_u &\ddots& & \rho_u^{L-2}  \nonumber\\
    \vdots &&\ddots & \vdots \\
   \rho_u^{L-1}&\rho_u^{L-2}&\hdots& 1 \end{pmatrix}, \nonumber\\
   &r_t = I_t, I_t \sim Ber(0.5) \text{ for } t = 1, 3, .., T-1; I_t = 1-I_{t-1} \text{ for } t= 2, 4, ..., T,\nonumber 
\end{align}
where the starred quantities are $c$ times unstarred ones with $c =\frac{1}{\sqrt{1+\sigma_u^2}}$.
The treatment assignments were generated to ensure an equal number of `on' and `off' periods.

We designed simulations considering the following aspects (refer to Tables ~\ref{tab:sim_design} and \ref{tab:sim_design_CO} for details):

(a) The magnitude of unmeasured confounding ($\rho$) was varied from 0.1 to 0.9. This allowed us to assess the impact of different levels of unmeasured confounding, which was of primary interest in our study.
(b) The duration of the trials (50, 200, or 1000 days) was another key factor we examined. Both (a) and (b) were our main areas of focus.
%% NOTE: 50 = 4 periods, each 13 days; 200 = 20 periods, each 10 days; 1000 = 100 periods, each 10 days. 
(c) The impact of IV strength, defined as the odds ratio of consuming alcohol $\frac{\Phi(\alpha_0^* +\alpha_1^*)}{1-\Phi(\alpha_0^* +\alpha_1^*)}\frac{\Phi(\alpha_0^*)}{1-\Phi(\alpha_0^*)}$ was also evaluated. A strong IV was represented by a large odds ratio of 6 and a weak IV by a relatively small odds ratio of 1.5. 
Values of $\alpha_0$ were chosen to reflect the drinking rate in the ISTOP study.  
(d) Auto-correlation (on confounding) strength ($\rho_u$) was explored at two levels: high ($\rho_u$ = 0.7) and moderate ($\rho_u = 0.3$).
(e) We also considered a large alcohol effect, represented by an odds ratio of 2.4 for having an AF event (i.e., log$(\tau(\mathbf 1, \mathbf 0)) = 0.86$ in Table \ref{tab:sim_design}), as well as a small effect, represented by an odds ratio of 1.5 (i.e., log$(\tau(\mathbf 1, \mathbf 0)) = 0.41$ in Table \ref{tab:sim_design}).  Values of $\beta_0$ were chosen to reflect the AF event rate in the ISTOP study. 
Odds ratios for alcohol effect differ between Table \ref{tab:sim_design_CO} and Table \ref{tab:sim_design} because of the inclusion of the carryover term $\beta_2^*$.
(f) We first considered scenarios without carryover effect 
(i.e.,  $\beta_2^*=0$ in equation (\ref{eq:sim_gen})) and then scenarios with 1-day carryover effect. There were fewer scenarios with carryover effect because results showed similar patterns to those from models with no carryover. 
Each scenario was simulated with 500 replicates. 

\textbf{Specification of prior distributions}

We imposed weakly informative priors on $\alpha^*_1,\beta^*_1$ in the individual-participant model, considering that adopting non-informative priors on parameters in a probit regression can lead to unstable posterior estimates of probabilities as well as odds ratios and risk ratios \citep{lemoine2019moving}. In particular, we assumed truncated Normal distributions for coefficients
    $\alpha^*_{0} \sim N(0,\sigma_{\alpha0}^2) [L_{0},U_{0}],$
    $\beta^*_{0} \sim N(0,\sigma_{\beta0}^2) [L_{0},U_{0}],$
    $\alpha^*_{1} \sim N(0,\sigma_{\alpha1}^2)[L_1,U_1],$
    $\beta^*_{1} \sim N(0,\sigma_{\beta1}^2)[L_1,U_1],$ and 
    $\beta^*_{2} \sim N(0,\sigma_{\beta2}^2)[L_1,U_1]$ for model CO,  where the notation $N(0,\sigma^2)[L,U]$ indicates a normal distribution truncated on the lower end at $L$ and the upper end at $U$.
   We assigned a uniform prior $U(0,1)$ to $\rho_u$ and $\rho^*$, where $\rho = \rho^{*2}$, to allow more weight on the smaller $\rho$ values (the mean of the prior for $\rho$ is approximately 0.3). 
    The assumptions of non-negative $\rho$ (s=1) and $\rho_u$  could be relaxed.
The truncation was used to further regularize the priors to avoid extreme values. 

Prior variances of $\sigma_{\alpha0}^2$ and $\sigma_{\beta0}^2$ were set to 1 to reflect the underlying probabilities and stabilize computations. 
A variance of $\sigma_{\beta1}^2=1$ suggests that 
$p(\beta^*_1\le2)=0.98$. 
Given a baseline probability of 0.15 (i.e., $\Phi(\beta^*_0) =0.15$) and assuming other covariates are zero, this translates to $p(\Phi(\beta^*_0+\beta^*_1)\le 0.83)=0.98$ and a corresponding odds ratio of 28.
A larger value of $\sigma_{\beta1}^2$ suggests a stronger belief in a higher probability of the event happening under intervention, as well as a greater odds ratio.  
Sensitivity analyses were conducted for $\sigma_{\alpha1}^2$ and $\sigma_{\beta1}^2$ to vary between 0.2 and 5 as well as for the bounds of the truncation. We set $\sigma_{\alpha1}^2=\sigma_{\beta1}^2 =\sigma_{\beta2}^2=1$ and $[-4,4]$ for $[L_0, U_0]$ and $[L_1, U_1]$ in the formal analysis.

We fit model NCO for scenarios without carryover effect and model CO for scenarios with carryover effect. For each scenario, we estimated the two estimands using the proposed IV method: the odds ratio of daily drinking effect (denoted by IV\textunderscore DD) and the odds ratio of usual drinking effect (denoted by IV\textunderscore UD).
We compared these estimators with ITT, conventional ``as treated'' (AT) and ``per protocol'' (PP) approaches.
AT analysis compares the AF risk among those days when a participant drinks with that among those when he/she does not drink, regardless of the treatment assignment. 
PP analyzes a subset of days when a participant complies with the treatment assignments \citep{hernan2012beyond} and ignores days of non-compliance. 
The AT and PP estimators did not account for AR(1) in outcomes and were not calculated for the scenarios with a carryover effect because of their poor performance for scenarios without a carryover effect. 

Accuracy of estimation for a parameter $\theta$ was evaluated using 
(a) percentage bias, computed as ${\sum_{n=1}^{N_{sim}} (\hat \theta_n - \theta)}/{(\theta N_{sim}})\times 100\%$, where $N_{sim}$ denotes the total number of replicates ($N_{sim} = 500)$; 
%R(b) root mean squared error, computed as $\frac{\sum_{n=1}^{N_{sim}}(\hat \theta_n - \theta)^2}{N_{sim}}$; 
and (b) coverage probability, i.e., the proportion of the truth covered by the posterior 95\% credible intervals. 

\begin{figure}[H]
     \centering
     \begin{subfigure}[b]{0.95\textwidth}
         \centering
         \includegraphics[width=0.95\textwidth]{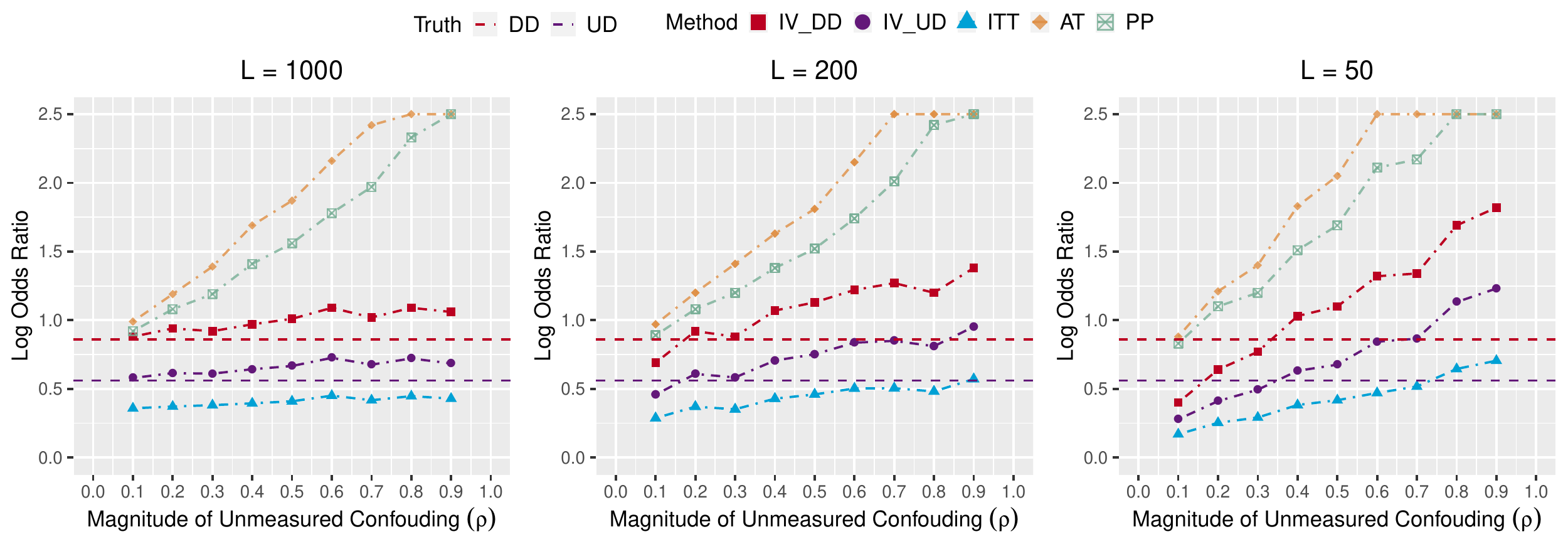}
        \vspace{0.1\baselineskip}
        \caption{}
         \label{fig:bias}
     \end{subfigure}
     \vfill
     \begin{subfigure}[b]{0.95\textwidth}
         \centering
         \includegraphics[width=0.95\textwidth]{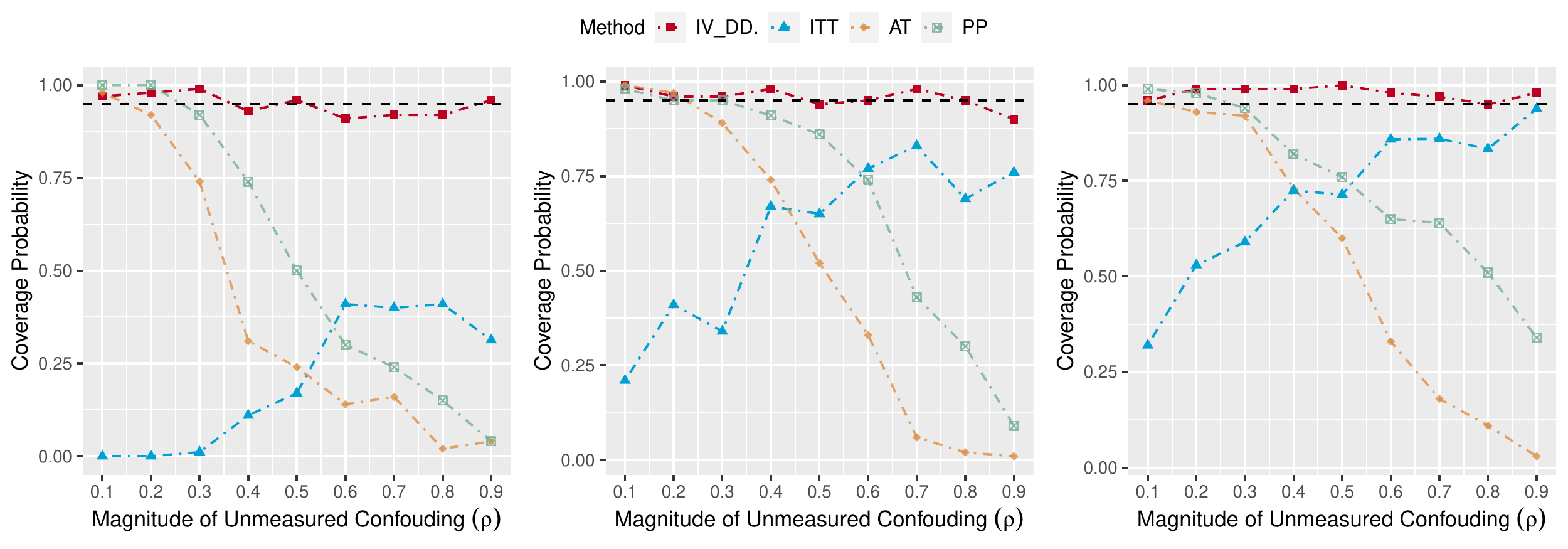}
         \vspace{0.1\baselineskip}
        \caption{}
         \label{fig:covp.alc}
     \end{subfigure}
      \vfill
     \begin{subfigure}[b]{0.95\textwidth}
         \centering
         \includegraphics[width=0.95\textwidth]{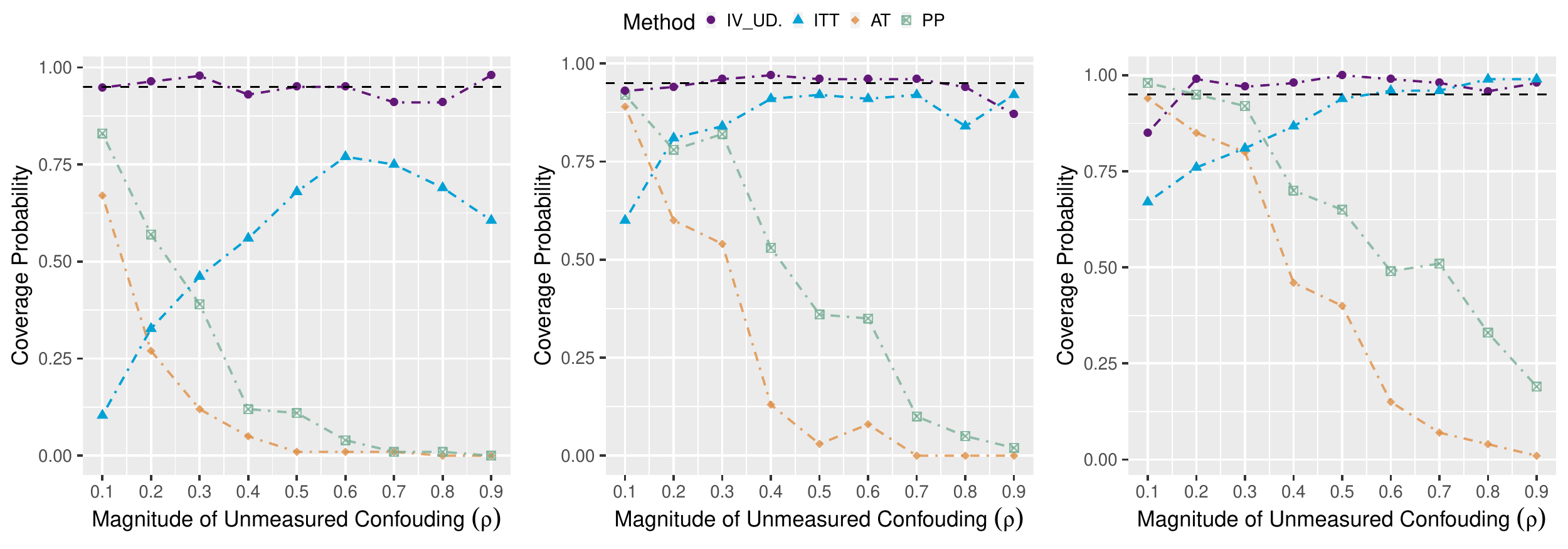}
         \vspace{-0.1\baselineskip}
        \caption{}
         \label{fig:covp.dhav}
     \end{subfigure}
        \caption{Performance for scenarios 1000L1-L9, 200L1-L9, and 50L1-L9 without carryover effect. 1000L1-L9 indicates scenarios 1-9 with $L=1000$.  (a) Average of 500 posterior means of odds ratios on the log scale (estimates capped at 2.5). The true values of daily drinking effect (log$(\overline \tau(\mathbf 1, \mathbf 0)$) and of usual drinking (log$(\overline\theta(\mathbf 1, \mathbf 0)$) are shown in red and purple dashed lines, respectively. (b) Coverage probability of 95\% CrI for the daily drinking effect. (c) Coverage probability of 95\% CrI for the usual drinking effect. The black dashed line represents the 95\% nominal level. $L$ represents the total number of measurements throughout one single N-of-1 trial. }
        \label{fig:simulation_results}
\end{figure}

\begin{figure}[H]
     \centering
     \begin{subfigure}[b]{0.95\textwidth}
         \centering
         \includegraphics[width=0.95\textwidth]{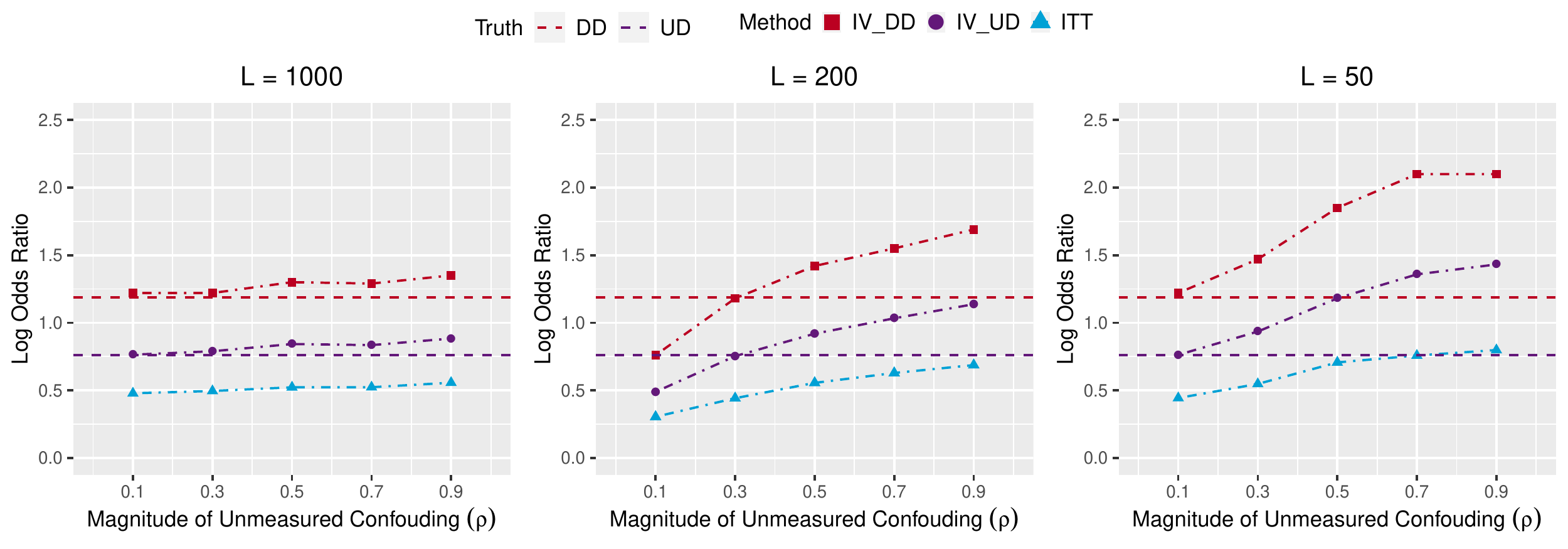}
        \vspace{0.1\baselineskip}
        \caption{}
         \label{fig:bias.CO}
     \end{subfigure}
     \vfill
     \begin{subfigure}[b]{0.95\textwidth}
         \centering
         \includegraphics[width=0.95\textwidth]{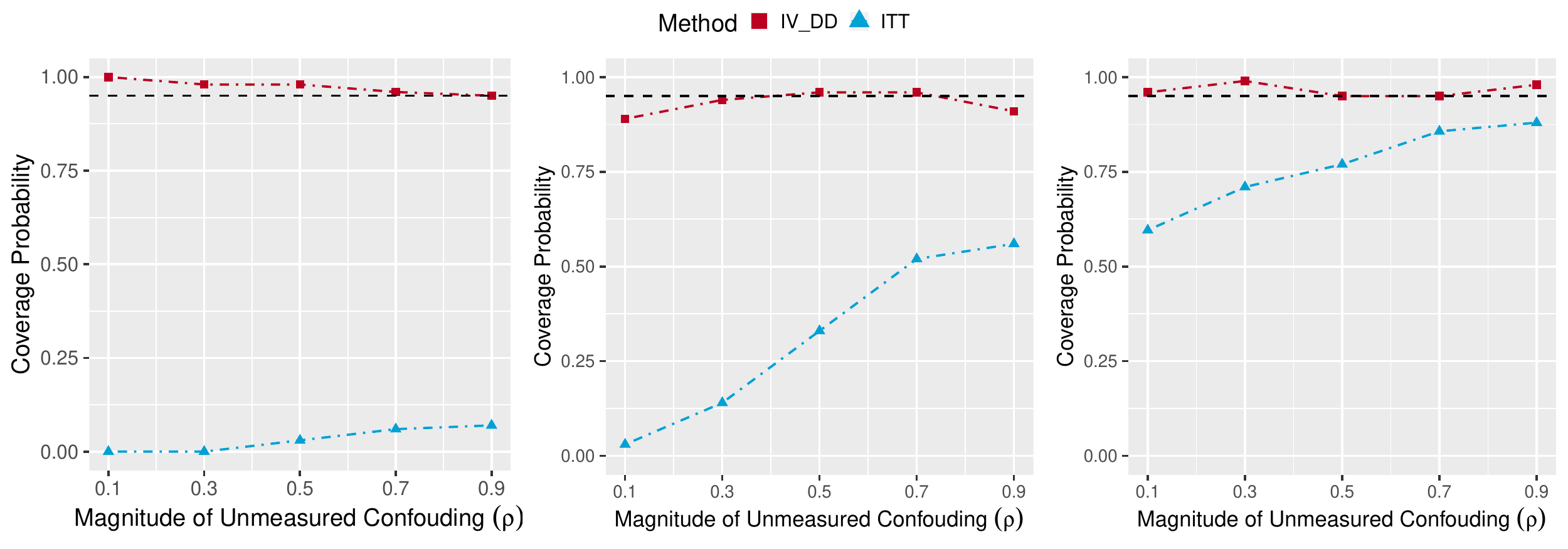}
         \vspace{0.1\baselineskip}
        \caption{}
         \label{fig:covp.alc.CO}
     \end{subfigure}
      \vfill
     \begin{subfigure}[b]{0.95\textwidth}
         \centering
         \includegraphics[width=0.95\textwidth]{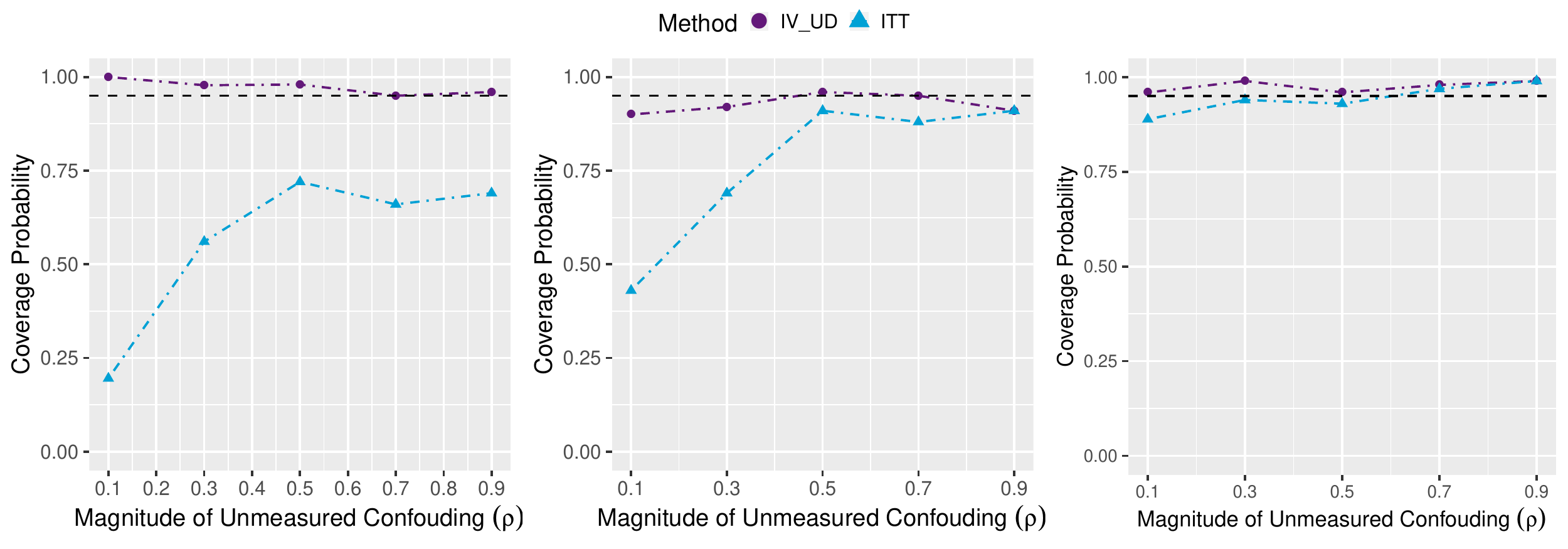}
         \vspace{-0.1\baselineskip}
        \caption{}
         \label{fig:covp.dhav.CO}
     \end{subfigure}
        \caption{Performance for scenarios 1000L1-L9, 200L1-L9, and 50L1-L9 without carryover effect. 
        1000L1-L9 indicates scenarios 1-9 with $L=1000$. (a) Average of 500 posterior means of odds ratios on the log scale. The true values of daily drinking effect (log$(\overline \tau(\mathbf 1, \mathbf 0)$) and of usual drinking (log$(\overline \theta(\mathbf 1, \mathbf 0)$) are shown in red and purple dashed lines, respectively. (b) Coverage probability of 95\% CrI for the daily drinking effect. (c) Coverage probability of 95\% CrI for the usual drinking effect. The black dashed line represents the 95\% nominal level. $L$ represents the total number of measurements throughout one single N-of-1 trial. }
        \label{fig:simulation_results_CO}
\end{figure}

Figures~\ref{fig:simulation_results} and \ref{fig:simulation_results_CO} show the odds ratio estimates on the log scale and the coverage probabilities for scenarios without carryover effect (1000L1-1000L9, 200L1-200L1, 50L1-50L9) and with 1-day carryover effect (1000L1CO-1000L9CO, 200L1CO-200L9CO, 50L1CO-50L9CO), respectively.
The actual values of the performance measures are included in Tables \ref{tab:sim_main_tb}, \ref{tab:sim_main_tb_db}, \ref{tab:sim_main_tb_co}, and \ref{tab:sim_main_tb_db_co}. 

In summary, using study randomization as IV allowed us to obtain causal effect estimates for continuous (daily) alcohol exposure and usual drinking behaviors, comparing them with complete abstinence from alcohol use. These causal effects cannot be revealed from the ITT, AT, or PP analyses. 
ITT had greater underestimation when confounding was small while AT or PP showed more over-estimation with increasing confounding. 
With a longer trial duration (which we denote by $L$, representing the total number of measurements for one participant), the performance of IV improved and became more stable, even in the presence of varying levels of unmeasured confounding effects. When the trial duration was short, IV estimates were not accurate unless one had a good idea of the prior distribution on the confounding.  
Minimal bias was observed at $\rho = 0.3$ for $L = 50$ or 200, where true confounding was equal to the mean of its prior (Figure \ref{fig:rho_prior}). 
When the true confounding $\rho$ was smaller than 0.3, the confounding tended to be overestimated, and thus the true treatment effect was underaccounted for.

The coverage probabilities of the credible intervals for IV estimates were generally close to the nominal level of 0.95. 

The ITT analyses contrasted the outcomes under two randomization conditions, hence revealing only the causal effect of the treatment randomizer, not the alcohol. Since a participant could choose to drink during the trigger OFF condition and not to drink during the ON condition (non-compliance), the ITT tended to underestimate the effects of alcohol due to the well-known `bias towards the null' phenomenon \citep{hernan2012beyond, neto2016towards}.
AT and PP generally do not have causal interpretations and showed substantial positive biases. These positive biases were also challenging for IV to completely correct when limited information was available due to a small sample size. 

Between the two IV estimates, IV\textunderscore UD performed comparably to IV\textunderscore DD but with better accuracy.
For instance, given a confounding magnitude of 0.4, the IV\textunderscore DD estimate exhibited a bias of 12.89\% (Table \ref{tab:sim_main_tb}), while IV\textunderscore UD showed a bias of -1.7\% (Table \ref{tab:sim_main_tb_db}).  
This difference could be attributed to the fact that IV\textunderscore UD was a contrast between usual drinking behaviors and complete abstinence, where the outcomes under usual drinking behaviors were partially observed (during trigger ON periods). Meanwhile, IV\textunderscore DD contrasted the potential outcomes of drinking daily and complete abstinence; both were counterfactual. 

There are scenarios in which the performance of IV estimates can be compromised, such as a weak IV, a small treatment effect, or a strong auto-correlation (see Tables \ref{tab:sim_bias} and \ref{tab:sim_covp} for results of all scenarios with varied prior variances on $\beta_1^*$), though IV approaches still provided the most trustable estimates. Specifically, in scenarios 1000L10 or 1000L11 where IV was weak and data were substantial, biases of -13.53\% and 23\% were observed, respectively, indicating larger biases compared to the two scenarios (L1 and L5) with strong IV and the same amount of auto-correlation, where the bias was 1.65\% and 17.69\% respectively.  Additionally, in scenarios with a strong auto-correlation ($\rho_u$ = 0.7 in L12 and L13), IV consistently underestimated the odds ratio, with more negative biases in smaller confounding effects. Conversely, in scenarios with a small alcohol effect (log$\tau(\mathbf 1, \mathbf 0)$ = 0.41 in L14 and L15), IV tended to overestimate the odds ratio, showing more positive biases in larger confounding effects. 
95\% credible intervals tended to be wide leading to overcoverage when IV was weak as less information was available to provide accurate estimates, or when autocorrelation was strong as the effective sample size was small. 
A larger variance on the priors resulted in more positive biases as it supported higher values of $\beta_1^*$ in the posteriors especially when the sample size was small. 

In scenarios with true 1-day carryover effect, model CO demonstrated similar patterns (Figure \ref{fig:simulation_results_CO}, Tables \ref{tab:sim_main_tb_co} and \ref{tab:sim_main_tb_db_co}) compared to the performance of model NCO in scenarios without carryover effect (Tables \ref{tab:sim_main_tb} and \ref{tab:sim_main_tb_db}, and Figure \ref{fig:simulation_results}). In particular, bias increased with a larger confounding effect and with a smaller total number of measurements. When $L=1000$, bias\% was less than 10\% for both IV\textunderscore DD and \textunderscore UD for confounding $\rho \le 0.7$. Compared to model NCO, model CO showed a slightly smaller bias in estimating the daily drinking effect (DD). 
This was attributed to a stronger drinking effect due to the carryover from drinking on the previous day.

Figure \ref{fig:predcheck_boxplot_sim} show examples of our posterior predictive checks of the Bayesian models. These checks showed good approximations of the empirical distributions of the deviance and non-extreme p-values for the three examined statistics, demonstrating good fits of IV models to the data.

\subsection{Application to I-STOP-AFib Study}

\begin{table}[H]
\resizebox{\textwidth}{!}{%
\begin{tabular}{lcccccccccc}
\toprule
\multicolumn{1}{c}{ID} & Event/Total & Estimand                  & \multicolumn{4}{c}{Single-participant Analysis}        & \multicolumn{4}{c}{Meta Analysis}                \\
\cmidrule(l){2-2}\cmidrule(l){4-7}\cmidrule(l){4-7}\cmidrule(l){8-11}  
                         &\multicolumn{1}{c}{n/N}   &  & \multicolumn{2}{c}{NCO}  & \multicolumn{2}{c}{CO}  & \multicolumn{2}{c}{NCO} & \multicolumn{2}{c}{CO} \\
 &
  \multicolumn{2}{l}{} & 
  \begin{tabular}[c]{@{}c@{}}Odds Ratio \\ Estimate (95\% Crl)\end{tabular} &
  p(OR$>$1) & 
  \begin{tabular}[c]{@{}c@{}}Odds Ratio \\ Estimate (95\% Crl)\end{tabular} &
  p(OR$>$1) & 
  \begin{tabular}[c]{@{}c@{}}Odds Ratio \\ Estimate (95\% Crl)\end{tabular} &
  p(OR$>$1) & 
  \begin{tabular}[c]{@{}c@{}}Odds Ratio \\ Estimate (95\% Crl)\end{tabular} &
  p(OR$>$1) \\
  \midrule
\multicolumn{1}{c}{1} & 2/42 & Daily Drinking (IV\textunderscore DD)    & 1.8 (0.2, $>10.0$)  & 0.62 & 0.4 (0.1, $>10.0$) & 0.23 & 0.9 (0.3,2.6)   & 0.47  & 1.4 (0.2,4.2)  & 0.57  \\
                      &  & Usual Drinking (IV\textunderscore UD)    & 1.4 (0.6,7)       & 0.61 & 1.0 (0.4,3.1)      & 0.4  & 1.0 (0.7,1.6)     & 0.48  & 1.1 (0.6,2)    & 0.6   \\
                      &  & Drinking Assignment (ITT) & 1.1 (0.7,2.4)     & 0.52 & 0.9 (0.5,1.8)    & 0.34 & 1.0 (0.8,1.3)     & 0.5   & 1.1 (0.7,1.6)  & 0.61  \\ 
 & &  &  &     & &     &    &   &  \\ 
\multicolumn{1}{c}{2} & 4/42 & Daily Drinking (IV\textunderscore DD)    & 0.1 (0,0.8)       & 0.01 & 0.2 (0,2.6)      & 0.09 & 0.8 (0.2,1.7)   & 0.32  & 1.0 (0.1,2.5)    & 0.43  \\
                      & & Usual Drinking (IV\textunderscore UD)    & 0.3 (0.1,1)       & 0.01 & 0.5 (0.1,2.2)    & 0.16 & 0.9 (0.4,1.5)   & 0.32  & 1.0 (0.3,2)      & 0.44  \\
                      &  & Drinking Assignment (ITT) & 0.4 (0.1,1)       & 0.01 & 0.5 (0.2,1.9)    & 0.14 & 0.9 (0.4,1.4)   & 0.32  & 1.0 (0.4,1.9)    & 0.44  \\
                        & &  &  &     & &    &    &   &  \\ 
\multicolumn{1}{c}{3} & 27/42 & Daily Drinking (IV\textunderscore DD)    & 0.8 (0.2,2.8)     & 0.34 & 0.8 (0.1, $>10.0$) & 0.4  & 0.9 (0.5,1.7)   & 0.41  & 1.2 (0.4,2.7)  & 0.54  \\
                      &  & Usual Drinking (IV\textunderscore UD)    & 0.8 (0.2,2.6)     & 0.34 & 0.8 (0.1,5.2)    & 0.38 & 0.9 (0.5,1.6)   & 0.41  & 1.1 (0.4,2.5)  & 0.53  \\
                      &  & Drinking Assignment (ITT) & 0.8 (0.3,2.5)     & 0.34 & 0.8 (0.2,4)      & 0.39 & 0.9 (0.5,1.6)   & 0.41  & 1.1 (0.4,2.3)  & 0.53  \\
                        & &  &  &     & &     &    &   &  \\ 
\multicolumn{1}{c}{4} & 10/42 & Daily Drinking (IV\textunderscore DD)    & 1.1 (0.1,8.3)     & 0.57 & 0.7 (0,8.3)      & 0.41 & 1.0 (0.5,2.2)     & 0.5   & 1.3 (0.4,3.2)  & 0.62  \\
                      &  & Usual Drinking (IV\textunderscore UD)    & 1.1 (0.3,4.4)     & 0.57 & 0.9 (0.2,4.1)    & 0.43 & 1.0 (0.6,1.7)     & 0.5   & 1.2 (0.6,2.2)  & 0.63  \\
                      &  & Drinking Assignment (ITT) & 1.0 (0.5,1.9)       & 0.56 & 0.9 (0.4,1.8)    & 0.41 & 1.0 (0.7,1.3)     & 0.51  & 1.1 (0.7,1.6)  & 0.63  \\
                        & &  &  &     & &     &    &   &  \\ 
\multicolumn{1}{c}{5} & 8/42 & Daily Drinking (IV\textunderscore DD)    & 0.3 (0,2.4)       & 0.15 & 0.9 (0.1, $>10.0$) & 0.49 & 0.9 (0.4,1.8)   & 0.4   & 1.3 (0.3,3.2)  & 0.58  \\
                      &  & Usual Drinking (IV\textunderscore UD)    & 0.6 (0.2,1.7)     & 0.15 & 1.2 (0.3,5)      & 0.62 & 0.9 (0.5,1.5)   & 0.4   & 1.1 (0.5,2.3)  & 0.59  \\
                      &  & Drinking Assignment (ITT) & 0.7 (0.2,1.5)     & 0.15 & 1.1 (0.4,3.5)    & 0.58 & 0.9 (0.6,1.4)   & 0.4   & 1.1 (0.6,1.9)  & 0.59  \\
                        & &  &  &     & &     &    &   &  \\ 
\multicolumn{1}{c}{6} & 0/41& Daily Drinking (IV\textunderscore DD)    & 1.1 (0.3, $>10.0$)  & 0.16 & 0.3 (0.1,5.2)    & 0.1  & 0.9 (0.2,3.1)   & 0.44  & 1.7 (0.1,7)    & 0.57  \\
                      &  & Usual Drinking (IV\textunderscore UD)    & 1.1 (0.6,7.8)     & 0.16 & 0.6 (0.2,3.5)    & 0.19 & 1.0 (0.4,2.2)     & 0.44  & 1.4 (0.4,4.1)  & 0.6   \\
                      &  & Drinking Assignment (ITT) & 1.1 (0.6,6.2)     & 0.16 & 0.7 (0.2,2.6)    & 0.18 & 1.0 (0.5,2)       & 0.44  & 1.3 (0.4,3.2)  & 0.6   \\
                        & &  &  &     & &     &    &   &  \\ 
\multicolumn{1}{c}{7} &4/41 & Daily Drinking (IV\textunderscore DD)    & 4.3 (0.4, $>10.0$)  & 0.88 & 1.1 (0.1, $>10.0$) & 0.5  & 1.1 (0.4,2.7)   & 0.55  & 1.5 (0.3,4.5)  & 0.64  \\
                     &   & Usual Drinking (IV\textunderscore UD)    & 1.9 (0.8, $>10.0$)  & 0.8  & 1.4 (0.6,5.4)    & 0.75 & 1.0 (0.7,1.6)     & 0.56  & 1.1 (0.7,2.1)  & 0.67  \\
                     &   & Drinking Assignment (ITT) & 1.8 (0.8,8.9)     & 0.79 & 1.3 (0.6,3.5)    & 0.72 & 1.0 (0.8,1.5)     & 0.56  & 1.1 (0.7,1.9)  & 0.67  \\
                        & &  &  &     & &     &    &   &  \\ 
\multicolumn{1}{c}{8} &9/40& Daily Drinking (IV\textunderscore DD)    & 1.8 (0.1, $>10.0$)  & 0.65 & 9.7 (0.2, $>10.0$) & 0.89 & 1.0 (0.5,2.3)     & 0.5   & 1.6 (0.5,4.9)  & 0.7   \\
                     &   & Usual Drinking (IV\textunderscore UD)    & 1.1 (0.7,3.1)     & 0.56 & 1.4 (0.9,4.2)    & 0.83 & 1.0 (0.8,1.2)     & 0.54  & 1.1 (0.9,1.6)  & 0.74  \\
                     &   & Drinking Assignment (ITT) & 1.1 (0.7,2.3)     & 0.46 & 1.1 (0.8,2.4)    & 0.56 & 1.0 (0.9,1.2)     & 0.59  & 1.1 (0.9,1.4)  & 0.73  \\
                        & &  &  &     & &    &    &   &  \\ 
\multicolumn{1}{c}{9} & 2/39 & Daily Drinking (IV\textunderscore DD)    & 27.5 (0.5, $>10.0$) & 0.93 & 3.8 (0.1, $>10.0$) & 0.71 & 1.0 (0.4,3.3)     & 0.52  & 1.8 (0.3,5.8)  & 0.65  \\
                     &   & Usual Drinking (IV\textunderscore UD)    & 1.8 (1, $>10.0$)    & 0.56 & 1.3 (0.9,3.4)    & 0.63 & 1.0 (0.9,1.3)     & 0.64  & 1.1 (0.9,1.5)  & 0.77  \\
                     &   & Drinking Assignment (ITT) & 1.6 (0.8, $>10.0$)  & 0.51 & 1.0 (0.6,1.8)      & 0.3  & 1.0 (0.9,1.2)     & 0.68  & 1.0 (0.9,1.4)    & 0.77  \\
                        & &  &  &     & &     &    &   &  \\ 
\multicolumn{1}{c}{10} & 8/39 & Daily Drinking (IV\textunderscore DD)    & 1.1 (0.2,5.7)     & 0.57 & 1.6 (0.1, $>10.0$) & 0.66 & 1.0 (0.4,2)       & 0.47  & 1.3 (0.4,3.1)  & 0.61  \\
                     &   & Usual Drinking (IV\textunderscore UD)    & 1.1 (0.4,3.5)     & 0.57 & 1.4 (0.3,6.3)    & 0.69 & 1.0 (0.6,1.6)     & 0.47  & 1.2 (0.6,2.2)  & 0.62  \\
                     &   & Drinking Assignment (ITT) & 1.1 (0.5,2.4)     & 0.57 & 1.2 (0.4,3.6)    & 0.67 & 1.0 (0.7,1.4)     & 0.47  & 1.1 (0.6,1.8)  & 0.62  \\
                        & &  &  &     & &     &    &   &  \\ 
\multicolumn{1}{c}{Summary} & 285/1432 & Daily Drinking (IV\textunderscore DD)    & --                & --   & --               & --   & 0.9 (0.5,1.7)   & 0.43  & 1.2 (0.5,2.3)  & 0.63  \\
                     &   & Usual Drinking (IV\textunderscore UD)    & --                & --   & --               & --   & 1.0 (0.7,1.3)     & 0.43  & 1.1 (0.7,1.6)  & 0.64  \\
                     &   & Drinking Assignment (ITT) & --                & --   & --               & --   & 1.0 (0.8,1.2)     & 0.43  & 1.1 (0.8,1.4)  & 0.64 \\
                        & &  &  &     & &     &    &   &  \\ \bottomrule
\end{tabular}%
}
\caption{Posterior odds ratio estimates of daily drinking, usual drinking,
and drinking assignment from single-participant and meta-analyses using model NCO and
model CO. }
\label{tab:istop_results}
\end{table}

% \begin{figure}[H]
%     \centering
%     \vspace{-2.5cm}
%     \includegraphics[width=0.95\textwidth]{iv_itt_habit_plot_1pt_2024.pdf}
%         \vspace{-0.5cm}
%     \caption{Forest plot for odds ratio estimates of alcohol (blue), drinking assignment (red), and drinking behavior (purple) from single-participant analyses using model NCO. }
%     \label{fig:frplot1pt}
% \end{figure}

% \begin{figure}[H]
%     \centering
%         \vspace{-2.5cm}
%     \includegraphics[width=0.95\textwidth]{iv_itt_habit_plot_1pt_carryover_2024.pdf}
%             \vspace{-0.5cm}
%     \caption{Forest plot for odds ratio estimates of alcohol (blue), drinking assignment (red), and drinking behavior (purple) from single-participant analyses using model CO. }
%     \label{fig:frplot1pt_CO}
% \end{figure}

We applied models NCO and CO to analyze the data of I-STOP-AFib Study participants who identified alcohol as the primary trigger ($N=51$). We did not account for measured confounders. 
We first performed individual assessments for each of the 51 participants, 
 then conducted a meta-analysis by combining the data from all participants. 

Results for the ten participants with the least missing treatment selections or outcomes (five participants had complete data, and the rest missed at most three days), as well as the overall averages from meta-analyzing 51 participants, are shown in Table \ref{tab:istop_results}. Results for the remaining participants shared similar trends but were less interesting because their missing data resulted in wide credible intervals. All missing values were explicitly passed in series to Jags and imputed from their posterior predictive distribution \citep{plummer2012jags}.

\textbf{Single Participant Analyses}

We chose priors of the truncated normal $N(0, 1)[-4,4]$ for $\alpha^*_1$ and $N(0, 4)[-4,4]$ for $\beta^*_1$ and $\beta_2^*$ (model CO), $N(\hat{\alpha}^*_0,1)[-4,4]$ for $\alpha_0$ and $N(\hat{\beta}^*_0,1)[-4,4]$ for $\beta_0$, where $\hat{\alpha}^*_0$ is the inverse normal of the probability of alcohol consumption under assignment to alcohol non-exposure throughout each trial in the ISTOP-AFib and $\hat{\beta}^*_0$ is the inverse normal of probability of AF occurrence under alcohol non-exposure throughout each trial.  
We borrowed the empirical information in specifying means as relying solely on non-empirical priors (e.g., setting the mean at 0) would result in unstable computations due to small sample size. 
Although data were involved in prior specification, this approach simulates the scenario where researchers consult an expert's opinion on the baseline event rate for prior specification, which would likely align closely with our choice.
Uniform priors $U(-1,1)$ and $U(0,1)$ were assumed for $\rho_u$ and $\rho$ respectively. $\rho$ is defined for non-negative values only as we assumed $s=1$.

 Analyses of the individual N-of-1 trials revealed considerable heterogeneity in the point estimates of odds ratios across the participants (Table \ref{tab:istop_results}). But the posterior probabilities were generally similar suggesting that the estimates and standard errors are probably moving together. 
 This indicates that the estimate and the standard error are probably moving together.
For some individuals, the three estimates differed, with IV\textunderscore DD usually being the largest and ITT the smallest, suggesting that the assignment had less effect on triggering AF compared to being exposed to alcohol. 
Although the three estimates for most participants indicated a slightly positive causal association with AF, credible intervals were too wide, primarily due to the limited duration of each data sequence,  to make decisions on the utility of alcohol as a trigger. 
Participant 2 showed significantly decreased odds of AF while drinking, as all four events occurred when they did not drink.
Assuming a 1-day carryover effect impacted the estimates differently. 
The differences between models CO and NCO, however, showed no clear patterns across individuals and were unstable due to the small amount of data observed.

\textbf{Meta-analysis}

We adopted a similar strategy for specifying prior distributions as used in one-participant analyses. 
In particular, $a^*_0 \sim N(\hat{a}^*_0, \pi^2_{a0})[L_0,U_0]$ and $b^*_0 \sim N(\hat{b}^*_0,\pi^2_{b0})[L_0,U_0]$, where $\hat a_0^*$ is the inverse normal of the population-level average alcohol consumption under assignment to alcohol abstinence throughout the trial and $\hat{b}^*_0$ is the inverse normal of the population average proportion of AF occurrence under alcohol non-exposure throughout the trial. ``Population'' refers to the I-Stop-AFib participants who were alcohol users and interested in testing alcohol as a trigger. 
$ a^*_1 \sim N(0,\pi_{a1}^2)[L_1,U_1]$ and $b^*_1 \sim N(0,\pi_{b1}^2)[L_1,U_1]$. We set $\pi_{a0} = \pi_{a1} = \pi_{b0} = \pi_{b1}=1$ in the formal analysis. 
 We chose a weakly informative half-Cauchy distribution for variance parameters $\sigma_{a_0},\sigma_{a_1},\sigma_{b_0},\sigma_{b_1}\sim \text{Half-Cauchy}(s)$. We tested the sensitivity of the scale parameter $s$ from 0.2 to 1 and selected $s = 0.4$ for the formal analysis. 
 A Half-Cauchy distribution with a scale parameter of 0.4 yielded a 95th quantile of 5, sufficient for capturing the heterogeneity of study effects.

In contrast to individual analyses, estimates for individuals derived from meta-analysis were shrunk towards the overall mean, resulting in narrower credible intervals and similar estimates across participants. This reflects the small amount of information contributed by each individual on account of the short trials, infrequent events, and missing data. Among 51 participants, there were 35 participants missing more than 7 days of outcome data or treatment selections, with a mean of 13 missing days (sd=12 days) for either treatment selection or outcome. 

In terms of the odds ratio for developing AF, 
none of the overall summary estimates for either of the models indicated evidence of any effect of alcohol. 
These cohort average estimates closely mirrored the majority of individual shrunk estimates and displayed narrower credible intervals.

The posterior predictive checks showed that the models fit the data adequately. 
As shown in Figure \ref{fig:predcheck_boxplot_real} which summarizes Bayesian p-values of all the individual analyses using model NCO in Figure \ref{fig:pcheck_10pt}, the three test quantities are distributed closely to 0.5, suggesting a reasonable fit between the observed data and the posterior draws based on our model. 
Figure \ref{fig:pcheck_10pt} compares the observed values to the posterior predictive distribution for each of the 10 individuals summarized in Table \ref{tab:istop_results}.
%Trace plots are provided in Figure \ref{fig:traceplot} to facilitate visual verification.

\section{Discussion}

Adopting study treatment randomization as an IV, we propose a two-stage Bayesian latent structural model to analyze data from the I-STOP-AF study and define two causal estimands that are relevant in the context of N-of-1 trials: causal effect of continuous alcohol exposure and the causal effect of a participant's drinking behavior. 
Our Bayesian IV structural model accounts for confounding due to self-selection of exposure (noncompliance), auto-correlation in study outcomes (interference among study outcomes), and is able to estimate funcions of model parameters to solve the issue of non-collapsibility. 
Through simulations, we demonstrate that the proposed method largely reduced bias in estimating the two causal effects and greatly improved the coverage probability, compared to the conventional ITT, AT, and PP approaches, especially with a moderate to large (un)measured confounding effect.  We also show that the single-participant Bayesian hierarchical model structure can be readily extended to analyze data from multiple N-of-1 studies to enable a meta-analysis. 

Although our method outperformed the conventional approaches, the performance of the model is compromised when the data sequence is short. 
In such cases, the model's performance in the simulation study is sensitive to the prior specification (Table \ref{tab:sim_bias}) where a larger variance generally leads to greater bias. 
The lack of information in the short series of the I-STOP data also led to inconclusive results. 
On the other hand,  when sample size is sufficiently large, i.e.,  a long sequence of individual data, a natural extension will be to relax parametric assumptions, which can be a future direction for analysis of single time-series data (e.g., \cite{conley2008semi,escobar1995bayesian,mcculloch2021causal,wiesenfarth2014bayesian,florens2012nonparametric,florens2016regularizing,kato2013quasi,liao2011posterior,wang2021scalable}). 

Our proposed method relies on a set of assumptions.  Some of these assumptions are not testable or not easy to verify. One such assumption is the AR(1) structure we impose on the errors within the latent structural model. While a first-order (AR(1)) model may suffice for many N-of-1 trial analyses \citep{wang2020meta,schmid2022bayesian,shadish2014using,moeyaert2014single}, verification of this assumption in our study proves complex due to its application on the latent scale and the relatively brief duration of the trials. As an alternative method to model correlated outcomes, apart from assuming auto-correlation on error terms, we proposed a model integrating a 1-lagged treatment to accommodate carryover effects. This 1-lag assumption suits our focus on the immediate impact of alcohol use on AF, considering alcohol generally clears from the body within 24 hours. While extending to L-lagged effects or modeling the cumulative effects of alcohol use is of interest, examining the long-term impact of alcohol in the paper is limited by the short duration of trials.
\cite{liao2023analysis} proposed a distributed lag model incorporating a general $L$-lagged covariate and a general AR(L) error process to estimate carryover effects in the analysis of N-of-1 trials. This approach holds promise for integration into our framework. 

While we specifically developed and applied our model to the I-STOP-AFib study, it is essential to highlight the generalizability of our approach to other N-of-1 trials with non-compliance issues, particularly those utilizing mobile or wearable devices for data collection.  We encourage further exploration and adoption of similar approaches in a wider range of N-of-1 trial settings.

\section{Acknowledgments}
The authors express our gratitude to Dr. Gregory M. Marcus and his team for providing access to the data in the I-STOP-AFib trials.
I-STOP-AFib trials were funded in part through a Patient-Centered Outcomes Research Institute (PCORI) Award (PPRND-1507- 31321).
Christopher H. Schmid and Tao Liu are partially supported by Institutional Development Award Number U54GM115677 from the National Institute of General Medical Sciences of the National Institutes of Health, which funds Advance Rhode Island Clinical and Translational Research (Advance RI-CTR). The content is solely the responsibility of the authors and does not necessarily represent the official views of the National Institutes of Health.

\section{Conflicts of Interest}
All authors declare that they have no conflicts of interest.

\bibliographystyle{plainnat}
\bibliography{main_arxiv_v5.7}

\newpage
\appendix

%\section*{Appendix}
\section{Appendix}
\beginsupplement

\subsection{Confounding Coefficient} \label{section:confoundingcoef}
\begin{figure}[H]
    \centering
\includegraphics[width=0.7\textwidth]{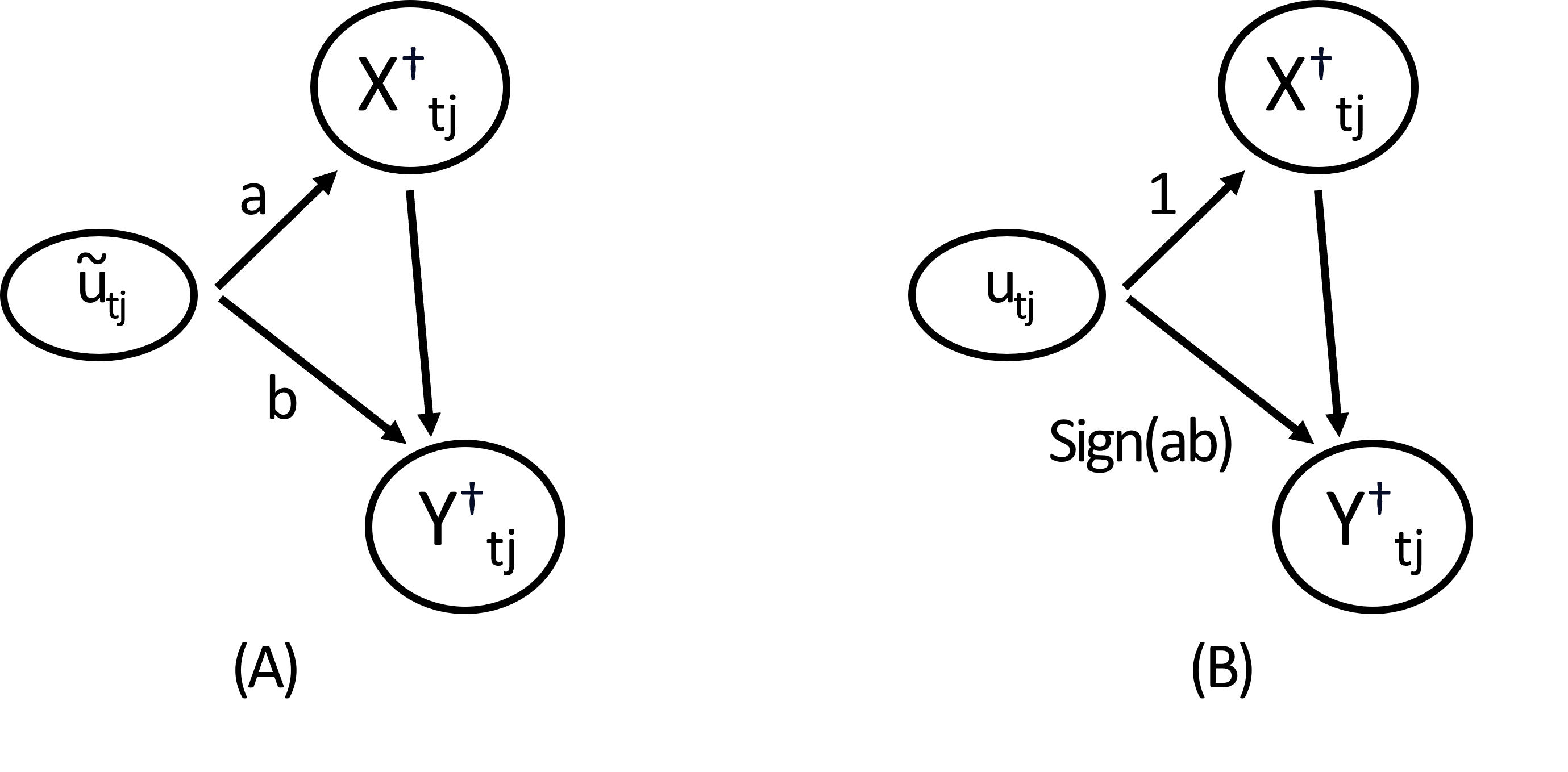}
    \caption{Two scenarios of unmeasured confounding affecting  latent outcomes $X_{tj}^\dagger$ and $Y_{tj}^\dagger$. Scenario (A) shows confounding may have different effects while in scenario (B) the effects may only be different in direction. }
    \label{fig:scenarios_AB}
\end{figure}

Consider two scenarios, where in scenario (A) (Figure \ref{fig:scenarios_AB}) we assume an unmeasured confounder $\tilde{u}_{tj} \sim N(0, \tilde\sigma_u^2)$ and its effect on $X_{tj}^\dagger$ and $Y_{tj}^\dagger$ are $a$ and $b$ respectively; in scenario (B) (Figure \ref{fig:scenarios_AB}) we assume an unmeasured confounder ${u}_{tj} \sim N(0, |ab|\tilde\sigma_u^2)$ and its effect on $X_{tj}^\dagger$ and $Y_{tj}^\dagger$ are 1 and $sgn(ab)$. 
We argue that the two scenarios cause the same magnitude of bias in estimating the causal effect parameter ($\beta_1$ in Eq.(\ref{appd_eq})). Given that the two scenarios cannot be distinguished without knowing $\tilde u_{tj}$ or $u_{tj}$, we propose to use $u_{tj}$ as a summary of the unconfounding effects in our simulation studies. 
Scenario (A) is represented by the following structural models: 
\begin{align}
     X_{tj}^\dagger &=  \alpha_{0} + \alpha_{1}r_t  + a\tilde{u}_{tj} +  \tilde{\epsilon}^x_{tj},\nonumber \\ 
     Y_{tj}^\dagger &=  \beta_{0} + \beta_{1}X_{tj}  +  b\tilde{u}_{tj} + \tilde{\epsilon}^y_{tj}. 
     \label{appd_m2}
\end{align}
We denote marginal variances $Var(X_{tj}^\dagger) = \sigma_x^2$ and $Var(Y_{tj}^\dagger) = \sigma_y^2$. 
The first equation in (\ref{appd_m2}) implies a bivariate normal
$( X_{tj}^\dagger, \tilde{u}_{tj})\sim N( 
\bigl( \begin{smallmatrix} \alpha_0 + \alpha_1r_t\\ 0\end{smallmatrix}\bigr), \bigl( \begin{smallmatrix} \sigma^2_x & a\tilde\sigma^2_u\\
a\tilde\sigma^2_u & \tilde\sigma_u^2\end{smallmatrix}\bigr))$, from which it can be shown that 
$$\E(\tilde{u}_{tj}|X_{tj}^\dagger) = \frac{a\tilde\sigma_u^2}{\sigma_x^2}(X_{tj}^\dagger - \alpha_0 - \alpha_1r_t ).$$
It follows that 
$$\E(Y_{tj}^\dagger|X_{tj}^\dagger) = \beta_0 + \beta_1\E I(X_{tj}^\dagger>0) + b\E(\tilde{u}_{tj}|X_{tj}^\dagger)$$

$$\E(Y_{tj}^\dagger|X_{tj}^\dagger) = \beta_0 + \beta_1\E I(X_{tj}^\dagger>0) + b\frac{a\tilde\sigma_u^2}{\sigma_x^2}(X_{tj}^\dagger - \alpha_0 - \alpha_1r_t).$$

Since $X_{tj}$ is deterministic given $X_{tj}^\dagger$, it follows that 
\begin{align}
&\E(Y_{tj}^\dagger|X_{tj}) \nonumber \\
&=\E_{X^\dagger_{tj}|X_{tj}} \E (Y_{tj}^\dagger|X_{tj},X_{tj}^\dagger) \nonumber \\
&= \E_{X^\dagger_{tj}|X_{tj}} \E (Y_{tj}^\dagger|X_{tj}^\dagger)\nonumber \\
 &= \beta_0 + \beta_1X_{tj} + b\frac{a\tilde\sigma_u^2}{\sigma_x^2}(\E (X_{tj}^\dagger|X_{tj}) - \alpha_0 - \alpha_1r_t).   
      \label{appd_eq}
\end{align}
If regressing $Y_{tj}^\dagger$ directly on $X_{tj}$, the bias of estimating $\beta_1$ due to correlation between $Y_{tj}^\dagger$ and $X_{tj}^\dagger$ is $\frac{ab\tilde\sigma_u^2}{\sigma_x^2}\E (X_{tj}^\dagger|X_{tj})$, a quantity dependent on the product of the coefficients $a$ and $b$, and the variance $\tilde\sigma_u^2$ given fixed $\sigma_x$ and $x_{tj}$. In other words, the effect $a$ or $b$ or the variance $\tilde\sigma_u^2$ do not affect the bias individually; instead, they contribute to it as a composite. 
Therefore we can replace $ a\tilde{u}_{tj}$ and $ b\tilde{u}_{tj}$ with a new term $u_{tj}$ = $\sqrt{|ab|}\tilde{u}_{tj}$ to represent a summary of the confounding effect, i.e., scenario (B) represented by the following structural models:
\begin{align}
     X_{tj}^\dagger &=  \alpha_{0} + \alpha_{1}r_t  + {u}_{tj} +  \tilde{e}^x_{tj},\nonumber \\ 
     Y_{tj}^\dagger &=  \beta_{0} + \beta_{1}X_{tj}  +  sgn(ab){u}_{tj} + \tilde{e}^y_{tj},\label{sB}
\end{align}
where ${u}_{tj} \sim N(0, ab\tilde\sigma_u^2).$
With Eq.(\ref{sB}), it follows that 
\begin{align}
&\E(Y_{tj}^\dagger|X_{tj}) \nonumber \\
 &= \beta_0 + \beta_1X_{tj} + \frac{ab\tilde\sigma_u^2}{\sigma_x^2}(\E (X_{tj}^\dagger|X_{tj}) - \alpha_0 - \alpha_1r_t),
      \label{sB_bias}
\end{align}
yielding the same bias as in Eq.(\ref{appd_eq}) of estimating $\beta_1$. 

\subsection{Additional Tables and Figures}

%\begin{figure}
%    \centering
%    \includegraphics[width = 0.6\textwidth]{DAG_full_picture.pdf}
%    \caption{Probability of taking the trigger under each treatment assignment for participants simulated using ISTOP study. }
%    \label{fig:pi0pi1}
%\end{figure}

%\begin{figure}[ht]
%    \centering
%    \includegraphics[width=0.8\textwidth]{simulation_rhoxy_performance.pdf}
%    \caption{Bias and RMSE of estimating confounding effect $\rho$ using IV in simulations.}
%    \label{fig:simulation_rho}
%\end{figure}

\begin{table}[H]
\resizebox{\textwidth}{!}{%
\begin{tabular}{ccccccccccc}
\toprule
Scenario & Trial Duration ($L$) & $\rho_u$ & IV Strength & $\alpha_0^*$ & $\alpha_1^*$ & $\beta_0^*$ & $\beta_1^*$ & \multicolumn{2}{c}{Estimand} & $\rho$ \\
 &  &  &  &  &  &  &  & log($\bar \tau (\mathbf 1, \mathbf 0)$) & log($\bar \theta (\mathbf 1, \mathbf 0)$) &  \\ \midrule
50L1 & 50 & 0.3 & 6.0 (strong) & -0.85 & 1.10 & -1.00 & 0.50 & 0.86 & 0.60 & 0.1 \\
50L2 & 50 & 0.3 & 6.0 (strong) & -0.85 & 1.10 & -1.00 & 0.50 & 0.86 & 0.60 & 0.2 \\
50L3 & 50 & 0.3 & 6.0 (strong) & -0.85 & 1.10 & -1.00 & 0.50 & 0.86 & 0.60 & 0.3 \\
50L4 & 50 & 0.3 & 6.0 (strong) & -0.85 & 1.10 & -1.00 & 0.50 & 0.86 & 0.60 & 0.4 \\
50L5 & 50 & 0.3 & 6.0 (strong) & -0.85 & 1.10 & -1.00 & 0.50 & 0.86 & 0.60 & 0.5 \\
50L6 & 50 & 0.3 & 6.0 (strong) & -0.85 & 1.10 & -1.00 & 0.50 & 0.86 & 0.60 & 0.6 \\
50L7 & 50 & 0.3 & 6.0 (strong) & -0.85 & 1.10 & -1.00 & 0.50 & 0.86 & 0.60 & 0.7 \\
50L8 & 50 & 0.3 & 6.0 (strong) & -0.85 & 1.10 & -1.00 & 0.50 & 0.86 & 0.60 & 0.8 \\
50L9 & 50 & 0.3 & 6.0 (strong) & -0.85 & 1.10 & -1.00 & 0.50 & 0.86 & 0.60 & 0.9 \\
50L19 &50 &0.3  & 1.5 (weak) & -0.25 & 0.25 & -1.00 & 0.50 & 0.86 & 0.50 & 0.1 \\
50L11 &50 &0.3  & 1.5 (weak) & -0.25 & 0.25 & -1.00 & 0.50 & 0.86 & 0.50 & 0.5 \\
50L12 &50 &0.7 & 6.0 (strong) & -0.85  & 1.10 & -1.00  & 0.50 & 0.86  & 0.60 & 0.1 \\
50L13 &50 &0.7 & 6.0 (strong) & -0.85  & 1.10 & -1.00  & 0.50 & 0.86  & 0.60 & 0.5  \\
50L14 &50 &0.3 & 6.0 (strong) & -0.85  & 1.10 & -0.25 & 0.25 & 0.40 & 0.24 & 0.1 \\
50L15 &50 &0.3 & 6.0 (strong) & -0.85  & 1.10 & -0.25 & 0.25 & 0.40 & 0.24 & 0.5 \\ \midrule
200L1 & 200 & 0.3 & 6.0 (strong) & -0.85 & 1.10 & -1.00 & 0.50 & 0.86 & 0.60 & 0.1 \\
200L2 & 200 & 0.3 & 6.0 (strong) & -0.85 & 1.10 & -1.00 & 0.50 & 0.86 & 0.60 & 0.2 \\
200L3 & 200 & 0.3 & 6.0 (strong) & -0.85 & 1.10 & -1.00 & 0.50 & 0.86 & 0.60 & 0.3 \\
200L4 & 200 & 0.3 & 6.0 (strong) & -0.85 & 1.10 & -1.00 & 0.50 & 0.86 & 0.60 & 0.4 \\
200L5 & 200 & 0.3 & 6.0 (strong) & -0.85 & 1.10 & -1.00 & 0.50 & 0.86 & 0.60 & 0.5 \\
200L6 & 200 & 0.3 & 6.0 (strong) & -0.85 & 1.10 & -1.00 & 0.50 & 0.86 & 0.60 & 0.6 \\
200L7 & 200 & 0.3 & 6.0 (strong) & -0.85 & 1.10 & -1.00 & 0.50 & 0.86 & 0.60 & 0.7 \\
200L8 & 200 & 0.3 & 6.0 (strong) & -0.85 & 1.10 & -1.00 & 0.50 & 0.86 & 0.60 & 0.8 \\
200L9 & 200 & 0.3 & 6.0 (strong) & -0.85 & 1.10 & -1.00 & 0.50 & 0.86 & 0.60 & 0.9 \\
200L10 &200 &0.3  & 1.5 (weak) & -0.25 & 0.25 & -1.00 & 0.50 & 0.86 & 0.50 & 0.1 \\
200L11 &200 &0.3  & 1.5 (weak) & -0.25 & 0.25 & -1.00 & 0.50 & 0.86 & 0.50 & 0.5 \\
200L12 &200 &0.7 & 6.0 (strong) & -0.85  & 1.10 & -1.00  & 0.50 & 0.86  & 0.60 & 0.1 \\
200L13 &200 &0.7 & 6.0 (strong) & -0.85  & 1.10 & -1.00  & 0.50 & 0.86  & 0.60 & 0.5  \\
200L14 &200 &0.3 & 6.0 (strong) & -0.85  & 1.10 & -0.25 & 0.25 & 0.40 & 0.24 & 0.1 \\
200L15 &200 &0.3 & 6.0 (strong) & -0.85  & 1.10 & -0.25 & 0.25 & 0.40 & 0.24 & 0.5 \\ \midrule
1000L1 & 1000 & 0.3 & 6.0 (strong) & -0.85 & 1.10 & -1.00 & 0.50 & 0.86 & 0.60 & 0.1 \\
1000L2 & 1000 & 0.3 & 6.0 (strong) & -0.85 & 1.10 & -1.00 & 0.50 & 0.86 & 0.60 & 0.2 \\
1000L3 & 1000 & 0.3 & 6.0 (strong) & -0.85 & 1.10 & -1.00 & 0.50 & 0.86 & 0.60 & 0.3 \\
1000L4 & 1000 & 0.3 & 6.0 (strong) & -0.85 & 1.10 & -1.00 & 0.50 & 0.86 & 0.60 & 0.4 \\
1000L5 & 1000 & 0.3 & 6.0 (strong) & -0.85 & 1.10 & -1.00 & 0.50 & 0.86 & 0.60 & 0.5 \\
1000L6 & 1000 & 0.3 & 6.0 (strong) & -0.85 & 1.10 & -1.00 & 0.50 & 0.86 & 0.60 & 0.6 \\
1000L7 & 1000 & 0.3 & 6.0 (strong) & -0.85 & 1.10 & -1.00 & 0.50 & 0.86 & 0.60 & 0.7 \\
1000L8 & 1000 & 0.3 & 6.0 (strong) & -0.85 & 1.10 & -1.00 & 0.50 & 0.86 & 0.60 & 0.8 \\
1000L9 & 1000 & 0.3 & 6.0 (strong) & -0.85 & 1.10 & -1.00 & 0.50 & 0.86 & 0.60 & 0.9 \\
1000L10 &1000 &0.3  & 1.5 (weak) & -0.25 & 0.25 & -1.00 & 0.50 & 0.86 & 0.50 & 0.1 \\
1000L11 &1000 &0.3  & 1.5 (weak) & -0.25 & 0.25 & -1.00 & 0.50 & 0.86 & 0.50 & 0.5 \\
1000L12 &1000 &0.7 & 6.0 (strong) & -0.85  & 1.10 & -1.00  & 0.50 & 0.86  & 0.60 & 0.1 \\
1000L13 &1000 &0.7 & 6.0 (strong) & -0.85  & 1.10 & -1.00  & 0.50 & 0.86  & 0.60 & 0.5  \\
1000L14 &1000 &0.3 & 6.0 (strong) & -0.85  & 1.10 & -0.25 & 0.25 & 0.40 & 0.24 & 0.1 \\
1000L15 &1000 &0.3 & 6.0 (strong) & -0.85  & 1.10 & -0.25 & 0.25 & 0.40 & 0.24 & 0.5 \\ \bottomrule
\end{tabular}%
}
\caption{Simulation scenarios assuming no carryover effect ($\beta_2 = 0$). 50L1-50L9, 200L1-200L9, and 1000L1-1000L9 are primary scenarios with results presented in the main text. IV strength is defined as the odds ratio of consuming alcohol, i.e., $\frac{\Phi(\alpha_0^* +\alpha_1^*)}{1-\Phi(\alpha_0^* +\alpha_1^*)}\frac{\Phi(\alpha_0^*)}{1-\Phi(\alpha_0^*)}$. }
\label{tab:sim_design}
\end{table}

% \usepackage{booktabs}
% \usepackage{graphicx}
\begin{table}[H]
\resizebox{\textwidth}{!}{%
\begin{tabular}{@{}cccccccccccc@{}}
\toprule
Scenario & Trial Duration ($L$) & $\rho_u$ & IV Strength & $\alpha_0^*$ & $\alpha_1^*$ & $\beta_0^*$ & $\beta_1^*$ & $\beta_2^*$ &\multicolumn{2}{c}{Estimand} & $\rho$ \\ \midrule
 &  &  &  &  &  &  & & & log($\bar \tau (\mathbf 1, \mathbf 0)$) & log($\bar \theta (\mathbf 1, \mathbf 0)$) &  \\
50L1CO & 50 & 0.3 & 6.0 (strong) & -0.85 & 1.10 & -1.00 & 0.50  &0.20 &1.19 & 0.76 & 0.1 \\
50L3CO &  &  &  &  &  & &  &  &  &  & 0.3 \\
50L5CO &  &  &  &  &  & &  &  &  &  & 0.5 \\
50L7CO &  &  &  &  &  & &  &  &  &  & 0.7 \\
50L9CO &  &  &  &  &  & &  &  &  &  & 0.9 \\ \midrule
200L1CO & 200 &  &  &  & &  &  &  &  &  & 0.1 \\
200L3CO &  &  &  &  &  & &  &  &  &  & 0.3 \\
200L5CO &  &  &  &  &  & &  &  &  &  & 0.5 \\
200L7CO &  &  &  &  &  & &  &  &  &  & 0.7 \\
200L9CO &  &  &  &  &  & &  &  &  &  & 0.9 \\ \midrule
1000L1CO & 1000 &  &  &  & &  &  &  &  &  & 0.1 \\
1000L3CO &  &  &  &  &  &  & &  &  &  & 0.3 \\
1000L5CO &  &  &  &  &  &  & &  &  &  & 0.5 \\
1000L7CO &  &  &  &  &  &  & &  &  &  & 0.7 \\
1000L9CO &  &  &  &  &  &  & &  &  &  & 0.9 \\ \bottomrule
\end{tabular}%
}
\caption{Simulation scenarios assuming 1-day carryover effect. IV strength is defined as the odds ratio of consuming alcohol, i.e., $\frac{\Phi(\alpha_0^* +\alpha_1^*)}{1-\Phi(\alpha_0^* +\alpha_1^*)}\frac{\Phi(\alpha_0^*)}{1-\Phi(\alpha_0^*)}$. All the parameters left blank are held constant.}
\label{tab:sim_design_CO}
\end{table}

\begin{table}[H]
\resizebox{\textwidth}{!}{%
\begin{tabular}{cccccccccccccc}
\toprule
Scenario                     & \multicolumn{1}{l}{Confounding ($\rho$)} & \multicolumn{4}{c}{Posterior Mean} & \multicolumn{4}{c}{Bias\%} & \multicolumn{4}{c}{Coverage} \\ \cmidrule(l){3-6}\cmidrule(l){7-10}\cmidrule(l){11-14}                      
\multicolumn{1}{l}{} & \multicolumn{1}{l}{}            & IV\textunderscore DD(Model NCO)     & ITT     & AT     & PP     & IV\textunderscore DD(Model NCO)   & ITT   & AT   & PP   & IV\textunderscore DD(Model NCO)    & ITT    & AT   & PP   \\ 
\midrule
1000L1 & 0.1 & 0.88 & 0.36 & 0.99 & 0.92 & 1.65   & -58.36 & 15.18  & 6.39   & 0.97 & 0.00    & 0.98 & 1.00    \\
1000L2 & 0.2 & 0.94 & 0.37 & 1.19 & 1.08 & 9.71   & -56.86 & 38.34  & 25.28  & 0.98 & 0.00    & 0.92 & 1.00    \\
1000L3 & 0.3 & 0.92 & 0.38 & 1.39 & 1.19 & 6.42   & -55.70  & 60.81  & 38.71  & 0.99 & 0.01 & 0.74 & 0.92 \\
1000L4 & 0.4 & 0.97 & 0.39 & 1.69 & 1.41 & 12.89  & -54.17 & 95.71  & 63.60  & 0.93 & 0.11 & 0.31 & 0.74 \\
1000L5 & 0.5 & 1.01 & 0.41 & 1.87 & 1.56 & 17.69  & -52.44 & 117.62 & 81.63  & 0.96 & 0.17 & 0.24 & 0.50  \\
1000L6 & 0.6 & 1.09 & 0.45 & 2.16 & 1.78 & 26.33  & -47.68 & 150.72 & 107.12 & 0.91 & 0.41 & 0.14 & 0.30  \\
1000L7 & 0.7 & 1.02 & 0.42 & 2.42 & 1.97 & 18.77  & -51.53 & 180.71 & 128.55 & 0.92 & 0.40  & 0.16 & 0.24 \\
1000L8 & 0.8 & 1.09 & 0.45 & 2.91 & 2.33 & 26.05  & -48.09 & 237.32 & 170.67 & 0.92 & 0.41 & 0.02 & 0.15 \\
1000L9 & 0.9 & 1.06 & 0.43 & 3.32 & 2.73 & 23.43  & -50.17 & 285.09 & 216.55 & 0.96 & 0.31 & 0.04 & 0.04 \\ \midrule
200L1  & 0.1 & 0.69 & 0.29 & 0.97 & 0.89 & -19.32 & -66.65 & 12.64  & 3.67   & 0.99 & 0.21 & 0.99 & 0.98 \\
200L2  & 0.2 & 0.92 & 0.37 & 1.20 & 1.08 & 7.25   & -56.98 & 38.98  & 25.45  & 0.96 & 0.41 & 0.97 & 0.95 \\
200L3  & 0.3 & 0.88 & 0.35 & 1.41 & 1.20 & 2.42   & -59.19 & 63.50   & 38.85  & 0.96 & 0.34 & 0.89 & 0.95 \\
200L4  & 0.4 & 1.07 & 0.43 & 1.63 & 1.38 & 24.37  & -50.20  & 89.05  & 59.85  & 0.98 & 0.67 & 0.74 & 0.91 \\
200L5  & 0.5 & 1.13 & 0.46 & 1.81 & 1.52 & 31.66  & -46.62 & 109.77 & 76.19  & 0.94 & 0.65 & 0.52 & 0.86 \\
200L6  & 0.6 & 1.22 & 0.50 & 2.15 & 1.74 & 41.76  & -41.62 & 149.09 & 102.58 & 0.95 & 0.77 & 0.33 & 0.74 \\
200L7  & 0.7 & 1.27 & 0.50 & 2.51 & 2.01 & 47.78  & -41.41 & 191.09 & 133.95 & 0.98 & 0.83 & 0.06 & 0.43 \\
200L8  & 0.8 & 1.20 & 0.48 & 3.01 & 2.42 & 39.25  & -44.12 & 249.22 & 181.28 & 0.95 & 0.69 & 0.02 & 0.30  \\
200L9   & 0.9 & 1.38 & 0.57 & 3.37 & 2.78 & 59.82  & -33.65 & 291.31 & 222.54 & 0.90  & 0.76 & 0.01 & 0.09 \\ \midrule
50L1   & 0.1 & 0.40 & 0.17 & 0.88 & 0.83 & -53.09 & -80.28 & 1.69   & -3.41  & 0.96 & 0.32 & 0.96 & 0.99 \\
50L2   & 0.2 & 0.64 & 0.25 & 1.21 & 1.10  & -26.17 & -70.62 & 39.96  & 27.49  & 0.99 & 0.53 & 0.93 & 0.98 \\
50L3   & 0.3 & 0.77 & 0.29 & 1.40  & 1.20  & -10.74 & -66.13 & 62.40   & 39.67  & 0.99 & 0.59 & 0.92 & 0.94 \\
50L4  & 0.4 & 1.03 & 0.38 & 1.83 & 1.51 & 19.33  & -55.56 & 112.33 & 75.70   & 0.99 & 0.72 & 0.73 & 0.82 \\
50L5  & 0.5 & 1.10 & 0.42 & 2.05 & 1.69 & 28.28  & -51.51 & 138.53 & 96.00     & 1.00    & 0.71 & 0.60  & 0.76 \\
50L6   & 0.6 & 1.32 & 0.47 & 2.56 & 2.11 & 53.57  & -45.36 & 197.52 & 145.31 & 0.98 & 0.86 & 0.33 & 0.65 \\
50L7  & 0.7 & 1.34 & 0.52 & 2.78 & 2.17 & 55.42  & -39.96 & 223.29 & 152.35 & 0.97 & 0.86 & 0.18 & 0.64 \\
50L8   & 0.8 & 1.69 & 0.65 & 3.53 & 2.81 & 96.13  & -25.09 & 309.86 & 226.77 & 0.95 & 0.83 & 0.11 & 0.51 \\
50L9   & 0.9 & 1.82 & 0.71 & 4.00  & 3.21 & 111.37 & -18.12 & 364.07 & 272.5  & 0.98 & 0.94 & 0.03 & 0.34 \\
\bottomrule
\end{tabular}%
}
\caption{Average of posterior means, average of bias\%, coverage proportions of the estimated log odds ratios using IV\textunderscore DD of Model NCO, ITT, PP, AT approaches for the daily drinking effect (i.e., log$\overline \tau(\mathbf 1,\mathbf 0)=0.86$).  }
\label{tab:sim_main_tb}
\end{table}

% \usepackage{graphicx}
\begin{table}[H]
\resizebox{\textwidth}{!}{%
\begin{tabular}{cccccccccccccc}
\toprule
Scenario                    & \multicolumn{1}{l}{Confounding ($\rho$)} & \multicolumn{4}{c}{Posterior Mean} & \multicolumn{4}{c}{Bias\%} & \multicolumn{4}{c}{Coverage} \\ \cmidrule(l){3-6}\cmidrule(l){7-10}\cmidrule(l){11-14}                      
\multicolumn{1}{l}{} & \multicolumn{1}{l}{}            & IV\textunderscore UD(Model NCO)     & ITT     & AT     & PP     & IV\textunderscore UD(Model NCO)      & ITT   & AT   & PP   & IV\textunderscore UD(Model NCO)     & ITT    & AT   & PP   \\ 
\midrule
1000L1 & 0.1 & 0.58 & 0.36 & 0.99 & 0.92 & -11.45 & -45.18 & 51.32  & 40.62  & 0.95 & 0.10  & 0.82 & 0.94 \\
1000L2 & 0.2 & 0.61 & 0.37 & 1.19 & 1.08 & -6.02  & -43.20  & 81.89  & 65.07  & 0.96 & 0.33 & 0.45 & 0.84 \\
1000L3 & 0.3 & 0.61 & 0.38 & 1.39 & 1.19 & -6.61  & -41.68 & 112.46 & 81.89  & 0.98 & 0.46 & 0.24 & 0.66 \\
1000L4 & 0.4 & 0.64 & 0.39 & 1.69 & 1.41 & -1.70  & -39.67 & 158.31 & 115.51 & 0.93 & 0.56 & 0.10  & 0.25 \\
1000L5 & 0.5 & 0.67 & 0.41 & 1.87 & 1.56 & 1.88   & -37.38 & 185.82 & 138.44 & 0.95 & 0.68 & 0.08 & 0.21 \\
1000L6 & 0.6 & 0.73 & 0.45 & 2.16 & 1.78 & 10.93  & -31.12 & 230.15 & 172.07 & 0.95 & 0.77 & 0.06 & 0.08 \\
1000L7 & 0.7 & 0.68 & 0.42 & 2.42 & 1.97 & 3.42   & -36.19 & 269.89 & 201.11 & 0.91 & 0.75 & 0.02 & 0.05 \\
1000L8 & 0.8 & 0.72 & 0.45 & 2.91 & 2.33 & 10.25  & -31.67 & 344.78 & 256.13 & 0.91 & 0.69 & 0.01 & 0.04 \\
1000L9 & 0.9 & 0.69 & 0.43 & 3.32 & 2.73 & 4.75   & -34.40  & 407.45 & 317.27 & 0.98 & 0.61 & 0.00    & 0.01 \\ \midrule
200L1  & 0.1 & 0.46 & 0.29 & 0.97 & 0.89 & -29.97 & -56.10  & 48.26  & 36.03  & 0.93 & 0.60  & 0.97 & 0.98 \\
200L2  & 0.2 & 0.61 & 0.37 & 1.20  & 1.08 & -6.99 & -43.36 & 83.42  & 65.07  & 0.94 & 0.81 & 0.77 & 0.84 \\
200L3  & 0.3 & 0.58 & 0.35 & 1.41 & 1.20  & -10.82 & -46.27 & 115.51 & 83.42  & 0.96 & 0.84 & 0.67 & 0.89 \\
200L4  & 0.4 & 0.71 & 0.43 & 1.63 & 1.38 & 7.99   & -34.44 & 149.14 & 110.93 & 0.97 & 0.91 & 0.26 & 0.67 \\
200L5   & 0.5 & 0.75 & 0.46 & 1.81 & 1.52 & 14.93  & -29.73 & 176.65 & 132.33 & 0.96 & 0.92 & 0.11 & 0.50  \\
200L6   & 0.6 & 0.84 & 0.50  & 2.15 & 1.74 & 27.82 & -23.14 & 228.62 & 165.95 & 0.96 & 0.91 & 0.13 & 0.45 \\
200L7  & 0.7 & 0.85 & 0.50  & 2.51 & 2.01 & 29.85 & -22.87 & 283.64 & 207.22 & 0.96 & 0.92 & 0.01 & 0.17 \\
200L8  & 0.8 & 0.81 & 0.48 & 3.01 & 2.42 & 23.89  & -26.44 & 360.07 & 269.89 & 0.94 & 0.84 & 0.00  & 0.09 \\
200L9  & 0.9 & 0.95 & 0.57 & 3.37 & 2.78 & 45.54  & -12.66 & 415.09 & 324.91 & 0.87 & 0.92 & 0.00  & 0.03 \\ \midrule
50L1   & 0.1 & 0.28 & 0.17 & 0.88 & 0.83 & -57.43 & -74.04 & 34.50   & 26.86  & 0.85 & 0.67 & 0.95 & 0.98 \\
50L2   & 0.2 & 0.41 & 0.25 & 1.21 & 1.10  & -37.16 & -61.32 & 84.94  & 68.13  & 0.99 & 0.76 & 0.87 & 0.96 \\
50L3   & 0.3 & 0.49 & 0.29 & 1.40  & 1.20  & -24.50  & -55.41 & 113.99 & 83.42  & 0.97 & 0.81 & 0.86 & 0.93 \\
50L4   & 0.4 & 0.63 & 0.38 & 1.83 & 1.51 & -3.48  & -41.50  & 179.71 & 130.80  & 0.98 & 0.87 & 0.51 & 0.73 \\
50L5   & 0.5 & 0.68 & 0.42 & 2.05 & 1.69 & 3.36   & -36.17 & 213.34 & 158.31 & 1.00    & 0.94 & 0.45 & 0.68 \\
50L6   & 0.6 & 0.84 & 0.47 & 2.56 & 2.11 & 28.96  & -28.07 & 291.29 & 222.51 & 0.99 & 0.96 & 0.2  & 0.52 \\
50L7  & 0.7 & 0.87 & 0.52 & 2.78 & 2.17 & 32.28  & -20.96 & 324.91 & 231.68 & 0.98 & 0.96 & 0.11 & 0.57 \\
50L8   & 0.8 & 1.13 & 0.65 & 3.53 & 2.81 & 73.00     & -1.38  & 439.55 & 329.50  & 0.96 & 0.99 & 0.07 & 0.40  \\
50L9  & 0.9 & 1.23 & 0.71 & 4.00  & 3.21 & 88.12  & 7.79   & 511.39 & 390.64 & 0.98 & 0.99 & 0.02 & 0.23 \\ \bottomrule
\end{tabular}%
}
\caption{Average of posterior means, bias\%, coverage proportions of the estimated log odds ratios using IV\textunderscore UD of Model NCO, ITT, PP, AT approaches for the usual drinking effect (i.e., log$\overline\theta(\mathbf 1,\mathbf 0)=0.65$).}
\label{tab:sim_main_tb_db}
\end{table}

% \usepackage{booktabs}
% \usepackage{graphicx}
\begin{table}[H]
\resizebox{\textwidth}{!}{%
\begin{tabular}{@{}cccccccc@{}}
\toprule
Scenario & Confounding ($\rho$) & \multicolumn{2}{c}{Posterior Mean} & \multicolumn{2}{c}{Bias\%} & \multicolumn{2}{c}{Coverage} \\
\cmidrule(l){3-4}\cmidrule(l){5-6}\cmidrule(l){7-8}   
 &  & IV\textunderscore DD(Model CO) & ITT & IV\textunderscore DD(Model CO) & ITT & IV\textunderscore DD(Model CO) & ITT \\ \midrule
1000L1CO & 0.1 & 1.22 & 0.48 & 2.40 & -59.77 & 1.00 & 0.00 \\
1000L3CO & 0.3 & 1.22 & 0.49 & 2.51 & -58.34 & 0.98 & 0.00 \\
1000L5CO & 0.5 & 1.30 & 0.52 & 9.77 & -56.04 & 0.98 & 0.03 \\
1000L7CO & 0.7 & 1.29 & 0.52 & 9.01 & -55.98 & 0.96 & 0.06 \\
1000L9CO & 0.9 & 1.35 & 0.56 & 13.78 & -53.18 & 0.95 & 0.07 \\ \midrule
200L1CO & 0.1 & 0.76 & 0.30 & -35.59 & -74.43 & 0.89 & 0.03 \\
200L3CO & 0.3 & 1.18 & 0.44 & -0.80 & -62.84 & 0.94 & 0.14 \\
200L5CO & 0.5 & 1.42 & 0.55 & 19.46 & -53.28 & 0.96 & 0.33 \\
200L7CO & 0.7 & 1.55 & 0.63 & 30.60 & -47.12 & 0.96 & 0.52 \\
200L9CO & 0.9 & 1.69 & 0.69 & 42.09 & -42.19 & 0.91 & 0.56 \\ \midrule
50L1CO & 0.1 & 1.22 & 0.44 & 2.41 & -62.72 & 0.96 & 0.6 \\
50L3CO & 0.3 & 1.47 & 0.55 & 24.04 & -53.95 & 0.99 & 0.71 \\
50L5CO & 0.5 & 1.85 & 0.71 & 55.75 & -40.52 & 0.95 & 0.77 \\
50L7CO & 0.7 & 2.10 & 0.76 & 76.64 & -36.24 & 0.95 & 0.86 \\
50L9CO & 0.9 & 2.10 & 0.80 & 77.15 & -32.81 & 0.98 & 0.88 \\ \bottomrule
\end{tabular}%
}
\caption{Average of posterior means, average of bias\%, coverage proportions of the estimated log odds ratios using IV\textunderscore DD of Model CO, ITT, PP, AT approaches for the daily drinking effect (i.e., log$\overline\tau(\mathbf 1,\mathbf 0)=1.19$).  }
\label{tab:sim_main_tb_co}
\end{table}

\begin{table}[H]
\resizebox{\textwidth}{!}{%
\begin{tabular}{@{}cccccccc@{}}
\toprule
Scenario & Confounding ($\rho$) & \multicolumn{2}{c}{Posterior Mean} & \multicolumn{2}{c}{Bias\%} & \multicolumn{2}{c}{Coverage} \\
\cmidrule(l){3-4}\cmidrule(l){5-6}\cmidrule(l){7-8}   
 &  & IV\textunderscore DD(Model CO) & ITT & IV\textunderscore DD(Model CO) & ITT & IV\textunderscore DD(Model CO) & ITT \\ \midrule
1000L1CO & 0.1 & 0.76 & 0.48 & 0.45 & -37.14 & 1.00 & 0.20 \\
1000L3CO & 0.3 & 0.79 & 0.49 & 3.79 & -34.90 & 0.98 & 0.56 \\
1000L5CO & 0.5 & 0.84 & 0.52 & 10.83 & -31.30 & 0.98 & 0.72 \\
1000L7CO & 0.7 & 0.83 & 0.52 & 9.78 & -31.22 & 0.95 & 0.66 \\
1000L9CO & 0.9 & 0.88 & 0.56 & 16.18 & -26.83 & 0.96 & 0.69 \\ \midrule
200L1CO & 0.1 & 0.49 & 0.30 & -35.75 & -60.04 & 0.90 & 0.43 \\
200L3CO & 0.3 & 0.75 & 0.44 & -0.96 & -41.94 & 0.92 & 0.69 \\
200L5CO & 0.5 & 0.92 & 0.55 & 20.90 & -26.99 & 0.96 & 0.91 \\
200L7CO & 0.7 & 1.03 & 0.63 & 35.87 & -17.36 & 0.95 & 0.88 \\
200L9CO & 0.9 & 1.14 & 0.69 & 49.60 & -9.66 & 0.91 & 0.91 \\
 \midrule
50L1CO & 0.1 & 0.76 & 0.44 & 0.28 & -41.74 & 0.96 & 0.89 \\
50L3CO & 0.3 & 0.93 & 0.55 & 22.97 & -28.04 & 0.99 & 0.94 \\
50L5CO & 0.5 & 1.18 & 0.71 & 55.37 & -7.06 & 0.96 & 0.93 \\
50L7CO & 0.7 & 1.36 & 0.76 & 78.66 & -0.36 & 0.98 & 0.97 \\
50L9CO & 0.9 & 1.43 & 0.80 & 88.36 & 4.99 & 0.99 & 0.99
\\
\bottomrule
\end{tabular}%
}
\caption{Average of posterior means, average of bias\%, coverage proportions of the estimated log odds ratios using IV\textunderscore UD of Model CO, ITT, PP, AT approaches for the usual drinking effect (i.e., log$\overline\theta(\mathbf 1,\mathbf 0)=0.76$).  }
\label{tab:sim_main_tb_db_co}
\end{table}

% \usepackage{booktabs}
% \usepackage{multirow}
% \usepackage{graphicx}
\begin{table}[H]
\vspace{-1cm}
\centering
\resizebox{0.6\textwidth}{!}{%
\begin{tabular}{@{}cccllllll@{}}
\toprule
\multirow{2}{*}{Scenario\#} & \multirow{2}{*}{Parameter} & \multirow{2}{*}{Truth} & \multicolumn{2}{c}{L = 50} & \multicolumn{2}{c}{L=200} & \multicolumn{2}{c}{L=1000} \\ \cmidrule(l){4-5}\cmidrule(l){6-7}\cmidrule(l){8-9}
   &                             &       & $\sigma_1^2=$1 & $\sigma_1^2=$3 & $\sigma_1^2=$1 & $\sigma_1^2=$3  & $\sigma_1^2=$1 & $\sigma_1^2=$3 \\ \midrule
1 &
  \multirow{15}{*}{log($\tau(\mathbf 1, \mathbf 0$))} &
  0.86 &
  -53.09 &
  -46.78 &
  -30.43 &
  -18.29 &
  
  1.65 &
  6.13 \\
2  &                             & 0.86  & -26.17 & -15.37 & -4.25  & 9.49                     & 9.71   & 17.81  \\
3  &                             & 0.86  & -10.74 & 2.65   & -10.57 & 5.24                      & 6.42   & 13.42  \\
4  &                             & 0.86  & 19.33  & 37.92  & 10.95  & 32.03                     & 12.89  & 20.61  \\
5  &                             & 0.86  & 28.28  & 51.77  & 16.79  & 38.94                     & 17.69  & 22.06  \\
6  &                             & 0.86  & 53.57  & 87.20   & 19.41  & 50.12                     & 26.33  & 31.89  \\
7  &                             & 0.86  & 55.42  & 91.32  & 25.44  & 61.71                     & 18.77  & 23.33  \\
8  &                             & 0.86  & 96.13  & 153.32 & 16.84  & 55.98                     & 26.05  & 29.77  \\
9  &                             & 0.86  & 111.37 & 180.04 & 32.60   & 84.42                    & 23.43  & 22.11  \\
10 &                             & 0.86  & -      & -      & -38.57 & -28.76                       & -13.53 & -2.48  \\
11 &                             & 0.86  & -      & -      & 26.87  & 66.16                        & 23.00     & 43.32  \\
12 &                             & 0.86  & -      & -      & -42.78 & -32.01                      & -13.06 & -11.63 \\
13 &                             & 0.86  & -      & -      & -32.24 & -16.64                        & -6.97  & -1.76  \\
14 &                             & 0.41   & -      & -      & -9.52  & -0.28                         & 7.36   & 9.46   \\
15 &                             & 0.41   & -      & -      & 74.67  & 100.11                         & 37.95  & 46.26  \\ \midrule
1  & \multirow{15}{*}{$\beta_0^*$} & -1.00    & -15.33 & -13.68 & -2.78  & 0.71                       & -1.42  & 0.02   \\
2  &                             & -1.00    & -13.60  & -11.03 & -2.99  & 1.20                        & 3.43   & 4.48   \\
3  &                             & -1.00    & -11.22 & -7.99  & -3.66  & 1.11                       & 1.12   & 2.17   \\
4  &                             & -1.00    & -27.93 & -23.65 & -3.85  & 2.43                       & 4.49   & 6.31   \\
5  &                             & -1.00    & -22.04 & -16.76 & -5.20   & 1.55                       & 2.43   & 3.42   \\
6  &                             & -1.00    & -11.29 & -3.07  & 13.66  & 23.26                    & 8.06   & 9.38   \\
7  &                             & -1.00    & -9.99  & -1.17  & 6.76   & 18.14                     & 5.36   & 6.58   \\
8  &                             & -1.00    & 25.24  & 40.07  & -1.92  & 10.77                      & 8.52   & 9.55   \\
9  &                             & -1.00    & 31.02  & 49.28  & 8.04   & 25.67                     & 7.63   & 6.44   \\
10 &                             & -1.00    & -      & -      & -14.69 & -11.37                        & -1.95  & 1.19   \\
11 &                             & -1.00    & -      & -      & -0.93  & 11.48                         & 5.04   & 11.14  \\
12 &                             & -1.00    & -      & -      & -12.6  & -9.25                         & -1.55  & -1.03  \\
13 &                             & -1.00    & -      & -      & -14.39 & -8.28                       & -1.28  & 0.66   \\
14 &                             & -0.25 & -      & -      & -11.72 & -7.24                        & 7.53   & 8.55   \\
15 &                             & -0.25 & -      & -      & 14.72  & 26.38                       & 15.4   & 19.18  \\ \midrule
1 &
  \multirow{15}{*}{$\beta_1^*$} &
  0.5 &
  -52.96 &
  -46.78 &
  -31.34 &
  -19.53 &
 
  1.91 &
  6.31 \\
2  &                             & 0.5   & -25.85 & -15.16 & -4.33  & 9.09                      & 9.23   & 17.5   \\
3  &                             & 0.5   & -10.56 & 2.41   & -11.04 & 4.37                      & 6.20    & 13.28  \\
4  &                             & 0.5   & 23.19  & 41.87  & 11.20   & 31.85                    & 12.12  & 19.76  \\
5  &                             & 0.5   & 31.82  & 55.32  & 17.47  & 39.25                     & 17.57  & 21.88  \\
6  &                             & 0.5   & 55.31  & 87.88  & 15.21  & 43.73                     & 25.04  & 30.44  \\
7  &                             & 0.5   & 56.88  & 91.55  & 22.97  & 57.08                     & 17.66  & 22.08  \\
8  &                             & 0.5   & 86.28  & 135.95 & 15.65  & 51.88                     & 24.32  & 27.87  \\
9  &                             & 0.5   & 99.50   & 158.09 & 28.90   & 73.83                    & 21.69  & 20.54  \\
10 &                             & 0.5   & -      & -      & -38.53 & -28.87                        & -14.08 & -3.15  \\
11 &                             & 0.5   & -      & -      & 25.83  & 63.53                        & 21.15  & 40.88  \\
12 &                             & 0.5   & -      & -      & -42.57 & -31.91                       & -13.38 & -12.00    \\
13 &                             & 0.5   & -      & -      & -32.32 & -17.34                       & -7.37  & -2.33  \\
14 &                             & 0.25  & -      & -      & -9.74  & -0.55                   & 7.22   & 9.32   \\
15 &                             & 0.25  & -      & -      & 73.24  & 98.25                     & 37.47  & 45.71  \\ \midrule
1  & \multirow{15}{*}{$\rho$}    & 0.1   & 102.51 & 97.77  & 94.16  & 70.88                    & 0.67   & -7.81  \\
2  &                             & 0.2   & 17.33  & 10.77  & 11.72  & -0.68                     & -11.90  & -27.86 \\
3  &                             & 0.3   & -18.66 & -22.87 & -5.81  & -16.34                   & -9.85  & -17.76 \\
4  &                             & 0.4   & -26.38 & -32.35 & -12.99 & -23.73                   & -6.31  & -12.06 \\
5  &                             & 0.5   & -31.39 & -38.01 & -20.61 & -29.26                   & -12.47 & -14.89 \\
6  &                             & 0.6   & -29.92 & -37.83 & -16.65 & -26.1                    & -12.29 & -14.70  \\
7  &                             & 0.7   & -29.22 & -36.88 & -13.29 & -22.31                   & -8.16  & -9.51  \\
8  &                             & 0.8   & -31.39 & -39.88 & -4.89  & -11.79                   & -5.87  & -6.74  \\
9  &                             & 0.9   & -27.90  & -36.37 & -5.80   & -11.22                   & -4.18  & -3.93  \\
10 &                             & 0.1   & -      & -      & 106.04 & 89.79                       & 30.05  & 5.42   \\
11 &                             & 0.5   & -      & -      & -19.03 & -33.58                        & -12.49 & -21.57 \\
12 &                             & 0.1   & -      & -      & 105.95 & 83.36                         & 37.24  & 36.11  \\
13 &                             & 0.5   & -      & -      & 7.47   & 1.96                         & 2.27   & 0.28   \\
14 &                             & 0.1   & -      & -      & 51.30   & 41.51                         & -0.18  & -2.45  \\
15 &                             & 0.5   & -      & -      & -18.81 & -24.45                     & -10.21 & -12.39 \\ \toprule
\end{tabular}%
}
\caption{Bias(\%) of log($\tau(\mathbf 1, \mathbf 0$)), $\beta_0^*$, $\beta_1^*$, and $\rho$ for IV\textunderscore DD approach under varying variances of $\beta_1$ prior for 15 simulation scenarios with 3 duration lengths. Scenario \# combining $L$ corresponds to scenario labels in the main text, e.g., 1000L1 corresponds to scenario 1 with $L = 1000$. Scenarios marked with `-' are not run.}
\label{tab:sim_bias}
\end{table}

\begin{table}[H]
\vspace{-1cm}
\centering
\resizebox{0.6\textwidth}{!}{%
\begin{tabular}{@{}ccccccccc@{}}
\toprule
\multirow{2}{*}{Scenario\#} & \multirow{2}{*}{Parameter} & \multirow{2}{*}{Truth} & \multicolumn{2}{c}{L = 50} & \multicolumn{2}{c}{L=200} & \multicolumn{2}{c}{L=1000} \\ \cmidrule(l){4-5}\cmidrule(l){6-7}\cmidrule(l){8-9}
   &                             &       & $\sigma_1^2=$1 & $\sigma_1^2=$3 & $\sigma_1^2=$1 & $\sigma_1^2=$3  & $\sigma_1^2=$1 & $\sigma_1^2=$3 \\ \midrule
1  & \multirow{15}{*}{log($\tau(\mathbf 1, \mathbf 0)$)}  & 0.86  & 0.96  & 0.96  & 0.96  & 0.99        & 0.97  & 0.97  \\
2  &                             & 0.86  & 0.99  & 0.99  & 0.94  & 0.95                   & 0.98  & 0.92  \\
3  &                             & 0.86  & 0.99  & 0.98  & 0.97  & 0.95                    & 0.99  & 0.97  \\
4  &                             & 0.86  & 0.99  & 0.98  & 0.96  & 0.98                   & 0.93  & 0.91  \\
5  &                             & 0.86  & 1.00     & 0.95  & 0.98  & 0.93                  & 0.96  & 0.92  \\
6  &                             & 0.86  & 0.98  & 0.94  & 0.96  & 0.93                   & 0.91  & 0.89  \\
7  &                             & 0.86  & 0.97  & 0.94  & 0.99  & 0.97                  & 0.92  & 0.92  \\
8  &                             & 0.86  & 0.95  & 0.9   & 0.96  & 0.89                    & 0.92  & 0.92  \\
9  &                             & 0.86  & 0.98  & 0.86  & 0.96  & 0.88                    & 0.96  & 0.94  \\
10 &                             & 0.86  & -     & -     & 1.00     & 1.00                      & 0.99  & 0.99  \\
11 &                             & 0.86  & -     & -     & 0.99  & 0.99                   & 0.99  & 0.97  \\
12 &                             & 0.86  & -     & -     & 0.96  & 0.96                     & 0.99  & 0.98  \\
13 &                             & 0.86  & -     & -     & 0.94  & 0.94                   & 0.98  & 0.97  \\
14 &                             & 0.41   & -     & -     & 0.99  & 0.99                   & 0.97  & 0.99  \\
15 &                             & 0.41   & -     & -     & 0.93  & 0.93                    & 0.96  & 0.96  \\ \midrule
1  & \multirow{15}{*}{$\beta_0^*$} & -1    & 0.99  & 0.98  & 1.00     & 1.00                        & 1.00     & 1.00   \\
2  &                             & -1    & 0.99  & 0.99  & 0.99  & 1.00                       & 0.95  & 0.93  \\
3  &                             & -1    & 1.00     & 0.99  & 0.98  & 0.96                 & 0.99  & 0.95  \\
4  &                             & -1    & 0.84  & 0.89  & 0.98  & 0.97                  & 0.94  & 0.93  \\
5  &                             & -1    & 0.94  & 0.94  & 0.96  & 0.97                  & 0.99  & 0.96  \\
6  &                             & -1    & 0.96  & 0.97  & 0.98  & 0.88                    & 0.89  & 0.89  \\
7  &                             & -1    & 0.98  & 0.98  & 0.99  & 0.98                    & 0.94  & 0.93  \\
8  &                             & -1    & 0.89  & 0.87  & 0.98  & 0.94                    & 0.93  & 0.91  \\
9  &                             & -1    & 0.93  & 0.85  & 0.98  & 0.92                   & 0.95  & 0.94  \\
10 &                             & -1    & -     & -     & 0.99  & 0.99                      & 1.00     & 0.99  \\
11 &                             & -1    & -     & -     & 0.99  & 0.99                    & 0.99  & 0.97  \\
12 &                             & -1    & -     & -     & 0.96  & 0.96                    & 0.99  & 0.98  \\
13 &                             & -1    & -     & -     & 0.93  & 0.93                    & 0.98  & 0.98  \\
14 &                             & -0.25 & -     & -     & 1.00     & 1.00                        & 1.00     & 1.00     \\
15 &                             & -0.25 & -     & -     & 0.95  & 0.95                    & 0.95  & 0.96  \\ \midrule
1  & \multirow{15}{*}{$\beta_1^*$} & 0.5   & 0.96  & 0.96  & 0.97  & 0.99                     & 0.97  & 0.97  \\
2  &                             & 0.5   & 0.99  & 0.99  & 0.94  & 0.95                   & 0.98  & 0.92  \\
3  &                             & 0.5   & 0.99  & 0.98  & 0.97  & 0.95                   & 0.99  & 0.97  \\
4  &                             & 0.5   & 0.99  & 0.98  & 0.96  & 0.97                   & 0.93  & 0.91  \\
5  &                             & 0.5   & 1.00     & 0.95  & 0.98  & 0.94                    & 0.96  & 0.9   \\
6  &                             & 0.5   & 0.98  & 0.94  & 0.96  & 0.94                   & 0.91  & 0.90   \\
7  &                             & 0.5   & 0.97  & 0.94  & 0.99  & 0.97                    & 0.92  & 0.91  \\
8  &                             & 0.5   & 0.95  & 0.91  & 0.96  & 0.89                   & 0.93  & 0.92  \\
9  &                             & 0.5   & 0.98  & 0.9   & 0.96  & 0.89                   & 0.96  & 0.94  \\
10 &                             & 0.5   & -     & -     & 1.00     & 1.00                         & 0.99  & 0.99  \\
11 &                             & 0.5   & -     & -     & 0.99  & 0.99                    & 0.99  & 0.97  \\
12 &                             & 0.5   & -     & -     & 0.95  & 0.95                      & 0.98  & 0.98  \\
13 &                             & 0.5   & -     & -     & 0.94  & 0.94                      & 0.97  & 0.97  \\
14 &                             & 0.25  & -     & -     & 0.99  & 0.99                     & 0.97  & 0.99  \\
15 &                             & 0.25  & -     & -     & 0.93  & 0.93                      & 0.96  & 0.96  \\ \midrule
1  & \multirow{15}{*}{$\rho$}     & 0.1   & 1.00     & 1.00     & 0.99  & 1.00                      & 1.00     & 0.99  \\
2  &                             & 0.2   & 1.00     & 1.00     & 0.98  & 0.99                  & 0.95  & 0.93  \\
3  &                             & 0.3   & 1.00     & 1.00     & 0.98  & 0.96                   & 0.99  & 0.92  \\
4  &                             & 0.4   & 0.99  & 0.99  & 0.99  & 0.96                  & 0.94  & 0.95  \\
5  &                             & 0.5   & 0.97  & 0.95  & 0.92  & 0.92                   & 0.93  & 0.92  \\
6  &                             & 0.6   & 0.97  & 0.95  & 0.96  & 0.90                     & 0.94  & 0.91  \\
7  &                             & 0.7   & 0.96  & 0.92  & 0.98  & 0.95                   & 0.88  & 0.88  \\
8  &                             & 0.8   & 0.95  & 0.91  & 0.99  & 0.98                    & 0.95  & 0.91  \\
9  &                             & 0.9   & 0.91  & 0.83  & 1.00     & 0.95                    & 0.97  & 0.97  \\
10 &                             & 0.1   & -     & -     & 0.99  & 0.99                    & 1.00     & 0.96  \\
11 &                             & 0.5   & -     & -     & 0.99  & 0.99                  & 1.00     & 0.98  \\
12 &                             & 0.1   & -     & -     & 0.98  & 0.98                    & 0.98  & 0.99  \\
13 &                             & 0.5   & -     & -     & 0.95  & 0.95                      & 0.95  & 0.96  \\
14 &                             & 0.1   & -     & -     & 1.00     & 1.00                        & 0.98  & 0.99  \\
15 &                             & 0.5   & -     & -     & 0.93  & 0.93                    & 0.95  & 0.93 \\
\toprule
\end{tabular}%
}
\caption{Coverage probabilities of log($\tau(\mathbf 1, \mathbf 0$)), $\beta_0^*$, $\beta_1^*$, and $\rho$ for IV\textunderscore DD approach under varying variances of $\beta_1$ prior for 15 simulation scenarios with 3 duration lengths.  Scenario \# combining $L$ corresponds to scenario labels in the main text, e.g., 1000L1 corresponds to scenario 1 with $L = 1000$. Scenarios marked with `-' are not run.}
\label{tab:sim_covp}
\end{table}

\begin{table}[H]
\resizebox{\textwidth}{!}{%
\begin{tabular}{cccccccccc}
\toprule
\multicolumn{1}{l}{\multirow{2}{*}{Scenario\#}} & \multicolumn{3}{c}{L = 50} & \multicolumn{3}{c}{L = 200} & \multicolumn{3}{c}{L = 1000} \\ 
\cmidrule(lr){2-4} 
\cmidrule(lr){5-7}
\cmidrule(lr){8-10}
\multicolumn{1}{l}{} & \multicolumn{1}{l}{Deviance} & \multicolumn{1}{l}{No. events} & \multicolumn{1}{l}{No. outcome changes} & \multicolumn{1}{l}{Deviance} & \multicolumn{1}{l}{No. events} & \multicolumn{1}{l}{No. outcome changes} & \multicolumn{1}{l}{Deviance} & \multicolumn{1}{l}{No. events} & \multicolumn{1}{l}{No. outcome changes} \\
\hline
1 & 0.52 & 0.46 & 0.46 & 0.55 & 0.46 & 0.52 & 0.54 & 0.46 & 0.50 \\
2 & 0.53 & 0.45 & 0.43 & 0.57 & 0.44 & 0.47 & 0.57 & 0.44 & 0.47 \\
3 & 0.58 & 0.47 & 0.49 & 0.61 & 0.41 & 0.46 & 0.61 & 0.42 & 0.46 \\
4 & 0.67 & 0.45 & 0.37 & 0.64 & 0.38 & 0.44 & 0.64 & 0.39 & 0.45 \\
5 & 0.68 & 0.43 & 0.39 & 0.68 & 0.36 & 0.40 & 0.68 & 0.36 & 0.42 \\
6 & 0.72 & 0.44 & 0.42 & 0.68 & 0.32 & 0.38 & 0.69 & 0.32 & 0.39 \\
7 & 0.72 & 0.44 & 0.42 & 0.71 & 0.29 & 0.35 & 0.72 & 0.30 & 0.35 \\
8 & 0.67 & 0.40 & 0.41 & 0.73 & 0.25 & 0.29 & 0.73 & 0.26 & 0.30 \\
9 & 0.66 & 0.37 & 0.36 & 0.75 & 0.22 & 0.25 & 0.75 & 0.22 & 0.25\\
\toprule
\end{tabular}%
}
\caption{Medians of Bayesian p values for the three test quantities in the posterior predictive checks
for simulation studies, shown as a boxplot in Figure \ref{fig:predcheck_boxplot_sim}.}
\label{tab:pB_sim}
\end{table}

%%%%%  Figures

\begin{figure}[H]
    \centering
    \includegraphics[width=0.4\textwidth]{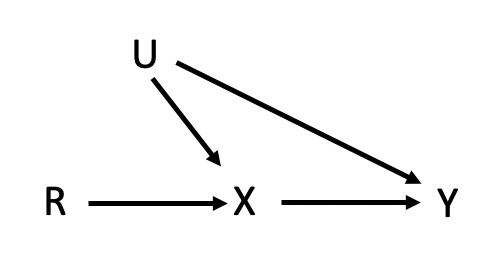}
    \caption{A randomized trial with treatment assignment $R$, observed treatment $X$, outcome $Y$, and unmeasured confounders $U$.}
    \label{fig:iv_DAG}
\end{figure}

\begin{figure}[H]
    \centering
    \includegraphics[width = 0.4\textwidth]{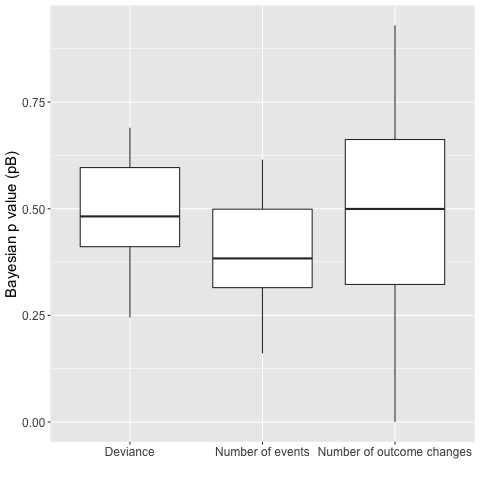}
    \caption{Bayesian p-values for the three test quantities in the posterior predictive checks for all ISTOP individual participant analyses. Predictive checks for the ten selected participants are presented in Figure \ref{fig:pcheck_10pt}. }
    \label{fig:predcheck_boxplot_real}
\end{figure}

\begin{figure}[H]
    \centering
    %\includepdf[pages={1-},scale=0.75]{Real_predictive_check_var4.pdf}
      \vspace{-1cm} 
           \includegraphics[width = 0.9\textwidth]{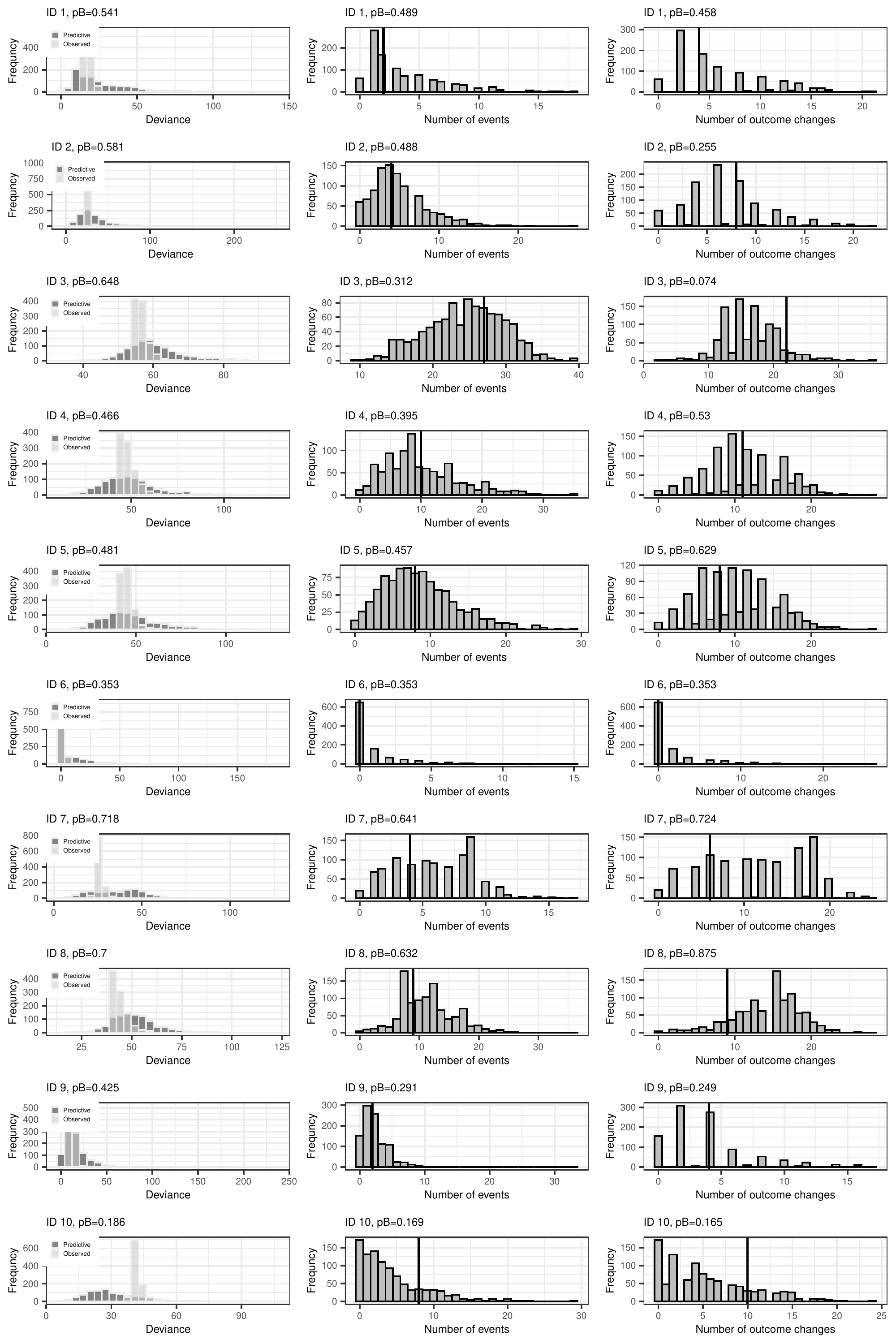}
    \caption{Posterior predictive checks for the ten selected individual-participant analysis. Starting from the left: Distribution of deviance from observed data plotted against the posterior draws; Observed number of events (AF) against the distribution of the number of events in the posterior draws; Observed number of outcome changes against the distribution of the number of outcome changes in the posterior draws. pB are Bayesian p values. }
           \label{fig:pcheck_10pt}
\end{figure}

\begin{figure}[H]
    \centering
    \includegraphics[width = 0.8\textwidth]{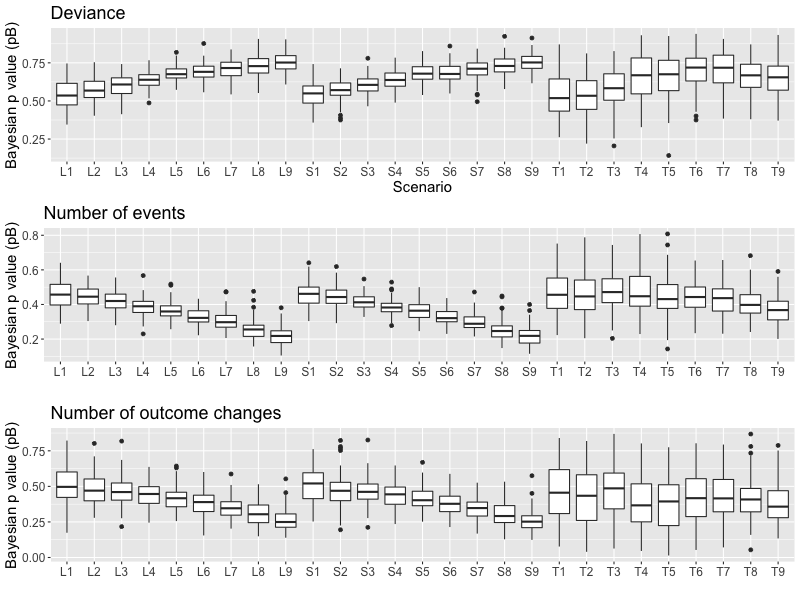}
    \caption{Bayesian p values for the three test quantities in the posterior predictive checks for simulation studies. }
    \label{fig:predcheck_boxplot_sim}
\end{figure}

\begin{figure}[H]
    \centering
    \includegraphics[width=0.6\textwidth]{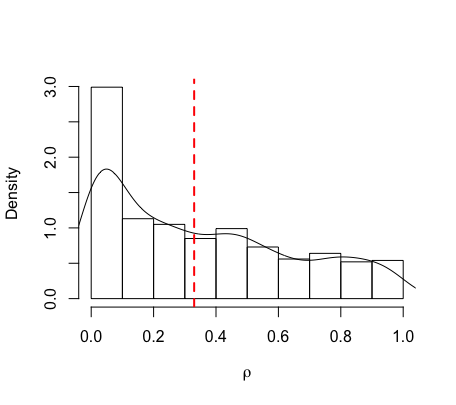}
    \caption{Prior distribution of $\rho$, where $\sqrt{\rho}\sim $Uniform(0,1). Mean of $\rho$ is 0.33, indicated by the red dashed line. }
    \label{fig:rho_prior}
\end{figure}

\subsection{JAGS Code for Model CO} \label{section: appd_code}
\begin{lstlisting}
model {
  
  #N = 1 for 1-patient model
  for(i in 1:N){
      prec_eta[i] <- 1/(1-rho[i])
      prec_e[i] <- prec_u[i]/(1-rho_u[i]^2)
      
      #j=1
      u[i,1] ~ dnorm(0, prec_u[i])
      for(j in 2:J){
        u[i,j] <- rho_u[i]*u[i,j-1] + e[i,j]
        e[i,j] ~ dnorm(0,prec_e[i])
      }
 
      #j = 1
      Xs[i,1] ~ dnorm(mu_x[i,1],prec_eta[I]) #latent Xs
      X[i,1] ~ dinterval(Xs[i,1],0)  #observed X
      mu_x[i,1] <- alpha0[i] + alpha1[i]*R[i,1] + u[i,1]
      
      mu_y[i,1] <- beta0[i] + beta1[i]*X[i,1]+ u[i,1]
      Ys[i,1] ~ dnorm(mu_y[i,j],prec_eta[I])  #latent Ys
      Y[i,1]  ~  dinterval(Ys[i,1],0)  #observed Y
      
    for(j in 2:J){
      Xs[i,j] ~ dnorm(mu_x[i,j],prec_eta[i])
      X[i,j] ~ dinterval(Xs[i,j],0)
      mu_x[i,j] <- alpha0[i] + alpha1[i]*R[i,j] + u[i,j]
    
      mu_y[i,j] <- beta0[i] + beta1[i]*X[i,j]+ beta2[i]*X[i,j-1]+ u[i,j]
      Ys[i,j] ~ dnorm(mu_y[i,j],prec_eta[i])
      Y[i,j]  ~  dinterval(Ys[i,j],0)
    }
    mu_y1[i] <- beta0[i] + beta1[i]
    mu_y0[i] <- beta0[i]
    p_y0[i] <- pnorm(mu_y0[i], 0, 1)
    p_y1[i] <-pnorm(mu_y1[i], 0, 1)
    log_CRR[i] <- log(p_y1[i])- log(p_y0[i])
    log_COR[i] <- log(p_y1[i]) +log(1-p_y0[i])-log(p_y0[i])- log(1-p_y1[i])
  }
  
  ########
  #priors#
  ########
  for(i in 1:N){
    alpha0[i] ~ dnorm(alpha0_mean_prior,1)T(-trunc.var,trunc.var)
    alpha1[i] ~ dnorm(alpha1_mean_prior,1)T(-trunc.var,trunc.var)
    beta0[i] ~ dnorm(beta0_mean_prior,1)T(-trunc.var,trunc.var)
    beta1[i] ~ dnorm(0,1/coef.var)T(-trunc.var,trunc.var)
    beta2[i] ~ dnorm(0,1/coef.var)T(-trunc.var,trunc.var)
    sigma_e[i] <- sqrt(1/prec_e[i])
    sqrt_rho[i] ~  dunif(-0.999, 0.999)
    rho[i]  <- (sqrt_rho[i])^2
    rho_u[i] ~  dunif(-0.99,0.99)
    prec_u[i] <- 1/(rho[i] + 0.001) #avoid boundary issues
  }
}
\end{lstlisting}

\end{document}

\section{Coefficient}
Replace $a\tilde u_{tj}$ or $b\tilde u_{tj}$ with $\sqrt{ab}\tilde u_{tj}$ in Eq.(15) (arx 5.5.pdf), and use $\tilde{e}^x_{tj}$ $\tilde{e}^y_{tj}$ to denote new error terms, i.e., 
\begin{align}
     X_{tj}^\dagger &=  \alpha_{0} + \alpha_{1}R_{tj}  + \sqrt{ab}\tilde u_{tj} +  \tilde{e}^x_{tj},\nonumber \\ 
     Y_{tj}^\dagger &=  \beta_{0} + \beta_{1}X_{tj}  +  sgn(ab)\sqrt{ab}\tilde u_{tj} + \tilde{e}^y_{tj},
\end{align}
Here Var($X_{tj}^\dagger$) = $\sigma_x^2$ and Var($Y_{tj}^\dagger$) = $\sigma_y^2$ remain the same as for $X_{tj}^\dagger$ and $Y_{tj}^\dagger$ in Eq.(15). The transformation should not change the marginal variances $\sigma_x^2$ and $\sigma_y^2$, but rather re-distribute error variance on $\sqrt{a}\tilde u_{tj} + \tilde{\epsilon}^x_{tj}$. After transformation, the variance of $\sqrt{ab}\tilde u_{tj}$ is just covariance between $X_{tj}^\dagger$ and $Y_{tj}^\dagger$. The difference in variances between $\sqrt{ab}\tilde u_{tj}$ and $\sqrt{a}\tilde u_{tj}$ goes to the new error term $\tilde{e}^x_{tj}$ . 
Then, 
$$\E(\tilde{u}_{tj}|X_{tj}^\dagger) = \frac{\sqrt{ab}\tilde\sigma_u^2}{\sigma_x^2}(X_{tj}^\dagger - \alpha_0 - \alpha_1R_{tj} ).$$

\begin{align}
&\E(Y_{tj}^\dagger|X_{tj}) \nonumber \\
 &= \beta_0 + \beta_1X_{tj} + \sqrt{ab}\frac{\sqrt{ab}\tilde\sigma_u^2}{\sigma_x^2}(\E (X_{tj}^\dagger|X_{tj}) - \alpha_0 - \alpha_1R_{tj}).   \\
 & = \beta_0 + \beta_1X_{tj} + \frac{ab\tilde\sigma_u^2}{\sigma_x^2}(\E (X_{tj}^\dagger|X_{tj}) - \alpha_0 - \alpha_1R_{tj}). 
      \label{appd_eq}
\end{align}

We argue that there is no bias for estimating $\beta_1$ which represents the treatment effect. But I am not sure in estimating odds ratio, which involves $\beta_0$, how or whether $\frac{ab\tilde\sigma_u^2}{\sigma_x^2}(\E (X_{tj}^\dagger|X_{tj}) - \alpha_0 - \alpha_1R_{tj})$ affects. Even if it affects, the bias would involve the product $\frac{ab\tilde\sigma_u^2}{\sigma_x^2}$ not $a,b,\tilde\sigma_u^2, \sigma_x^2$ by itself.
 $\frac{\alpha_0}{\sigma_x^2}$, $\frac{\alpha_1}{\sigma_x^2}$, and $\frac{ab\tilde\sigma_u^2}{\sigma_x^2}$ can all be estimated from the model.